\pdfoutput=1
\documentclass[11pt]{article}

.5 scaled \magstep4
\font\medio=cmr9.5 scaled \magstep2
\outer\def\beginsection#1\par{\medbreak\bigskip
      \message{#1}\leftline{\bf#1}\nobreak\medskip
\vskip-\parskip
      \noindent}
\usepackage{graphicx} 
\begin{document}
\bibliographystyle{unsrt}

\begin{center}
{\Large {\bf Relic gravitons and pulsar timing arrays: a  theoretical viewpoint}}\\
\vspace{15mm}
 Massimo Giovannini 
 \footnote{Electronic address: massimo.giovannini@cern.ch} \\
\vspace{1cm}
{{\sl  INFN, Section of Milan-Bicocca, 20126 Milan, Italy}}
\vspace*{1cm}
\end{center}
    
\centerline{\medio  Abstract}
\vspace{5mm}
During the last three years the pulsar timing arrays reported a series of repeated evidences of gravitational radiation (with stochastically distributed Fourier amplitudes) at a benchmark frequency of the order of $30$ nHz and characterized by spectral energy densities (in critical units) ranging between $10^{-8}$ and $10^{-9}$. While it is still unclear whether or not these effects 
are just a consequence of the pristine variation of the space-time curvature, the nature of the underlying physical processes would suggest that the spectral energy density of the relic gravitons in the nHz domain may only depend on the evolution of the comoving horizon at late, intermediate and early times. Along this systematic perspective we first consider the most conventional option, namely a post-inflationary modification of the expansion rate. Given the present constraints on the relic graviton backgrounds, we then show that such a late-time effect is unable to produce the desired hump in the nHz region. We then analyze a modified exit of the relevant wavelengths as it may happen when the gravitons inherit an effective refractive index from the interactions with the geometry. A relatively short inflationary phase leads, in this case, to an excess in the nHz region even if the observational data coming from competing experiments do not pin down exactly the same regions in the parameter space. We finally examine an early stage of increasing curvature and argue that it is not compatible with the observed spectral energy density unless the wavelengths crossing the comoving horizon at early times reenter in a decelerated stage not dominated by radiation. 
\vskip 0.5cm

\nonumber
\noindent

\vspace{5mm}

\vfill
\newpage

\tableofcontents
\newpage
\renewcommand{\theequation}{1.\arabic{equation}}
\setcounter{equation}{0}
\section{Introduction}
\label{sec1}
The low-frequency gravitons notoriously affect the the propagation of electromagnetic signals so that the temperature and the polarization anisotropies of the cosmic microwave background (CMB) are in fact detectors of relic  gravitational waves in the aHz range\footnote{The scale factor is normalized at the present conformal time as $a(\tau_{0}) = a_{0} =1$ so that comoving and physical frequencies coincide for $\tau\to\tau_{0}$. Moreover the standard prefixes of the International System of Units are consistently employed, e.g.   $1\,\mathrm{aHz} = 10^{-18}, \mathrm{Hz}$, 
$1\, \mathrm{nHz} = 10^{-9}\, \mathrm{Hz}$ and similarly for all the other frequency domains of the spectrum. } (see, for instance, \cite{wein}).  Even though the physical nature of the effects remains practically the same, the nHz range exceeds the Cosmic Microwave Background (CMB) frequencies by roughly $10$ orders of magnitude and, as suggested 
long ago \cite{PP1a,PP1b,PP1c}, the millisecond pulsars can be employed as effective detectors of random gravitational waves for a typical domain that corresponds to the inverse of the observation time during which the pulsar timing has been monitored. More specifically, Sazhin \cite{PP1a} suggested that the arrival times of pulsar's pulses could be used for direct searches of gravitational radiation and shortly after Detweiler \cite{PP1b} made a similar point by deriving one of  the first upper limits on the relic gravitational waves in the nHz range. This aspect has been more accurately reinstated in Ref. \cite{PP1c} where the authors also derived an important property of the isotropic backgrounds of gravitational radiation stipulating that the signal coming, in our context, from relic gravitons should be correlated across the baselines while that from others noises will not.  This effect depends on the angle between a pair of Earth-pulsars baselines and it is often dubbed by saying that the correlation signature of an isotropic and random gravitational wave background follows the so-called Hellings-Downs curve \cite{PP1c}. If the gravitational waves {\em are not} characterized by stochastically distributed Fourier amplitudes the corresponding signal {\em does not necessarily follow} the Hellings-Downs correlation.

By using this logic, since the mid 1990s a series upper 
limits on the spectral energy density of the relic gravitons in the nHz range has been obtained \cite{PP2a,PP2b,PP2c,PP2d} and during the last three years 
the pulsar timing arrays (PTA) reported an evidence that could be attributed to isotropic backgrounds of gravitational radiation \cite{NANO1,PPTA1,EPTA1,IPTA1}. The PTA collaborations\footnote{A generic pulsar timing array is in fact a  series of millisecond pulsars that are monitored with a rhythm that  depends on the 
specific experiment. We refer here, in particular, to the NANOgrav collaboration \cite{NANO1}, to the Parkes Pulsar Timing array (PPTA) \cite{PPTA1}, to the European Pulsar Timing array (EPTA) \cite{EPTA1}; the PTA data have been also combined in the consortium named International Pulsar Timing array (IPTA) \cite{IPTA1}. } customarily assign the chirp amplitude at a reference frequency $\nu_{ref} = 31.68\,\, \mathrm{nHz}$ that corresponds to $\mathrm{yr}^{-1}$:
\begin{equation}
h_{c}(\nu,\tau_{0}) = \,Q \,\biggl(\frac{\nu}{\nu_{ref}}\biggr)^{\beta}, \qquad \qquad \nu_{ref} = \frac{1}{\mathrm{yr}}=   31.68\,\, \mathrm{nHz}.
\label{NOTT1}
\end{equation}
The pivotal models analyzed in Refs. \cite{NANO1,PPTA1,EPTA1,IPTA1} assumed $\beta = -2/3$ or, which is the same\footnote{Even if the various collaborations use indifferently $\overline{\gamma}$ and $\beta$ we shall refrain from this practice since $\overline{\gamma}$ could be confused with a different physical quantity (denoted simply by $\gamma$) that shall be introduced later on in the discussion. For the record we recall that the relation between $\overline{\gamma}$ and $\beta$ is simply given by since $\beta = (3-\overline{\gamma})/2$.}, $\overline{\gamma} = 13/3$.  In Eq. (\ref{NOTT1}) the value of $Q$ inferred from the observational data should be ${\mathcal O}(10^{-15})$ so that we find it natural to parametrize the amplitude at $\nu_{ref}$ as $Q= q_{0} \times 10^{-15}$. In the previous data releases the  $q_{0}$ ranged between $1.92$ and $5.13$ depending on the values of $\beta$ \cite{NANO1,PPTA1,EPTA1,IPTA1}. Even more recently the new data of the PTA collaborations have been released \cite{PPTA2,NANO2,EPTA2} together with the first determinations of the Chinese Pulsar Timing array (CPTA) \cite{CPTA}. The Parkes PTA collaboration considered $30$ millisecond pulsars spanning $18$ years of observations; they estimated 
$q_{0}= 3.1^{1.3}_{-0.9}$ with a spectral index $\beta = -0.45\pm 0.20$ \cite{PPTA2}; for a spectral 
index $\beta =-2/3$ the collaboration obtains instead $q_{0} =2.04_{-0.22}^{0.25}$ which is compatible with 
the determinations of the previous data releases \cite{PPTA1}.  The Parkes PTA collaboration did not claim the detection 
of the Hellings-Downs correlation \cite{PPTA2} and carefully considered the issues related to time-dependence of the common noise. The conclusions of the Parkes PTA  seem significantly more conservative than the one of the NANOgrav collaboration examining $67$ millisecond pulsars in the last $15$ years. The NANOgrav experiment  claimed the detection of the Hellings-Downs correlation \cite{NANO2} but the inferred values of the spectral 
parameters are slightly different from the ones of PPTA since $q_{0} = 6.4^{+4.2}_{-2.7}$
and $\beta = -0.10 \pm 0.30$ \cite{NANO2}. The chirp amplitude of Eq. (\ref{NOTT1}) can be
related to the spectral energy density evaluated for typical frequencies larger than the Hubble rate at the present time \cite{MG1}
\begin{equation}
h_{0}^2 \, \Omega_{gw}(\nu_{ref},\tau_{0}) = 6.287 \times 10^{-10} \, \, q_{0}^2,
\label{NOTT2}
\end{equation}
implying $h_{0}^2 \, \Omega_{gw}(\nu_{ref},\tau_{0}) = {\mathcal O}(2.57) \times 10^{-8}$ in the case of Ref. \cite{NANO2} (for $q_{0} =6.4$) and $h_{0}^2 \, \Omega_{gw}(\nu_{ref},\tau_{0}) = {\mathcal O}(6.04)\times 10^{-9}$ for Ref. \cite{PPTA2} (for $q_{0} =3.1$). The connection between $h_{0}^2 \, \Omega_{gw}(\nu_{ref},\tau_{0})$ and $h_{c}(\nu,\tau_{0})$ has been explicitly discussed in Eqs. (\ref{APP11})--(\ref{APP12}) and (\ref{APP13}) of appendix \ref{APPA}.

Even if the interpretation of various collaborations preferentially focuses on a background produced by diffuse astrophysical sources at the present time\footnote{This situation would actually correspond to the benchmark case $\beta \to -2/3$ in Eq. (\ref{NOTT1}) as it happens in the case of binary systems formed by black-holes.}, in this paper we are going to examine the qualitatively different hypothesis  that the excesses summarized by Eqs. (\ref{NOTT1})--(\ref{NOTT2}) are caused by the parametric amplification of the quantum fluctuations of the tensor modes of the geometry. Since the relic gravitons are only coupled to the evolution of the space-time curvature Grishchuk \cite{gr1,gr2}, Ford and Parker \cite{par1} and some others first outlined that the formation of a diffuse backgrounds of gravitational radiation are a generic consequence of quantum mechanics in cosmological space-times.  Depending on the early variation the space-time curvature, relic gravitons are then expected in different dynamical situations and, in particular, during a stage of de Sitter and quasi-de Sitter expansion \cite{star1}. 

Considering that, at early times, the conventional expanding stages of the concordance paradigm are complemented by an inflationary epoch, $h_{0}^2 \, \Omega_{gw}(\nu,\tau_{0})$ is a monotonically decreasing function of the comoving frequency from the aHz range to the MHz domain. More precisely between few aHz and $100$ aHz the spectral energy density in critical units scales approximately as $\nu^{-2}$ \cite{LL1,LL2,LL3}; for higher frequencies $h_{0}^2 \Omega_{gw}(\nu,\tau_{0})$ exhibits a quasi-flat plateau whose precise slope depends in fact upon the parameter $r_{T}$ measuring the ratio between the tensor power spectrum and its scalar counterpart. In the case of single-field inflationary scenarios (often taken as the benchmark class of models by some observers \cite{TS1,TS2,TS3}) the consistency relations stipulate that the tensor spectral index $m_{T}$ and the slow-roll parameter $\epsilon_{k}$ are both determined by the value of $r_{T}$ according to the following (approximate) chain  of equalities $ m_{T} \simeq - r_{T}/8 \simeq - 2 \epsilon_{k}$ that are customarily referred to as the consistency conditions.  The simplest version of the concordance scenario includes only one further free parameter, namely the ratio $r_{T}(k_{p})$  describing the tensor component of the large-scale inhomogeneity at a conventional pivot scale that coincides, in what follows, with $k_{p} = 0.002\,\, \mathrm{Mpc}^{-1}$.  The addition of a single tensor component (only described by $r_{T}$) allows for an accurate set of limits implying that $r_{T} \leq 0.06$ or even $r_{T}\leq 0.03$ \cite{TS1,TS2,TS3}.  Moreover, the pivot wavenumber $k_{p}$ corresponds to $\nu_{p} = k_{p}/(2\pi) = 3.09\,\,\mathrm{aHz}$ and this is why the limits on $r_{T}(k_{p})$ may be translated into constraints on the spectral energy density of the relic gravitons in the aHz range. 

For typical comoving frequencies of the order of the nHz the current bounds on $r_{T}$ suggest that in the concordance paradigm $h_{0}^2 \Omega_{gw}(\nu, \tau_{0})$ must be ${\mathcal O}(10^{-17})$ (or smaller); in this case all the relevant wavelengths exit in the inflationary stage and reenter during the radiation phase. Furthermore, the accurate estimate of $h_{0}^2 \Omega_{gw}(\nu, \tau_{0})$ in the nHz region involves a number of late-time effects that may reduce even further the overall 
amplitude of the spectral energy density: the corrections due to neutrino free-streaming \cite{NU1} (see also, for instance, \cite{NU2,NU3,NU4}) suppress $h_{0}^2 \Omega_{gw}(\nu, \tau_{0})$ for $\nu <\nu_{bbn}$ where $\nu_{bbn} = {\mathcal O}(10^{-2})\,\, \mathrm{nHz}$ is the typical frequency associated with big-bang nucleosynthesis. Thus in the nHz range the relic gravitons produced within the conventional lore are nine or ten orders of magnitude smaller than the figures reported in Eq. (\ref{NOTT2}).  This means that if the nHz excess is caused by relic gravitons amplified by the evolution of the space-time curvature the relevant time-scale of the 
problem is primarily given by $\tau_{k}$ defining the moment at which the wavelength associated with $\nu_{ref}$ crossed the comoving Hubble radius {\em after} the end of inflation. A simple estimate suggests that  $\tau_{k}$ is just a fraction of the time-scale of big-bang nucleosynthesis\footnote{We are here 
considering a standard thermal history where $g_{s,\,eq} = 3.94$ while $g_{\rho,\, bbn} = 
g_{s,\,bbn}= 10.75$; as usual $g_{\rho}$ and $g_{s}$ are, respectively, the relativistic degrees of freedom associated with the energy density and with the entropy density of the plasma; $\Omega_{R0}$ and  $T_{bbn}$  denote, respectively, the energy density of the relativistic species in the concordance paradigm and the big-bang nucleosynthesis temperature.}
\begin{equation}
\frac{\tau_{k}}{\tau_{bbn}} = 3.40 \times 10^{-2} \biggl(\frac{\nu_{PTA}}{31.68 \, \mathrm{nHz}}\biggr)^{-1}\, \biggl(\frac{T_{bbn}}{\mathrm{MeV}}\biggr) \, \biggl(\frac{h_{0}^2 \Omega_{R0}}{4.15 \times 10^{-5}}\biggr)^{1/4}.
\label{NOTT3}
\end{equation}
In what follows $\nu_{PTA}$ denote the frequency of the PTA and it can be either 
slightly larger or smaller than $\nu_{ref}$; when $\nu_{PTA} > \nu_{ref}$  the corresponding wavelength crossed the comoving 
horizon even earlier. The second relevant scale of the problem is given by the ratio between $\nu_{PTA}$ 
and the expansion rate at the end of inflation:
\begin{equation}
\frac{\nu_{PTA}}{a_{1} \, H_{1}} = 2.05 \times 10^{-17} \biggl(\frac{\nu_{PTA}}{31.68 \, \mathrm{nHz}}\biggr)\,\biggl(\frac{h_{0}^2 \Omega_{R0}}{4.15 \times 10^{-5}}\biggr)^{-1/4} \, \biggl(\frac{r_{T}}{0.03}\biggr)^{-1/4} \, \biggl(\frac{{\mathcal A}_{{\mathcal R}}}{2.41\times 10^{-9}}\biggr)^{-1/4},
\label{NOTT4}
\end{equation}
where ${\mathcal A}_{{\mathcal R}}$ denotes the amplitude of the curvature inhomogeneities 
at the pivot scale $k_{p}$.  From Eqs. (\ref{NOTT3})--(\ref{NOTT4}) it follows that any modification of the post-inflationary evolution is unlikely to produce a hump for frequencies ${\mathcal O}(\nu_{ref})$; interesting enhancements may arise for $\nu > \mathrm{mHz} \gg \nu_{ref}$ implying that the post-inflationary evolution is indeed 
able to increase the spectral energy density between the mHz and the MHz \cite{MG2,MG3,MG4} but not in the 
nHz range. As we shall see Eqs. (\ref{NOTT3})--(\ref{NOTT4}) ultimately imply that an excess comparable 
with Eq. (\ref{NOTT2}) cannot be related to a post-inflationary modification of the comoving horizon.

Before plunging into the analysis it is probably useful to stress that already after the first data releases of the PTA a large number of different (and 
sometimes opposite) explanations for the nHz excess have been proposed. A common characteristic distinguishing the hypothesis 
pursued in this paper and the ones propounded in the current literature is that the nHz excess is typically due 
the presence of late-time sources of anisotropic stress. An example along this direction is represented by cosmic strings whose 
oscillating loops may emit gravitational waves at different epochs ultimately producing a stochastic background \cite{STT3}  with quasi-flat spectral energy density which is typically larger than the inflationary signal. 
For the largest values of the string tension in Planck units there is the possibility of an excess in the nHz region (see e.g. \cite{STT4,STT5}). Various subsequent analyses 
based on these  observations have been proposed more recently with various degrees of success. Another late-time source of gravitational 
radiation is represented by strongly first-order phase transitions that are known to produce spikes at low and intermediate frequencies because of the partial breaking 
of homogeneity due to the nucleation of bubbles of the new phase. The amount of gravitational radiation 
produced by the phase transition depends chiefly on the 
difference between the energy density in the broken and in the symmetric phase. This 
energy density may be comparable with the energy density of the ambient plasma 
(and in this case the phase transition experiences a strong supercooling) or smaller 
than the energy density of the surrounding radiation (and in this case the phase transition 
is mildly supercooled). When the gravitational radiation is produced from the 
collisions of the bubbles of the new phase \cite{CCC1,CCC2,CCC3} the spectral energy density scales like $\nu^3$, reaches a
maximum and then decreases with a power that may be faster than $\nu^{-1}$. The spectral energy density inherits also contribution from the sound waves of the plasma  
\cite{CCC4} and this second component may be even larger than the one due to bubble collisions. There are two points that make this 
explanation difficult. The first one is that the powers of the hump are typically steeper than the ones suggested by the PTA observations.
The second observation concerns the physics of the phase transition. The bursts of gravitational radiation coming from the TeV
are absent at frequencies much lower than the $\mu$Hz also because the electroweak phase transition is {\em not} 
strongly first-order given the values of the Higgs mass. This implies the necessity of more contrived explanations (see e.g. \cite{CCC5} and references 
therein).  There exist also more exotic possibilities where the nHz excess is caused  by modifications of gravity in alternative cosmological scenarios; see, in this respect, \cite{CCC6} and references therein.

The layout of this investigation is the following. In section \ref{sec2} the post-inflationary 
modifications of the expansion rate (related to the reentry of $\lambda_{PTA} \simeq 1/\nu_{PTA}$) are analyzed and confronted with 
the nHz excesses. In section \ref{sec3} we instead examine 
the exit of the PTA scale during inflation and focus on the possibility of a refractive index
 coming from the interactions of the relic gravitons with the background geometry. 
In section \ref{sec4} the general features of the bouncing scenarios (modifying both the exit and the reentry of 
$\lambda_{PTA}$) are specifically examined in the light of a potential nHz signal. 
Section \ref{sec5} contains our concluding remarks. Appendix \ref{APPA} covers the main notations 
employed in the paper and appendix \ref{APPB} is instead focussed on the general form of the effective action of the relic gravitons and on its different parametrizations.

\renewcommand{\theequation}{2.\arabic{equation}}
\setcounter{equation}{0}
\section{The comoving horizon after inflation}
\label{sec2}
\subsection{General considerations}
When all the wavelengths of the relic gravitons are shorter than the comoving Hubble radius at the present time, the spectral energy density in critical units can be expressed as:
\begin{equation}
\Omega_{gw}(k, \tau) = \frac{ 2\, k^4 }{3 \,\pi \, M_{P}^2\, H^2 \, a^4 } \biggl(\frac{a_{re}}{a_{ex}}\biggr)^2 \bigl|{\mathcal Q}_{k}(\tau_{ex}, \tau_{re})\bigr|^2 
\biggl[ 1 + \biggl(\frac{{\mathcal H}_{re}}{k}\biggr)^2 + {\mathcal O}({\mathcal H}/k) + {\mathcal O}({\mathcal H}^2/k^2) \biggr],
\label{STWO1}
\end{equation}
where ${\mathcal H} = a \, H$ and ${\mathcal H}= a^{\prime}/a$; the notations employed here are summarized in appendix \ref{APPA} and, in particular,  in Eqs. (\ref{APP1a}) and (\ref{APP9})--(\ref{APP10}).
Equation (\ref{STWO1}) has been deduced in the limit $a_{re} \gg a_{ex}$ which is valid for all conventional and unconventional inflationary scenarios even if, in some cases, $\Omega_{gw}(k,\tau)$ 
may also depend on the intermediate evolution between $\tau_{ex}$ and $\tau_{re}$ since ${\mathcal Q}_{k}(\tau_{ex}, \tau_{re})$ (whose square modulus enters Eq. (\ref{STWO1})) contains an integral over the conformal 
time coordinate $\tau$:
 \begin{equation}
 {\mathcal Q}_{k}(\tau_{ex}, \tau_{re}) = 1 - ({\mathcal H}_{ex} + i k) \int_{\tau_{ex}}^{\tau_{re}} \frac{a_{ex}^2}{a^2(\tau)} \, \,d\tau. 
 \label{STWO2}
 \end{equation}
 From the evolution of the mode functions of Eq. (\ref{PPSS3a}), $\tau_{ex}$ and $\tau_{re}$ are the roots of the equation
 \begin{equation}
 k^2 = a^2 \, H^2 [ 2 - \epsilon(a)], \qquad \qquad \epsilon = - \dot{H}/H^2,
 \label{STWO3}
 \end{equation}
that ultimately defines the different regimes of the evolution of the mode functions. In a stricter mathematical perspective, $\tau_{re}$ and $\tau_{ex}$ define the turning points 
where the solutions of Eq. (\ref{PPSS3a}) change their asymptotic behaviour. Defining, for the sake of simplicity, $\epsilon_{re}= \epsilon(\tau_{re})$ and $\epsilon_{ex} = \epsilon(\tau_{ex})$, the turning points are regular whenever $\epsilon_{re} \neq 2$ and $\epsilon_{ex} \neq 2$. 
For instance, if the exit occurs during a conventional stage of inflationary expansion (i.e. $\epsilon_{ex} \ll 1$) we have that, by definition,
 $k \simeq a_{ex} \, H_{ex}$ and $\tau_{ex} \simeq 1/k$. Conversely, if the reentry takes place close to the radiation-dominated stage of expansion we should have $\epsilon_{re} \to 2$ in Eq. (\ref{STWO3}) so that  $k \ll a_{re}\, H_{re}$.
 
 The corrections appearing inside the squared bracket of Eq. (\ref{STWO1}) (of the order ${\mathcal H}/k$ and 
 ${\mathcal H}^2/k^2$) are both negligible in the limit $k \tau \gg 1$ when all the relevant wavelengths are inside the comoving 
 horizon. Therefore, if all the wavelengths exit during inflation and reenter in a stage dominated by radiation  (as it happens in the concordance paradigm and in Fig. \ref{FIG1}) Eq. (\ref{STWO2}) implies that ${\mathcal Q}_{k}(\tau_{ex}, \tau_{re}) \to 1$. There can be however physical contingencies where the contribution of the integral in Eq. (\ref{STWO2}) gets larger than $1$ and, as long as the {\em exit} is a regular turning point that the integrals appearing in Eq. (\ref{STWO2}) 
 can  be estimated  as: 
 \begin{equation}
 {\mathcal Q}_{k}(\tau_{ex}, \tau_{re}) = 1 - \int_{\tau_{ex}}^{-\tau_{1}} \frac{a_{ex}^2}{a_{inf}^2(x)} \, d\, x -  \int_{\tau_{-1}}^{\tau_{re}} \frac{a_{ex}^2}{a_{post}^2(x)} \, d\, x,
 \label{STWO2a}
 \end{equation}
where $ x = \tau/\tau_{ex}$ and ${\mathcal H}_{ex} = a_{ex} \, H_{ex} \simeq k$. In Eq. (\ref{STWO2a}) 
we also included two subscripts in the integrand with the purpose of stressing that the corresponding 
contributions may arise during the inflationary and in the post-inflationary stage; we also conventionally assumed, for the sake of concreteness, that the accelerated stage lasts up to $-\tau_{1}$ and it is replaced for $\tau > - \tau_{1}$ by a decelerated evolution.

\subsection{The post-inflationary evolution of the comoving horizon} 
\begin{figure}[!ht]
\centering
\includegraphics[height=7.5cm]{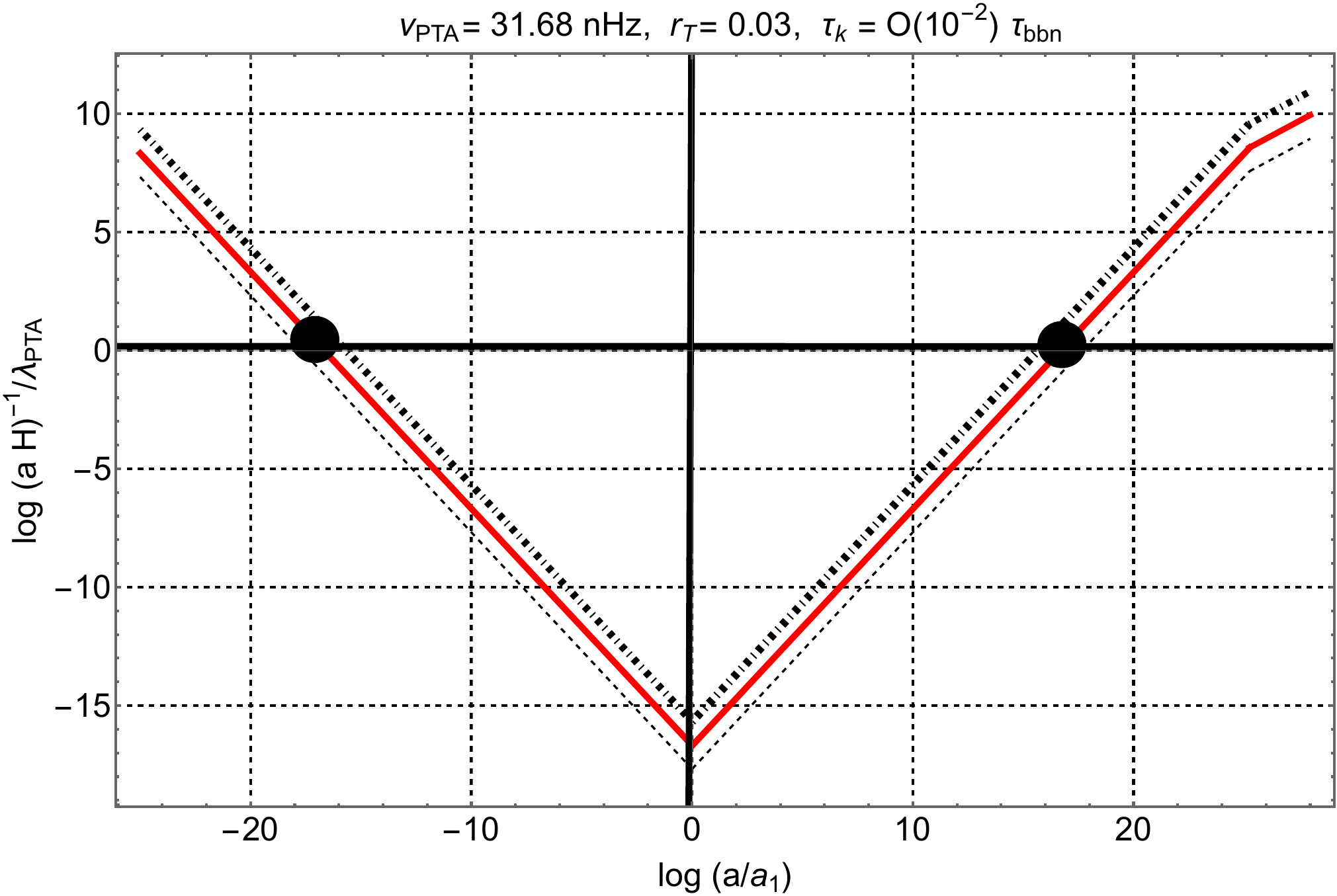}
\caption[a]{The ratio between the comoving horizon and the wavelength $\lambda_{PTA} = {\mathcal O}(0.30) \, \mathrm{pc}$ is illustrated; as already mentioned $\lambda_{PTA}$ corresponds to the comoving 
frequency  $\nu_{PTA} = 31.68 \, \mathrm{nHz}$. Common logarithms are employed on both axes and the two blobs indicate the exit and to the reentry of the corresponding wavelength. The full line is associated with $\nu_{PTA}$ while the dot-dashed and the dashed lines illustrate, respectively, the situations where the frequency is slightly larger (i.e. $10\, \nu_{PTA}$) or slightly smaller (i.e. $0.1\, \nu_{PTA}$) than the benchmark value provided by $\nu_{ref}$. The timeline reported in this cartoon is characteristic of the concordance paradigm where all the wavelengths (including $\lambda_{PTA}$) do their first crossing during inflation (i.e. for $a< a_{1}$) and reenter in the radiation-dominated stage.}
\label{FIG1}      
\end{figure}

\begin{figure}[!ht]
\centering
\includegraphics[height=7.7cm]{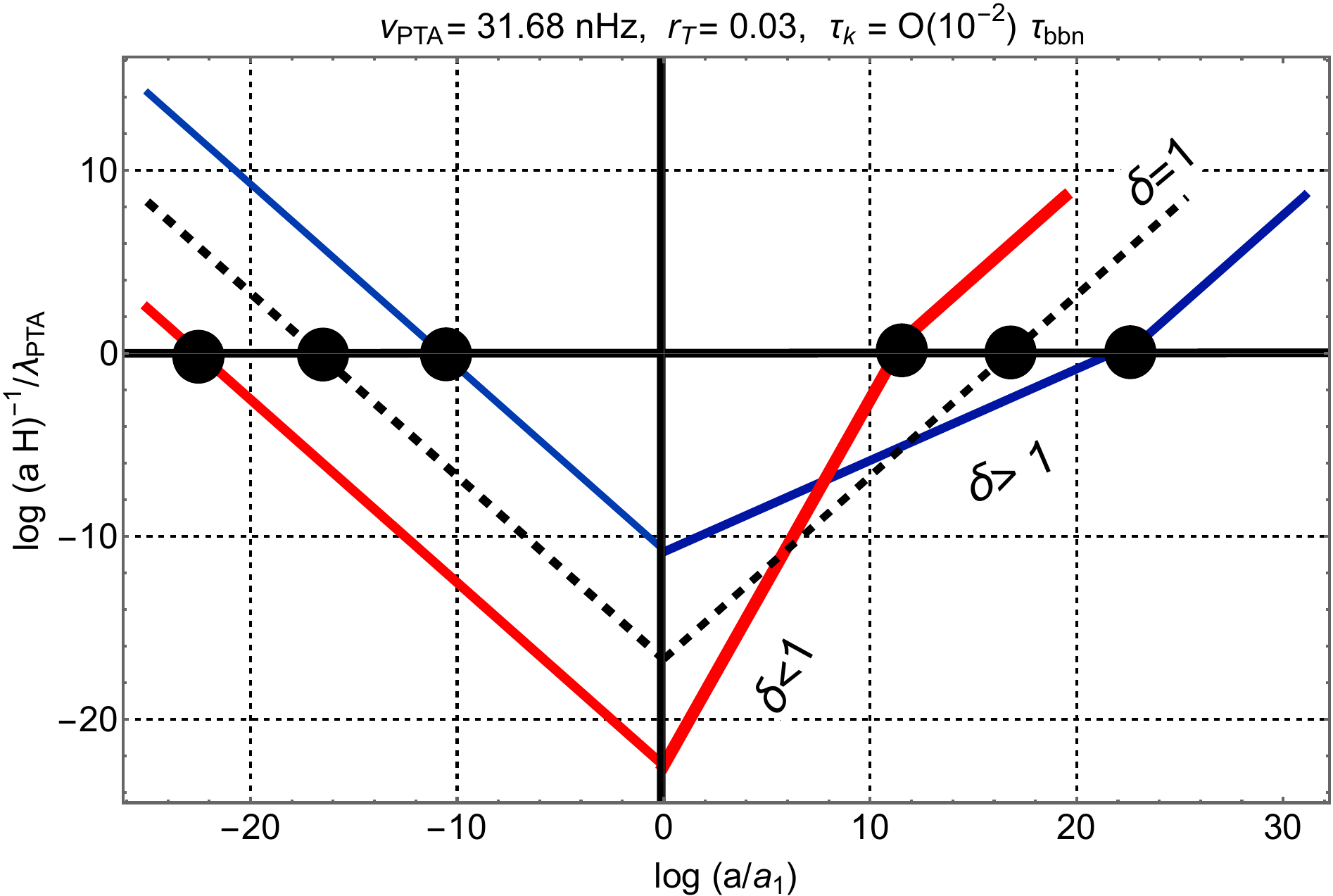}
\caption[a]{As in the case of Fig. \ref{FIG1} the ratio between the comoving horizon and  $\lambda_{PTA}$ is illustrated with the difference that, in the present cartoon, the post-inflationary evolution is {\em not} immediately dominated by radiation; the curve $\delta \to 1$ (dashed line) is also reported for comparison. The post-inflationary stage with $\delta \neq 1$ 
lasts until the crossing time of $\lambda_{PTA}$. On the one hand this choice maximizes the deviations from the standard 
form of the spectral energy density, on the other hand the modified post-inflationary evolution cannot last much longer; this is because the crossing 
time of $\lambda_{PTA}$ is just a fraction of $\tau_{bbn}$ (see Eq. (\ref{NOTT3}) and discussion therein).}
\label{FIG2}      
\end{figure}

 The ratio between the comoving horizon and $\lambda_{PTA}$ is illustrated in Fig. \ref{FIG1} when the relevant bunch of wavelengths exits the comoving horizon during inflation (i.e. $\epsilon_{ex} \ll 1$) and reenters in a radiation-dominated stage of expansion (i.e. $\epsilon_{re} \to 2$); in this case the spectral slope of $\Omega_{gw}(\nu, \tau_{0})$ follows from Eqs. (\ref{STWO1})--(\ref{STWO2a}) 
and by just focussing, for simplicity, on the $k$-dependence 
we can deduce the spectral slope
\begin{equation}
\frac{k^4}{a_{1}^4 \, H_{1}^4} \biggl(\frac{a_{re}}{a_{ex}}\biggr)^2 \biggl(\frac{k}{a_{re}\, H_{re}}\biggr)^{-2} \bigl|{\mathcal Q}_{k}(\tau_{ex}, \tau_{re})\bigr|^2 = \frac{a_{re}^4 \, H_{re}^2}{a^4 \, H^2} \, \biggl(\frac{k}{a_{1} \, H_{1}}\biggr)^{m_{T}}.
\label{STWO2b}
\end{equation}
In Eq. (\ref{STWO2b}) we used that $a_{ex}\, H_{ex} = - 1/[(1 -\epsilon_{ex}) \tau_{ex}]$ 
and $ k \tau_{re} \ll 1$; moreover, by enforcing  the validity of the consistency relations (i.e. $r_{T} = 16 \epsilon_{k}$), the spectral index $m_{T}$ is given by\footnote{When the bunch of wavelengths ${\mathcal O}(k_{p}^{-1})$ exit the comoving horizon during inflation
the slow-roll parameter acquire a specific numerical value that shall be conventionally denoted by $\epsilon_{k}$.}
\begin{equation}
m_{T} = - \frac{2 \epsilon_{k}}{1 - \epsilon_{k}} = - 2 \epsilon_{k} + {\mathcal O}(\epsilon_{k}^2) \simeq - \frac{r_{T}}{8} + {\mathcal O}(r_{T}^2) \ll 1.
\label{STWO2c}
\end{equation}
Equation (\ref{STWO2c}) is a consequence of Fig. \ref{FIG1} where the first crossing of $\lambda_{PTA}$ occurs during inflation and the reentry takes place for $\epsilon_{re} \to 2$; this is however not the most general situation 
even assuming that the left portion of Fig. \ref{FIG1} (i.e. $a< \,a_{1}$) is not modified. 
Indeed for $a\,>\,a_{1}$, the post-inflationary expansion rate can be modified as  $(a\, H)^{-1} \propto a^{1/\delta}$ with $\delta \neq 1$ and this case is illustrated in Fig. \ref{FIG2} when the usual radiation epoch (with $\delta \to 1$) is either preceded by a stage expanding faster than radiation (i.e. $\delta > 1$) or by one expanding at a rate slower than radiation (i.e. $\delta<1$). While more complicated possibilities are swiftly examined at the end of the present section, already these two opposite situations lead to a modified $\Omega_{gw}(k,\tau)$ above $\nu_{bbn}$ \cite{MG2,MG3,MG4} and may potentially 
lead to an excess in the nHz range.

The dashed curve of Fig. \ref{FIG2} corresponds to the comoving horizon in the 
limit $\delta \to 1$ while for the two remaining profiles the radiation phase is preceded by a stage where $\delta \neq 1$. In the three cases the minima depend both on the length of the post-inflationary stage and on its expansion rate:
\begin{equation}
\frac{(a_{1} \, H_{1})^{-1}}{\lambda_{PTA}} = 2.05 \times 10^{-17} \, \, \xi^{(1 - \delta)/[ 2 (\delta +1)]}, \qquad\qquad \xi = H_{r}/H_{1},
\label{STWO4}
\end{equation}
where $H_{r}$ and $H_{1}$ denote, respectively, the Hubble rates at the onset of the radiation stage and at the end of inflation (see e.g. Eq. (\ref{APP1a})). Since, by definition, $\xi < 1$ the comoving horizon at its minimum is comparatively larger for $\delta > 1 $ than for $\delta \to 1$. For the same reason the opposite is true when $\delta < 1$ and $(a_{1}\, H_{1})^{-1}/\lambda_{PTA}$ gets smaller than in the limit $\delta \to 1$. In Fig. \ref{FIG2} the $\lambda_{PTA}$ crosses the comoving Hubble radius when the expansion rate is 
different from radiation and this is why, according to Eq. (\ref{STWO1}) the spectral 
energy density gets modified. 

If the $\lambda_{PTA}$ reenters the Hubble radius when $\delta \neq 1$ (as it may happen in Fig. \ref{FIG2}) 
$\epsilon_{re} = {\mathcal O}(1)$ while, as before, $\epsilon_{ex} \ll 1$. For this reason, unlike the standard case 
illustrated in Fig. \ref{FIG1}, $|k \tau_{re}|^{-2}= {\mathcal O}(1)$ in Eq. (\ref{STWO1}) and the spectral slope  is not given by $m_{T}$ (as in Eq. (\ref{STWO2c})) but rather by the intermediate spectral index $n_{T}$:
\begin{equation}
n_{T} = \frac{1 - 3 \epsilon_{k}}{1 - \epsilon_{k}} - 2 \biggl|\delta -\frac{1}{2}\biggr| \simeq \frac{16 - 3 r_{T}}{16 - r_{T}} - 2 \,\biggl|\delta -\frac{1}{2}\biggr|.
\label{STWO4a}
\end{equation}
The consistency relations remain valid since, for $a< a_{1}$, Figs. \ref{FIG1} and \ref{FIG2} 
share the same inflationary evolution. We note however that for $\delta > 1/2$  the integral appearing in Eq. (\ref{STWO2a}) gives a subleading contribution and ${\mathcal Q}_{k}(\tau_{ex}, \tau_{re}) \to 1$. Conversely, when $\delta < 1/2$, we have instead ${\mathcal Q}_{k}(\tau_{ex}, \tau_{re}) \simeq 1 + |k\,\tau_{1}|^{ 1 - 2\delta}$ so that the second contribution dominates for all the frequencies of the spectrum (i.e. for $k \tau_{1} < 1$).  Finally, when $ \delta \to 1/2$ the integral of Eq. (\ref{STWO2a}) inherits a logarithmic correction which is relevant in specific models \cite{MG2,MG3,MG4}.

\subsection{Enhanced spectrum at intermediate frequencies}
It is useful to remark that, for conventional sources driving the background geometry (e.g. perfect 
irrotational fluids or scalar fields), the condition $\delta \geq 1/2$ is generally verified and when the wavelengths $\lambda= {\mathcal O}(\lambda_{PTA})$ cross the comoving Hubble radius the spectral slope of $h_{0}^2 \Omega_{gw}(\nu, \tau_{0})$ can be deduced from Eq. (\ref{STWO4a}):
\begin{equation}
n_{T} = \frac{32 - 4 \, r_{T}}{16 - r_{T}} - 2 \delta, \qquad \qquad r_{T} \leq 0.03.
\label{STWO5}
\end{equation}
It then follows from Eq. (\ref{STWO5}) that $n_{T} = 2 ( 1 - \delta) + {\mathcal O}(r_{T})$.
Thus $h_{0}^2 \Omega_{gw}(\nu, \tau_{0})$ increases for $\delta < 1$ while it decreases 
for $\delta >1$; when $\delta \to 1$, Eq. (\ref{STWO5}) 
reduces to Eq. (\ref{STWO2c}). To 
compare the slopes of Eq. (\ref{STWO5}) with the potential signals 
suggested by the PTA we recall that the relation between the chirp amplitude 
$h_{c}(\nu, \tau_{0})$ of Eq. (\ref{NOTT1}) follows from Eqs. (\ref{APP11})--(\ref{APP12}) of appendix \ref{APPA} and the final result is 
\begin{equation}
h_{0}^2 \Omega_{gw}(\nu, \tau_{0}) = 6.287 \times 10^{-10} \, q_{0}^2 \, \biggl(\frac{\nu}{\nu_{ref}}\biggr)^{2 + 2 \beta}, \qquad 
\qquad \nu_{ref} = 31.68\, \, \mathrm{nHz}.
\label{STWO6}
\end{equation}
For $\nu = \nu_{ref}$ Eq. (\ref{STWO6}) coincides with Eq. (\ref{NOTT2}) and since the values of $q_{0}$ and $\beta$ are determined observationally \cite{PPTA2,NANO2} (see also Refs. \cite{NANO1,PPTA1} for the preceding data releases) we can preliminarily 
identify the slopes of Eqs. (\ref{STWO5}) and (\ref{STWO6}). Therefore, when $\delta > 1/2$, 
the relation between $\delta$, $\beta$ and $r_{T}$ is simply given by:
 \begin{equation}
\delta = - \beta - \frac{r_{T}}{r_{T} +1} >\frac{1}{2}, \qquad \beta < 0.
\label{STWO7}
\end{equation}
Both in the previous \cite{NANO1,PPTA1} and in the most recent \cite{PPTA2,NANO2} data releases the value of $2 (1 + \beta)$ is always positive definite (i.e. $1 +  \beta >0$) while $\beta$ itself is generally smaller than $1$. In the special case $\beta \to -2/3$, Eq. (\ref{STWO7}) 
implies $ \delta = 2/3 + {\mathcal O}(r_{T})$. If we would now assume that the post-inflationary evolution is driven by a relativistic and irrotational fluid we would have $\delta= 2/(3 w+1)$ implying that $\beta \to- 2/3$ for $w \to 2/3$. Another possibility would be that the effective expansion rate is dictated by an oscillating scalar field (like the inflaton) with potential $V(\varphi) = V_{0} (\varphi/\overline{M}_{P})^{2 q}$; in this case the expansion rate during the oscillating phase would be given by $\delta = (q+1)/(2q -1)$ \cite{osc1,osc2,osc3,osc4,osc5,osc6,osc7} suggesting that $q = {\mathcal O}(5)$ for $\beta = {\mathcal O}(-2/3)$. 

Unless the relic gravitons would lead exactly to the same slope of the astrophysical foregrounds associated with black-hole binary systems, the value $\beta =-2/3$ is not particularly compelling in a cosmological setting. In the general case (i.e. when the special value $\beta = -2/3$ is not preliminarily selected) the Parkes PTA collaboration \cite{PPTA2} suggests that $\beta = - 0.45 \pm 0.20$. This determination is only marginally compatible with the value of Eq. (\ref{STWO7}) in the limit $\delta \geq 1/2$. The discrepancy between the observational determination of $\beta$ and the values predicted by Eq. (\ref{STWO7}) becomes even more significant if we look at that NANOgrav data suggesting \cite{NANO2} $\beta = -0.10 \pm 0.30$. Thus if we stick to the general situation suggested 
by the observational collaborations the limit $\delta<1/2$ should also be adequately considered in Eq. (\ref{STWO4a}) so that Eq. (\ref{STWO7}) is ultimately replaced by:
\begin{equation}
\delta = \beta + \frac{16}{16 - r_{T}} < \frac{1}{2}, \qquad \beta < 0.
\label{STWO7a}
\end{equation}
From Eq. (\ref{STWO7}) we have $\delta = -\beta + {\mathcal O}(r_{T})$ while in the case $\delta < 1/2$  Eq. (\ref{STWO7a}) implies $\delta = 1 + \beta + {\mathcal O}(r_{T})$.
From the profile of Fig. \ref{FIG2} with the help of Eqs. (\ref{STWO1})--(\ref{STWO2}) and (\ref{STWO4a}) the spectral energy density at high-frequencies becomes:
\begin{equation}
h_{0}^2 \, \Omega_{gw}(\nu, \tau_{0}) = {\mathcal N}(r_{T})\, \biggl(\frac{\nu}{\nu_{r}}\biggr)^{n_{T}}, \qquad \nu> \nu_{r},
\label{STWO8}
\end{equation}
where ${\mathcal N}(r_{T})$ includes the effect of the suppressions associated with the low-frequency transfer function, with the neutrino free-streaming \cite{NU1,NU2,NU3} and with the other late-time sources of damping (like the one associated with the dark-energy dominance \cite{MG1}). For $r_{T} =0.03$ we can numerically estimate that ${\mathcal N}(r_{T})= 10^{-16.8}$ \cite{est1}; we then safely consider\footnote{Above $\nu_{bbn}$ the value of ${\mathcal N}(r_{T})$ has a mild frequency dependence controlled by the value of $m_{T}$ (see Eq. (\ref{STWO2c})) and by the low-frequency transfer function; overall we have that for $\nu\geq \nu_{bbn}$ the low-frequency transfer function goes to $1$ and $\partial \ln{{\mathcal N}}/\partial\ln{\nu} \simeq m_{T} = - r_{T}/8 \ll 1$ \cite{est1}.} for the present ends ${\mathcal N}= {\mathcal O}(10^{-17})$ with a 
 theoretical amplitude at $\nu_{ref}$ that ultimately depends upon $\nu_{r}$ (see Eq. (\ref{STWO8}))
\begin{equation}
\nu_{r} =  \overline{\nu}_{max} \sqrt{\xi} = 271.88 \biggl(\frac{g_{s,\, eq}}{g_{s\,r}}\biggr)^{1/3} \biggl(\frac{g_{\rho,\, r}}{g_{\rho\,eq}}\biggr)^{1/4} \,\, \sqrt{\xi} \,\, \mathrm{MHz},
\label{STWO9}
\end{equation}
where the ratio $\xi = H_{r}/H_{1}$ has been already introduced in Eq. (\ref{STWO4}) while 
 $\overline{\nu}_{max}$ given by:
\begin{equation}
\overline{\nu}_{max} = 271.88 \biggl(\frac{g_{s,\, eq}}{g_{s\,r}}\biggr)^{1/3} \biggl(\frac{g_{\rho,\, r}}{g_{\rho\,eq}}\biggr)^{1/4} 
\biggl(\frac{h_{0}^2 \Omega_{R0}}{4.15 \times 10^{-5}}\biggr)^{1/4} \, \biggl(\frac{r_{T}}{0.03}\biggr)^{1/4} \, \biggl(\frac{{\mathcal A}_{{\mathcal R}}}{2.41\times 10^{-9}}\biggr)^{1/4}\, \, \mathrm{MHz}. 
\label{STWO10}
\end{equation}
In spite of the specific value given in Eq. (\ref{STWO9}), $\nu_{r}$ cannot be smaller than $\nu_{bbn}$ and this remark turns out to be quite relevant for the comparison between the observed excesses and the theoretical expectations.  Note that 
${\mathcal A}_{\mathcal R}$ is the amplitude of curvature inhomogeneities at the pivot scale $k_{p}$ (see 
also Eq. (\ref{NOTT4}) and discussion thereafter).

\subsection{Theoretical expectations and observed excesses}
\begin{figure}[!ht]
\centering
\includegraphics[height=7.5cm]{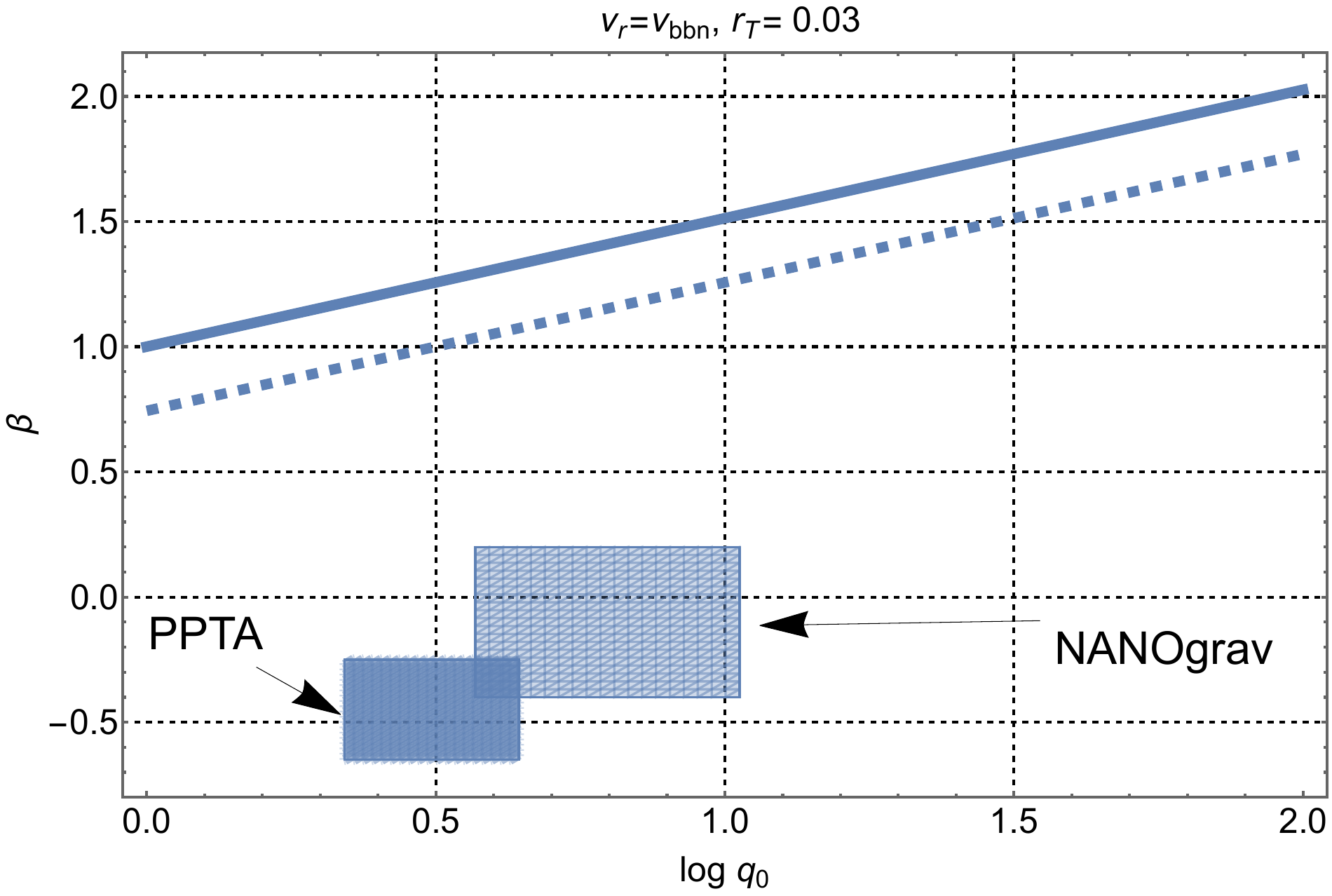}
\caption[a]{The two straight lines illustrate Eq. (\ref{STWO13}) for ${\mathcal N}(r_{T}) =10^{-17}$ 
(full line) and for ${\mathcal N}(r_{T}) =10^{-16}$ (dashed line). The two filled rectangles represent the regions probed by the Parkes PTA and by NANOgrav in the plane $(\log{q_{0}}, \, \beta)$.  Since the two diagonal lines do not overlap with the shaded areas appearing in the lower 
portion of the plot, the amplitudes and the slopes of the theoretical signal cannot be simultaneously matched with the corresponding observational determinations. Common logarithms have been employed on the horizontal axis since the range of the observational determinations of $q_{0}$ spans nearly one order of magnitude. }
\label{FIG3}      
\end{figure}
Even if the theoretical and the observed slopes can be compatible for specific values of $\delta$, the corresponding amplitudes involve rather different orders of magnitude and to analyze this essential aspect we may impose that Eqs. (\ref{STWO6}) and (\ref{STWO8}) coincide at $\nu_{ref}$
\begin{equation}
{\mathcal N}(r_{T}) \bigl(\nu_{ref}/\nu_{r}\bigr)^{n_{T}}= 6.287\,\times 10^{-10}\, q_{0}^2.
\label{STWO11}
\end{equation}
In spite of the equality sign, the left-hand side of Eq. (\ref{STWO11}) 
turns out to be systematically smaller than the right-hand side and the two 
sides of the equation are in agreement only if $\nu_{r}$ is much smaller than $\nu_{ref}$ while, at the same time, $n_{T}= 2 + 2 \beta$ is sufficiently large and positive. A large enough value of $n_{T}$ 
guarantees a sharp increase of the spectral energy density while the condition $\nu_{r}\ll \nu_{ref}$ makes wider the frequency range of the potential growth. Since the minimal value of $\nu_{r}$ is provided by $\nu_{bbn}$
\begin{equation}
\nu_{bbn} = 0.35\,\, \biggl(\frac{T_{bbn}}{10\,\mathrm{MeV}}\biggr) \, \biggl(\frac{h_{0}^2 \, \Omega_{R\,0}}{4.15\times 10^{-5}}\biggr)^{-1/4} \, \, \mathrm{nHz},
\label{STWO12}
\end{equation}
 we can select the most favourable situation and posit $\nu_{r} = {\mathcal O}(\nu_{bbn})$. 
For different values of ${\mathcal N}(r_{T})$ (see Eq. (\ref{STWO8}) and discussion thereafter) Eq. (\ref{STWO11}) leads therefore to a specific relation between $\beta$ and $\log{q_{0}}$: 
\begin{equation}
\beta = -1 + \frac{ 2 \log{q_{0}} - \log{{\mathcal N}(r_{T})} - 9.201}{2 \log{(\nu_{ref}/\nu_{bbn})}}.
\label{STWO13}
\end{equation}
To close the circle, the result of Eq. (\ref{STWO13}) must then be compared in the plane $(\log{q_{0}}, \, \beta)$ with the ranges of $\beta$ and $q_{0}$ determined by the PTA collaborations \cite{PPTA2,NANO2}. The two filled rectangles in Fig. \ref{FIG3} correspond, in this respect, to the 
observational ranges of $q_{0}$ and $\beta$; in the same plot the relation between $\beta$ and $\log{q_{0}}$ has been illustrated as it follows from Eq. (\ref{STWO13}) for two neighbouring values of ${\mathcal N}(r_{T})$. The two diagonal lines in Fig. \ref{FIG3} imply that the values of $\beta$ required to obtain $h_{0}^2 \Omega_{gw}(\nu_{ref}, \tau_{0})$ of the order of $10^{-8}$ or $10^{-9}$ should be much larger than the ones determined observationally and represented by the two shaded regions. Since the full and dashed lines of Fig. \ref{FIG3} do not overlap with the two rectangles in the lower part of the cartoon, we can conclude that the excess observed by the PTA collaborations cannot be explained by the modified post-inflationary evolution suggested of Fig. \ref{FIG2}. For the sake of accuracy we may separately analyze the case $\beta= -2/3$ and, in this situation, Eq. (\ref{STWO11}) becomes
\begin{equation}
{\mathcal N}(r_{T}) \bigl(\nu_{ref}/\nu_{r}\bigr)^{2/3}= 6.287\,\times 10^{-10}\, q_{0}^2.
\label{STWO11b}
\end{equation}
Again to maximize the potential growth of the spectral energy density we set $\nu_{r} = {\mathcal O}(\nu_{bbn})$ and obtain that the left-hand side 
of Eq. (\ref{STWO11b}) is $2.015\times 10^{-15}$ whereas the right-hand side 
is always larger than ${\mathcal O}(10^{-9})$. As in the previous case, larger values of $\nu_{r}$
only reduce the left-hand side of Eq. (\ref{STWO11b}) and ultimately increase the mismatch between Eqs. (\ref{STWO6}) and (\ref{STWO8}).

\begin{figure}[!ht]
\centering
\includegraphics[height=5.5cm]{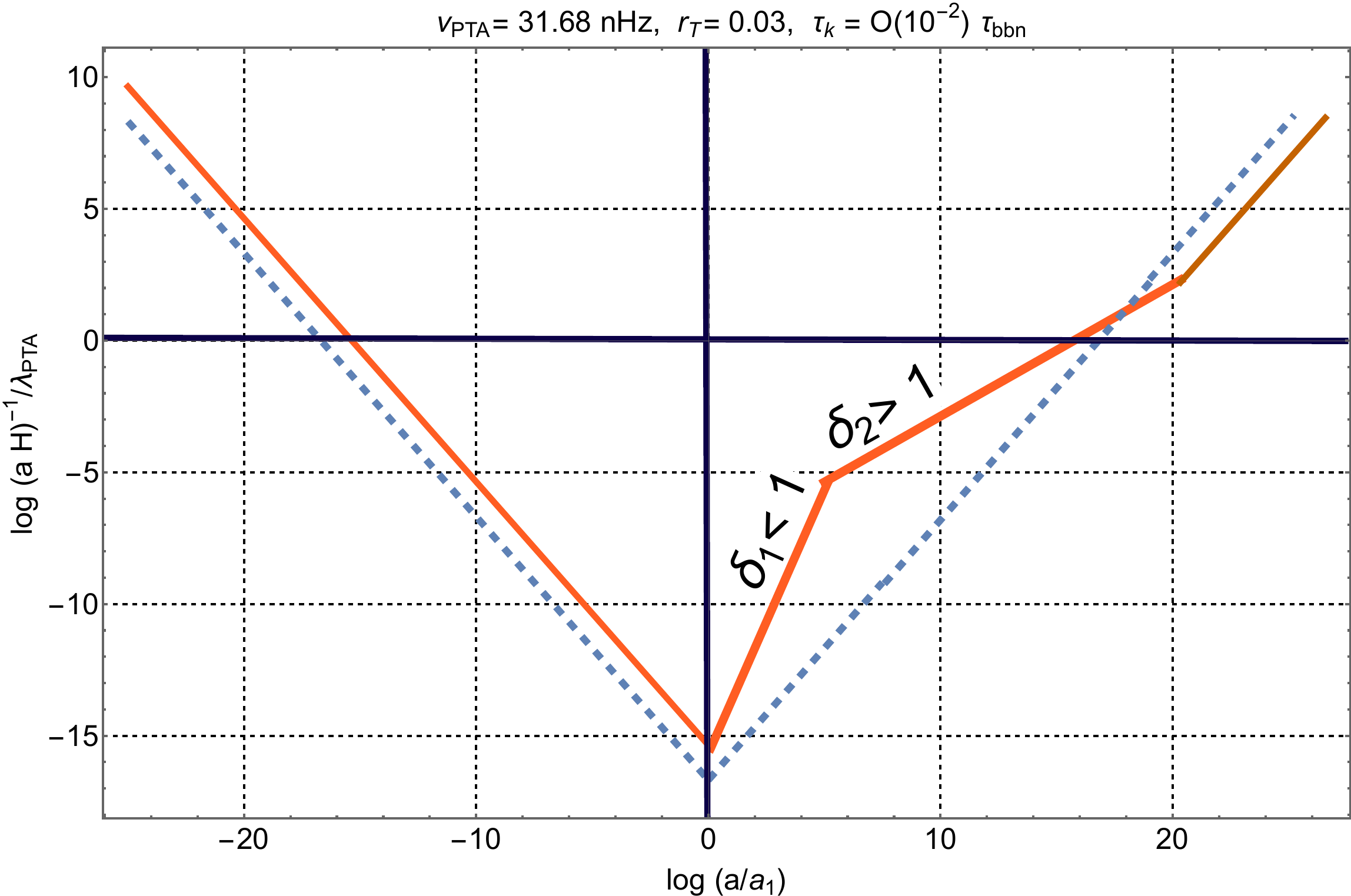}
\includegraphics[height=5.5cm]{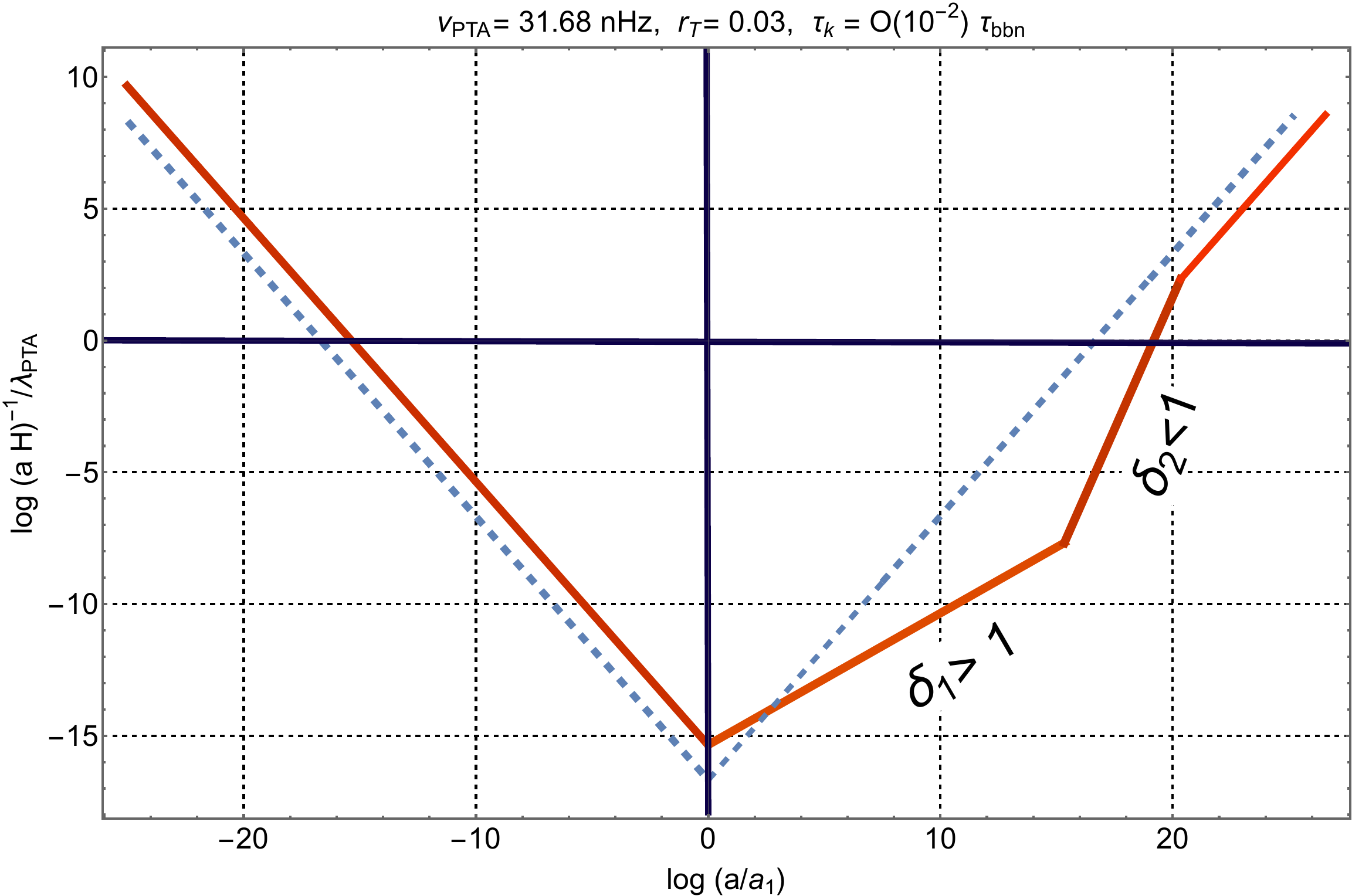}
\caption[a]{The evolution of the comoving horizon is again illustrated in units 
of $\lambda_{PTA}$ but the post-inflationary evolution 
is now characterized by two different expansion rates prior to the radiation-dominated phase.
In the plot at the left we have that the expansion rate is initially slower than radiation 
(i.e. $\delta_{1} <1$) and then faster than radiation (i.e. $\delta_{2} > 1$). In the right 
plot the timeline is inverted and $\delta_{1} >1$ while $\delta_{2}< 1$. In the left plot $\lambda_{PTA}$ 
crosses the comoving horizon when the plasma expands faster than radiation and this is 
why, for $\nu= {\mathcal O}(\nu_{ref})$ the spectral energy density decreases in frequency. Conversely, in the right plot, the expansion rate is slower than radiation when $\lambda_{PTA}$ crosses the comoving horizon and, in this case, $h_{0}^2 \Omega_{gw}(\nu,\tau_{0})$ inherits a growing frequency spectrum. }
\label{FIG4}      
\end{figure}

The argument leading to Eqs. (\ref{STWO13}) and (\ref{STWO11b}) follows from the profile  
of Fig. \ref{FIG2} when the post-inflationary expansion rate is slower 
than radiation since only in this case the slope of $h_{0}^2 \, \Omega_{gw}(\nu,\tau_{0})$ 
increases, as required by Eq. (\ref{STWO6}). There is however a complementary 
possibility stipulating that the comoving horizon prior to radiation dominance 
consists of {\em two successive stages expanding at different rates} and this situation
is illustrated in Fig. \ref{FIG4} where the two consecutive post-inflationary stages (characterized by the rates $\delta_{1}$ and $\delta_{2}$) precede the ordinary radiation-dominated epoch. Even the standard sources considered before (satisfying $\delta_{i} \geq 1/2$ for $i =1,\,2$) imply that 
when $\delta_{1}< 1$ and $\delta_{2} > 1$ (see the left plot of Fig. \ref{FIG4}) the spectral energy density $h_{0}^2 \Omega_{gw}(\nu,\tau_{0})$ develops a trough  
at the intermediate frequency $\nu_{2} \gg \nu_{r} \geq \nu_{bbn}$. In other words for $\nu< \nu_{2}$ the spectral energy density decreases while it increases above $\nu_{2}$. 
The presence of a trough in $h_{0}^2\, \Omega_{gw}(\nu,\tau_{0})$ corresponds to an expansion rate that is first slower (i.e. $\delta_{1} <1)$ 
and then faster (i.e. $\delta_{2} >1$) than radiation. In this case, however, $\lambda_{PTA}$ crosses the comoving horizon 
when $\delta_{2} > 1$ and the spectral index for the corresponding bunch of wavelengths 
is {\em negative} whereas Eq. (\ref{STWO6}) suggests it should be positive. Instead of a single spectral index $n_{T}$ there will now be two different 
spectral indices $n_{1}$ and $n_{2}$. For the profile appearing in the {\em left plot} 
of Fig. \ref{FIG4} the spectral index $n_{1}$ is positive 
\begin{equation}
n_{1} = \frac{32 - 4 \, r_{T}}{16 - r_{T}} - 2 \delta_{1} = 2( 1 - \delta_{1}) + {\mathcal O}(r_{T}) > 0,
\label{STWO12a}
\end{equation}
since $\delta_{1} <1$. However for $\nu< \nu_{2}$ the spectral index will be instead 
given by:
\begin{equation}
n_{2} = \frac{32 - 4 \, r_{T}}{16 - r_{T}} - 2 \delta_{2} = 2( 1 - \delta_{2}) + {\mathcal O}(r_{T}) < 0,
\label{STWO13a}
\end{equation}
since $\delta_{2} >1$. This means that the spectral index is {\em positive for $\nu> \nu_{2}$} and 
{\em negative when $\nu< \nu_{2}$} with a trough in $\nu= {\mathcal O}(\nu_{2})$.
Consequently the spectral energy density {\em is even more suppressed than in the case of the concordance paradigm} and the corresponding timeline of the comoving horizon is not relevant  for the present discussion where would instead aim at a large signal in the nHz range.

For the present ends the relevant timeline is illustrated in the {\em right plot} in Fig. \ref{FIG4}
where $\delta_{1} > 1$ and $\delta_{2} < 1$: the expansion rate is initially faster than radiation and then it gets slower so that 
$h_{0}^2 \, \Omega_{gw}(\nu,\tau_{0})$ decreases above $\nu_{2}$ while it increases 
for $\nu< \nu_{2}$. This means that the spectral energy density in critical units develops a hump
for $\nu = {\mathcal O}(\nu_{2})$; the signs of the spectral 
indices appearing in Eqs. (\ref{STWO12a}) and (\ref{STWO13a}) are exchanged: since $\delta_{1}> 1$ 
we have that $n_{1}< 0$ while $ n_{2} >0$ because $\delta_{2}< 1$ (see always the right plot 
of Fig. \ref{FIG4}). In this case the wavelengths smaller than $\lambda_{PTA}$ cross the comoving horizon when 
the the expansion rate is controlled by $\delta_{2} < 1$ and, for frequencies 
$\nu= {\mathcal O}(\nu_{ref})$, the spectral energy density increases as:
\begin{equation}
h_{0}^2 \, \Omega_{gw}(\nu, \tau_{0}) = {\mathcal N}(r_{T}) (\nu/\nu_{r})^{n_{2}}, \qquad \qquad \nu_{r} < \nu< \nu_{2}
\label{STWO14}
\end{equation}
where, as already stressed, $n_{2} > 0$. The  frequency $\nu_{2}$ follows from profile of the comoving 
horizon and it is given by 
\begin{equation}
\nu_{2} = \sqrt{\xi_{1}} \, \,\xi_{2}^{\frac{(\delta_{2} -1)}{2 (\delta_{2}+1)}}\,\,\overline{\nu}_{max}, \qquad \xi_{1} = H_{2}/H_{1}, \qquad \xi_{2} = H_{r}/H_{2},
\label{STWO14a}
\end{equation}
where, by definition, $\xi_{1} <1$ and $\xi_{2} <1$ denote the ratios of the curvature scales 
at the end and at the beginning of each expanding stage that precedes the ordinary radiation phase. The expressions of $\nu_{r}$ and  $\overline{\nu}_{max}$ coincide, respectively, with Eqs. (\ref{STWO9}) and   (\ref{STWO10}) since, by definition, $\xi= \xi_{1}\, \xi_{2} = H_{1}/H_{r} <1$. As before the 
 analytic expression of $\nu_{2}$ is not essential for the present purposes. Indeed, in spite of the expression of $\nu_{2}$ the largest signal will be obtained when $\nu_{2} = \nu_{ref}$ since for $\nu > \nu_{2}$ the spectral energy density decreases as: 
\begin{equation}
h_{0}^2 \, \Omega_{gw}(\nu, \tau_{0}) = {\mathcal N}(r_{T},\nu) (\nu_{2}/\nu_{r})^{n_{2}}\,\, (\nu/\nu_{2})^{- |n_1|}, \qquad \qquad \nu_{2} < \nu< \nu_{max},
\label{STWO15}
\end{equation}
where we introduced the absolute value of the spectral index $n_{1}$ to stress that $h_{0}^2 \, \Omega_{gw}(\nu,\tau_{0})$ is suppressed for $\nu> \nu_{2}$ with a negative spectral index $n_{1} = 2 (1 - \delta_{1}) < 0$  (i.e. $\delta_{1} > 1$). We can now try to match the amplitude 
and the slope of Eq. (\ref{STWO14}) with Eq. (\ref{STWO6}). If we set $\nu_{2}= \nu_{ref}$ and
$n_{2} = 2 ( 1 - \delta_{2}) + {\mathcal O}(r_{T}) = 2 + 2 \beta$ the condition on the amplitude becomes:
\begin{equation}
 {\mathcal N}(r_{T}) (\nu_{ref}/\nu_{r})^{n_{2}} = 6.287 \times 10^{-10} \, q_{0}^2.
 \label{STWO16}
 \end{equation}
 Again the widest frequency range corresponds to the case when $\nu_{r} \to \nu_{bbn}$ and after some algebra we get back exactly to the same condition of Eq. (\ref{STWO13}).  Due to the smallness of the potential signal, the direct bounds coming 
from wide-band interferometers \cite{LIGO1,LIGO2} and the indirect constraints from big-bang nucleosynthesis \cite{bbn1,bbn2,bbn3} do not play a relevant r\^ole in the argument of this section\footnote{There are however situations, as we shall see in the forthcoming sections, where 
the spectral energy density gets larger already in the nHz range and while these situations can be used to address 
the PTA excesses, they are comparatively more constrained also at higher frequencies by the current bounds in the audio band. }. The results obtained so far can then be summarized 
in the following manner:
\begin{itemize}
\item{} if the post-inflationary evolution is modified prior to radiation dominance 
 $h_{0}^2 \Omega_{gw}(\nu,\tau_{0})$ may increase in comparison with the
  concordance paradigm for typical frequencies $\nu = {\mathcal O}(\nu_{ref})$; this happens only if
   the wavelengths ${\mathcal O}(0.3)$ pc (roughly corresponding to comoving frequencies ${\mathcal O}(30)$ nHz) 
   cross the comoving horizon when the expansion rate is slower than radiation;
 \item{} the  PTA signals imply a growing spectral energy density (i.e. $2 + 2 \beta>0$) 
 and this is consistent with an expansion rate that is slower than radiation at least when 
 the wavelengths ${\mathcal O}(\lambda_{PTA})$ reenter the comoving horizon;
 \item{} however the amplitudes and the slopes of the theoretical signal do not match 
 simultaneously the observational determinations of the PTA in the $(\log{q_{0}}, \, \beta)$ 
 plane.
 \end{itemize}
A hump in $ h_{0}^2 \Omega_{gw}(\nu,\tau_{0})$ for $\nu= {\mathcal O}(\nu_{ref})$ indeed follows when
the post-inflationary expansion rate is slower than radiation but the theoretical signal is not 
 consistent with the slope and with the amplitude of the observational spectrum. In
 specific cases the theoretical and the observed slopes are compatible but
 the corresponding amplitudes differ by $9$ or even $10$ orders of magnitude. As usual the agreement between theoretical and observed slopes is just a mere indication that is, per se, irrelevant for the final conclusion.

\renewcommand{\theequation}{3.\arabic{equation}}
\setcounter{equation}{0}
\section{The comoving horizon during inflation}
\label{sec3}
\subsection{Modified evolution of the comoving horizon} 
\begin{figure}[!ht]
\centering
\includegraphics[height=7cm]{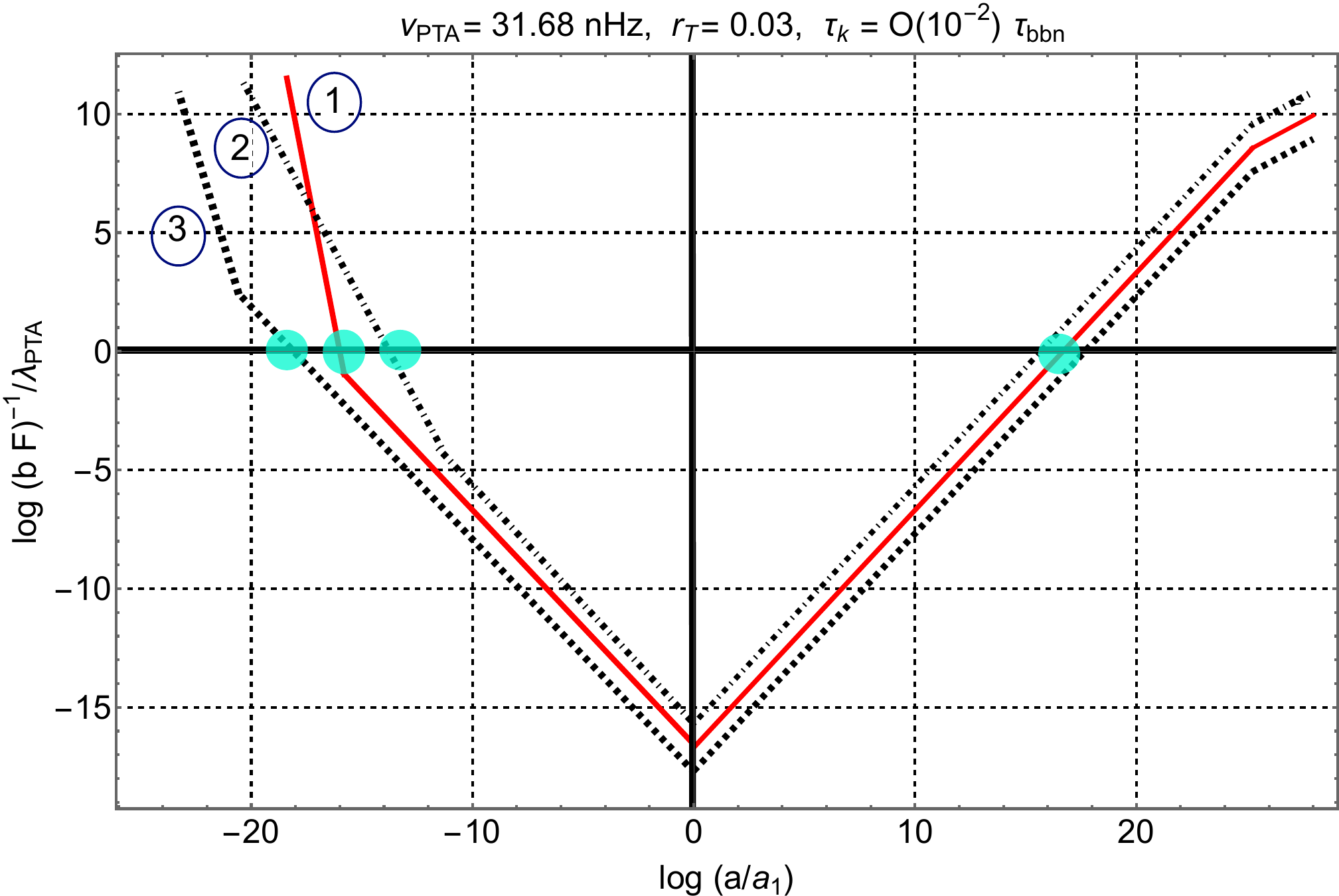}
\caption[a]{The comoving horizon is illustrated  in units 
of $\lambda_{PTA}$  when its early evolution is modified during the inflationary stage; as usual common logarithms are employed on both axes. The dynamics of the refractive index of the relic gravitons is responsible for the modified evolution illustrated in this cartoon.  If the crossing of $\lambda_{PTA}$ occurs during the refractive phase (or at the end of it) the spectral energy density inherits a blue or violet spectrum that may eventually explain, under very specific conditions, the PTA excess.}
\label{FIG5}      
\end{figure}

If the refractive index $n(a)$ of the relic gravitons is dynamical 
the condition of Eq. (\ref{STWO3}) defining the exit of a given wavelength is now replaced by:
\begin{equation}
k^2 = n^3 \, b^2 \, F^2 \biggl[ 2 - \overline{\epsilon}+ \frac{3\, \dot{n}}{2 \,n \,F}\biggr], \qquad F = \frac{\dot{b}}{b}, \qquad \overline{\epsilon} = - \frac{\dot{F}}{F^2},
\label{STHR0}
\end{equation}
where the overdot denotes, as usual, a derivation with respect to the cosmic time coordinate; in Eq. (\ref{STHR0})
$a(\tau)$ is actually rescaled as $b = a/\sqrt{n}$ and $\overline{\epsilon}$ is, 
in practice, the generalization of the slow-roll parameter $\epsilon$ already introduced in the previous section and in the appendices. 
The details on the connection between $F$ and $H$ are also discussed in Eqs. (\ref{APPB5})--(\ref{APPB6}) of appendix \ref{APPB} and the analog of $(a\, H)^{-1}$ 
is now represented by $(b\, F)^{-1}$ whose evolution in units of $\lambda_{PTA}$ is illustrated in Fig. \ref{FIG5}. By looking at the profiles of Fig. \ref{FIG5} there are in fact two complementary possibilities:
\begin{itemize}
\item{} if $\lambda_{PTA}$ crosses 
the comoving horizon when the refractive index is dynamical $h_{0}^2 \, \Omega_{gw}(\nu,\tau_{0})$ may inherit a growing spectrum comparable with the PTA excess and in Fig. \ref{FIG5} the two curves (labeled, respectively, by $1$ and $2$) illustrate this possibility; while in the case $2$ the first crossing occurs during the refractive phase, for the curve labeled by $1$ the crossing occurs nearly at the end of it;
\item{}  if the first crossing takes place {\em after} the end of the 
refractive phase (see the curve labeled by $3$ in Fig. \ref{FIG5}) the spectral energy density does not show any appreciable excess and the resulting spectral energy density is quasi-flat.
\end{itemize}
Even if the curve $3$ in Fig. \ref{FIG5} does not lead to a growing spectrum for $\lambda_{PTA}$ it is anyway 
relevant for shorter wavelengths $\lambda \ll \lambda_{PTA}$; for the corresponding frequencies $h_{0}^2 \Omega_{gw}(\nu, \tau_{0})$ may have a flatter slope but also further spikes caused by the post-inflationary evolution.  Both structures are significantly constrained in the audio and in the MHz bands.
\subsection{Refractive index and effective action}
When the refractive index of the relic gravitons increases during a conventional stage of inflationary expansion the spectral energy density is blue at intermediate frequencies (typically above the fHz) and then flattens out after a knee that is generally smaller than the mHz \cite{CC2}. The general shapes of $h_{0}^2 \, \Omega_{gw}(\nu,\tau_{0})$ suggest that this possibility is particularly interesting in the light of the PTA excesses. When the refractive index is dynamical the action of the relic gravitons (see e.g. Eq. (\ref{APP2}) and discussion therein) is modified as described in appendix \ref{APPB} and it can be written as:
\begin{equation}
S = \frac{\overline{M}_{P}^2}{8} \int \, d^{3} x\, \int d\tau \,\, a^2(\tau) \,\biggl[ \partial_{\tau} h_{i\, j} \partial_{\tau} h^{i\, j} - \frac{1}{n^2(\tau)} 
\partial_{k} h_{i\, j} \partial^{k} h^{i\, j} \biggr].
\label{STHR1}
\end{equation}
The analysis of Eq. (\ref{STHR1}) is greatly simplified if the conformal time coordinate is redefined from $\tau$ to $\eta$ where the relation between the new and the old time parametrizations implicitly follows from  $n(\eta) \, d\eta = d \tau$. Equation (\ref{STHR1}) becomes then canonical in terms of a redefined scale factor conventionally denoted by $b(\eta)$ \cite{CC2}: 
\begin{equation}
S = \frac{\overline{M}_{P}^2}{8} \int \, d^{3} x\, \int d\eta \, b^2(\eta)\, \biggl[ \partial_{\eta} h_{i\, j} \partial_{\eta} h^{i\, j} - 
\partial_{k} h_{i\, j} \partial^{k} h^{i\, j} \biggr], \qquad b(\eta) = a(\eta)/\sqrt{n(\eta)}.
\label{STHR2}
\end{equation}
The result of Eq. (\ref{STHR2}) generalizes the standard Ford-Parker action discussed in appendix \ref{APPA} to the case of a dynamical refractive index and it explains how and why the evolution of the tensor modes is modified even during a conventional 
stage of inflationary expansion. A number of different physical reasons may lead to an effective index of refraction of gravitational waves in curved space-times \cite{CC2,CC1a,CC1b}. For instance  the effective action of single-field inflationary models involves all the terms that include four derivatives\footnote{When parity breaking terms are included in the effective action \cite{TWO,THREE}, the relic graviton background may be polarized but this possibility has been already discussed elsewhere \cite{FOUR}.} and are suppressed by the negative powers of a large mass scale \cite{AAAW}. Another possible origin of a refractive index are non-generic models of inflation where the higher-order corrections assume a specific form since the inflaton has some particular symmetry (like a shift symmetry $\varphi \to \varphi + c$) or because the rate of inflaton roll remains
constant (and possibly larger than $1$), as it happens in certain fast-roll scenarios \cite{NON1,NON2,NON3}. There are also the cases where the higher-order curvature corrections are given in terms of the Gauss-Bonnet combination weighted by some inflaton dependent-coupling \cite{NON4,NON5,NON6}. In \cite{CC2} (see also \cite{CC4,CC5,CC3}) it has been argued that in all these situations the effective action of the relic gravitons can be modified and ultimately assumes the general form discussed in appendix \ref{APPB}.

For the present purposes what matters is not so much the origin of the refractive index but 
rather the possibility that its dynamical evolution {\em could} lead to a nHz excess.
We then assign $n(a)$ and even if the phase velocity of the relic gravitons is not required to be sub-luminal we impose, for consistency, that $n(a) \geq 1$. Moreover, since the contributions to $n(a)$ arise from diverse physical considerations we prefer to reverse the problem by focussing the attention on those profiles that eventually lead to potential excesses in the nHz range. Along this perspective we are led to consider an appreciable change of the refractive index during inflation with the concurrent requirement that $n(a)$ reaches $1$ in the standard decelerated stage of expansion\footnote{Some other possibilities have been considered in \cite{CC2} and these cases can be easily added but they will not be examined here for the sake of conciseness.  In this sense we shall regard the profile of Eq. (\ref{STHR3}) as the minimal example that potentially leads to a nHz excess.}:
\begin{equation}
n(a) = n_{\ast} \frac{ (a/a_{\ast})^{\alpha} \,\,e^{- \overline{d}\,(a/a_{1})}}{(a/a_{*})^{\alpha} + 1} + 1, \qquad\qquad
n_{\ast} = n_{i} (a_{\ast}/a_{i})^{\alpha} = n_{i} e^{\alpha \, N_{\ast}},
\label{STHR3}
\end{equation}
where $a_{i}$ and $a_{1}$ denote, respectively, the beginning and the end of the 
inflationary epoch; $a_{*}$ indicates the boundary of the refractive stage. 
The three successive physical regimes described by Eq. (\ref{STHR3}) are, in some sense, more relevant than the specific analytic form that is however quite useful for numerical estimates. When 
$a\gg a_{1}$ we have that $n(a) \to 1$  and the sharpness of the transition is controlled by the parameter $\overline{d} \geq 1$.  In the range $a_{*} < a < a_{1}$ $n(a)$ is constant but still larger than $1$ (i.e. $n(a)\simeq n_{\ast} > 1$) and, finally, when $a< a_{\ast}$ the refractive index is truly dynamical since $n (a) \simeq n_{\ast} (a/a_{\ast})^{\alpha}$. 
 
\subsection{The spectral energy density in critical units}
We start by examining the spectral energy density for typical wavenumbers $ k < a_{\ast}\, H_{\ast}$
where $\Omega_{gw}(k,\tau)$ is actually increasing. Using Eqs. (\ref{STHR2})--(\ref{STHR3})  the spectral energy density in critical units can be obtained from Eqs. (\ref{STWO1})--(\ref{STWO2})
with few relevant modifications:
\begin{equation}
\Omega_{gw}(k,\tau) = \frac{k^4}{12 \,\pi^2\, H^2\, \overline{M}_{P}^2\, a^{4}}\,\, \bigl|{\mathcal Q}_{k}(\eta_{ex}, \eta_{re})\bigr|^2 \,\, \biggl(\frac{b_{re}}{b_{ex}}\biggr)^2 \biggl(
1 + \frac{1}{k^2 \tau_{re}^2} \biggr).
\label{STHR5}
\end{equation}
Equation (\ref{STHR5}) is valid under the assumption that $\eta_{re} = \tau_{re}$ so that the reentry 
of the relevant wavelengths occurs when the refractive index is not dynamical; furthermore
if $\tau_{re}$ falls within the radiation phase (i.e. $a^{\prime\prime} \to 0$) we also have $k \tau_{re}\ll 1$ in Eq. (\ref{STHR5}). Since any wavelength  
exiting for $\eta < - \eta_{\ast}$ does its first crossing during the inflationary phase, the corresponding refractive index is  $n = n_{\ast} (a/a_{\ast})^{\alpha}$; moreover the explicit expression of ${\mathcal Q}_{k}(\eta_{ex}, \eta_{re})$
is also slightly more general than in the case of Eq. (\ref{STWO2})
\begin{equation}
 {\mathcal Q}_{k}(\eta_{ex}, \eta_{re}) = 1 - ({\mathcal F}_{ex} + i k) \int_{\eta_{ex}}^{\eta_{re}} \frac{b_{ex}^2}{b^2(\tau)} \, d\eta, \qquad\qquad {\mathcal F}= \partial_{\eta}b/b.
 \label{STHR5a}
 \end{equation}
Finally, from Eq. (\ref{STHR2}) $b(\eta)$ can be expressed in the following manner:
\begin{equation}
b(\eta) = b_{\ast} (- \eta/\eta_{\ast})^{-\zeta}, \qquad b_{\ast} = a_{\ast}/\sqrt{n_{\ast}}, \qquad 
\zeta = \frac{ 2 - \alpha}{2 ( 1 - \epsilon + \alpha)}.
 \label{STHR6}
\end{equation}
The result of Eq. (\ref{STHR6}) is valid if all the wavelengths 
${\mathcal O}(\lambda_{PTA})$ exit while the refractive index is still dynamical 
(as illustrated in the curves $1$ and $2$ of Fig. \ref{FIG5}). 
Inserting now Eq. (\ref{STHR6}) into Eq. (\ref{STHR5}) a more explicit form of the spectral energy density can be deduced:
\begin{eqnarray}
h_{0}^2 \,\Omega_{gw}(\nu,\tau_{0}) &=&  \biggl(\frac{H_{1}}{M_{P}}\biggr)^2  \,{\mathcal D}_{\ast}(\alpha,n_{T}) \biggl(\frac{\nu}{\nu_{\ast}}\biggr)^{n_{T}}, \qquad \qquad \nu_{eq} < \nu< \nu_{\ast},
 \label{STHR7}\\
{\mathcal D}_{\ast}(\alpha,n_{T}) &=& \frac{4 n_{\ast}^3\, h_{0}^2 \Omega_{R0}  }{3 \pi} \, \biggl(1 + \frac{\alpha}{1-\epsilon_{k}}\biggr)^2 \, \biggl(\frac{g_{\rho, \, r}}{g_{\rho, \, eq}}\biggr) \biggl(\frac{g_{s,\, eq}}{g_{s,\, r}}\biggr)^{4/3} \, \biggl(\frac{\Omega_{M0}}{\Omega_{\Lambda}}\biggr)^2,\, \,
\label{dstar}
\end{eqnarray}
where $g_{\rho}$ and $g_{s}$ are, respectively, the effective number of relativistic species associated 
with the energy and with the entropy density. As usual $\Omega_{M0}$ and $\Omega_{\Lambda}$ denote the present critical fractions of matter and dark energy.  It is well known that the dominance of dark energy suppresses the spectrum by a factor $(\Omega_{M0}/\Omega_{\Lambda})^2 = {\mathcal O}(0.1)$ (see, for instance, \cite{MG1}). In Eq. (\ref{STHR7}) $\nu_{\ast}$ denotes the frequency of the spectrum associated 
with $\eta_{\ast}$ and since $k_{\ast} = 1/\eta_{\ast}$ the corresponding 
comoving frequency is:
\begin{equation}
\nu < \nu_{\ast} = \biggl(1 + \frac{\alpha}{1 - \epsilon_{k}}\biggr) e^{\alpha N_{\ast} - \Delta N} \, \overline{\nu}_{max},\qquad 
\Delta N= N_{t} - N_{\ast}.
\label{STHR8}
\end{equation}
In Eq. (\ref{STHR8}) $N_{\ast} = \ln{(a_{\ast}/a_{i})}$ is the number 
of $e$-folds during the refractive stage while $N_{t} = \ln{(a_{1}/a_{i})}$ denotes the {\em total} number of $e$-folds; finally  $\overline{\nu}_{max}$ indicates the maximal frequency of the spectrum and it coincides with Eq. (\ref{STWO10}) since, so far, the radiation dominance starts right after the end of inflation. The spectral index $n_{T}$ appearing in Eq. (\ref{STHR7})  depends on $\alpha$ and on $\epsilon_{k}$ and it is:
\begin{eqnarray}
n_{T} = 2 - 2 \zeta &=& \frac{3 \alpha - 2\epsilon_{k}}{(1 + \alpha - \epsilon_{k})}
\nonumber\\
&=& \frac{3 \alpha}{1 + \alpha} + 
 \frac{\epsilon_{k}\, (\alpha - 2)}{(1 + \alpha)^2} + {\mathcal O}(\epsilon_{k}^2).
\label{STHR9}
\end{eqnarray}
The tensor spectral index of Eq. (\ref{STHR9}) applies in the intermediate 
frequency range when the corresponding wavelengths exit during inflation and reenter in the radiation phase; in Eq. (\ref{STHR9}) $\alpha$ is always much larger than $\epsilon_{k}\simeq r_{T}/16 \leq 0.03/16 \ll 1$ so that the exact result can be accurately evaluated in the limit $\epsilon_{k} \ll 1$. While Eqs. (\ref{STHR7}) and (\ref{STHR9}) hold for $\nu<\nu_{\ast}$, 
the spectral energy density can also be evaluated in the range $\nu_{\ast} < \nu < \overline{\nu}_{max}$ (i.e.  $a_{\ast} \, H_{\ast} < k < a_{1}\, H_{1}$) corresponding to wavelengths that exited the comoving horizon when the refractive index was already constant (see the curve 3 in Fig. \ref{FIG5}). 
Since the corresponding wavelengths do their 
first crossing when the refractive index is already constant (i.e. $n\to n_{\ast}$) we have that  $\eta_{ex} = \tau_{\ast}/n_{\ast}$ and the spectral energy density becomes:
 \begin{equation}
h_{0}^2 \Omega_{gw}(\nu,\tau_{0}) = \biggl(\frac{H_{1}}{M_{P}}\biggr)^2\, {\mathcal D}_{max}(\alpha, m_{T}) \, \, \biggl(\frac{\nu}{\overline{\nu}_{max}}\biggr)^{m_{T}}, \qquad \qquad \nu_{\ast}< \nu< \overline{\nu}_{max},
\label{STHR10}
\end{equation}
where the spectral index is given by $m_{T} = - 2 \epsilon_{k} = - r_{T}/8$ and 
\begin{equation}
{\mathcal D}_{max}(\alpha, m_{T}) =
\frac{4 \,h_{0}^2 \Omega_{R0} }{3 \pi} \, e^{(3 -m_{T})\alpha\,N_{\ast}} e^{ m_{T} \Delta N} \biggl(1 + \frac{\alpha}{1-\epsilon_{k}}\biggr)^{2-m_{T}}\,\ \,  \, \biggl(\frac{g_{\rho, \, r}}{g_{\rho, \, eq}}\biggr) \biggl(\frac{g_{s,\, eq}}{g_{s,\, r}}\biggr)^{4/3} \, 
\biggl(\frac{\Omega_{M0}}{\Omega_{\Lambda}}\biggr)^2.
\label{dmax}
\end{equation}
Equation (\ref{STHR10}) evaluated for $\nu = \nu_{\ast}$ corresponds exactly to Eq. (\ref{STHR7}) computed at the same reference frequency and the equivalence of the two expressions ultimately follows from Eq. (\ref{STHR8}). Furtheremore, in Eqs. (\ref{STHR7}) and (\ref{STHR9}) $(H_{1}/M_{P})^2$ can be traded for $\pi \, \epsilon_{k}\, {\mathcal A}_{{\mathcal R}}$ where 
${\mathcal A}_{{\mathcal R}}$ is the amplitude of curvature inhomogeneities at the pivot scale $k_{p}$ (see also Eqs. (\ref{NOTT4}) and (\ref{STWO10})). It is finally worth recalling that, for a standard thermal history, $g_{s,\,eq} = 3.94$ while $g_{\rho,\, r} = g_{s,\, r}=106.75$ in Eqs. (\ref{dstar}) and (\ref{dmax}).

Before proceeding further it is useful to comment on the possible range of variation of $N_{t}$ and $N_{\ast}$ appearing in Eq. (\ref{STHR8}). The value of $N_{\ast}$ measures the range 
of variation of the refractive index during inflation and, for this reason, $N_{\ast} < N_{t}$. The total number of $e$-folds is, in principle, arbitrary but a useful benchmark value is 
notoriously given by $N_{t} = {\mathcal O}(60)$. This figure coincides, approximately, with the number of $e$-folds elapsed from the moment where the CMB wavelengths 
crossed the Hubble radius during inflation. If we estimate these wavelengths with $k_{p}^{-1}$ (where $k_{p} = 0.002\,\, \mathrm{Mpc}^{-1}$) then we have that 
$N_{k}$ (i.e. the number of $e$-folds elapsed since the crossing of $k_{p}^{-1}$) is, approximately, ${\mathcal O}(60)$ for $r_{T}= 0.06$; more precisely it can be 
shown that 
\begin{equation}
N_{k} = 59.4 + \frac{1}{4}\ln{\biggl(\frac{\epsilon_{k}}{0.001}\biggr)} - \ln{\biggl(\frac{k}{0.002 \,\, \mathrm{Mpc}^{-1}}\biggr)}
\label{NNk}
\end{equation}
where, as before, $\ln{}$ denotes the natural logarithm and $\epsilon_{k}$ is the value of the slow-roll parameter at the crossing of the given 
set of wavelengths. Equation (\ref{NNk}) assumes that after inflation the evolution is always dominated by radiation even if in Fig. \ref{FIG10} 
this assumption will be relaxed.
The value of $N_{k}$ is also of the order of $N_{max}$ (i.e. the maximal number of $e$-folds presently accessible to large-scale 
observations); see in this respect the appendices of Ref. \cite{MM2}. This value follows by requiring that the redshifted inflationary event horizon fits within the present Hubble patch; in practice this means 
that $H_{i}^{-1} (a_{0}/a_{i}) \simeq H_{0}^{-1}$ where $H_{i}$ denotes the expansion rated during the initial stages of inflation. Neglecting 
for simplicity the evolution of the relativistic species of the plasma we get that  $N_{max} = 61.55 $ for the same fiducial set 
of parameters employed in Eq. (\ref{NNk}). Typical values of $N_{t}$ of the order of $N_{k}$ and $N_{max}$ define, in practice, the 
{\em minimal} duration of the inflationary stage. Conversely values of $N_{t}$ {\em smaller} than $N_{k}$ (or $N_{max}$) 
characterize the durations of inflationary stages that are comparatively shorter than the benchmark value of Eq. (\ref{NNk}).
As we shall see in a moment, the relatively short inflationary stages (where $N_{t} \leq {\mathcal O}(61)$) seem to be preferred for a potential 
explanations of the PTA excesses.

\begin{figure}[!ht]
\centering
\includegraphics[height=7.5cm]{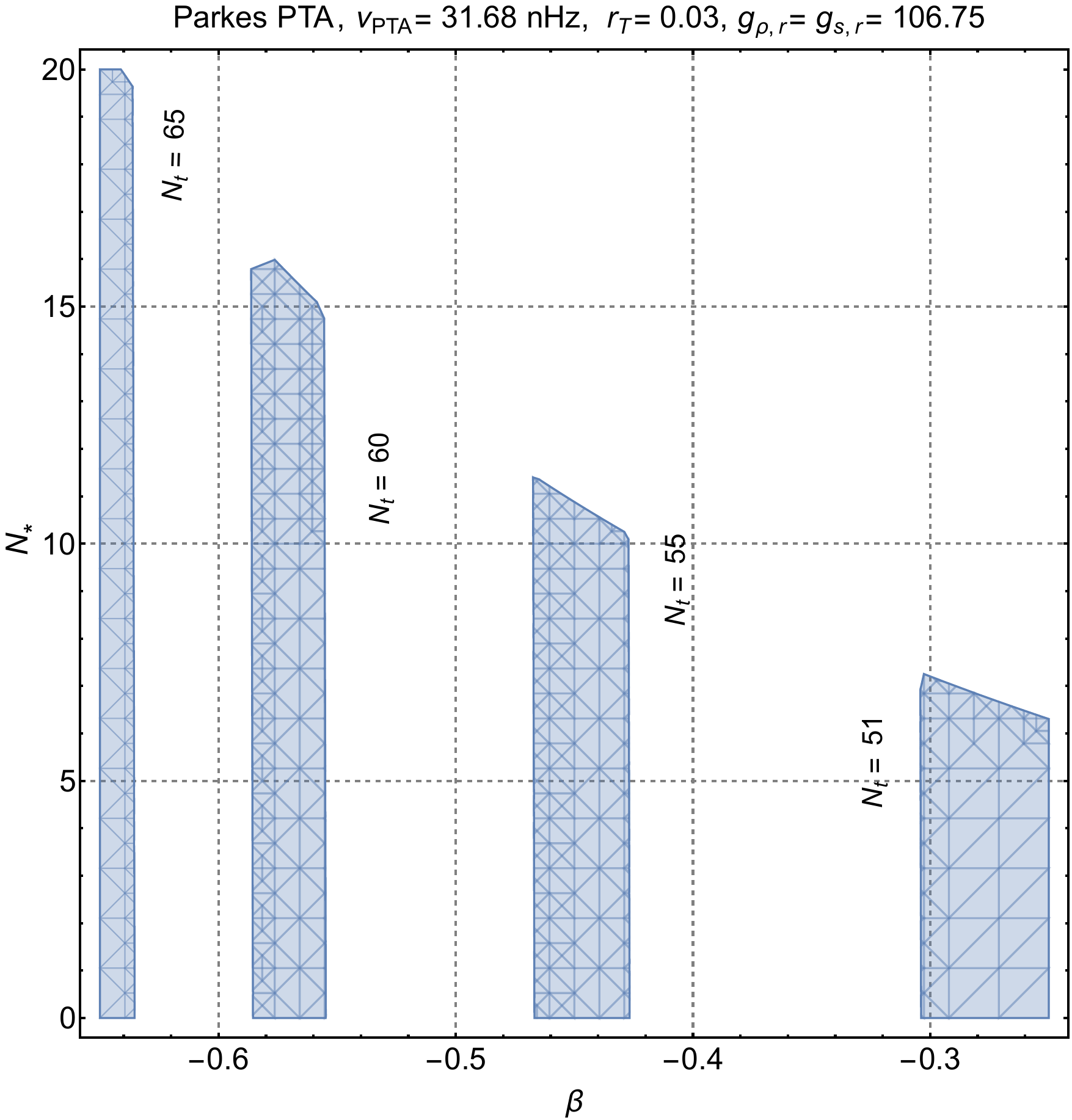}
\includegraphics[height=7.5cm]{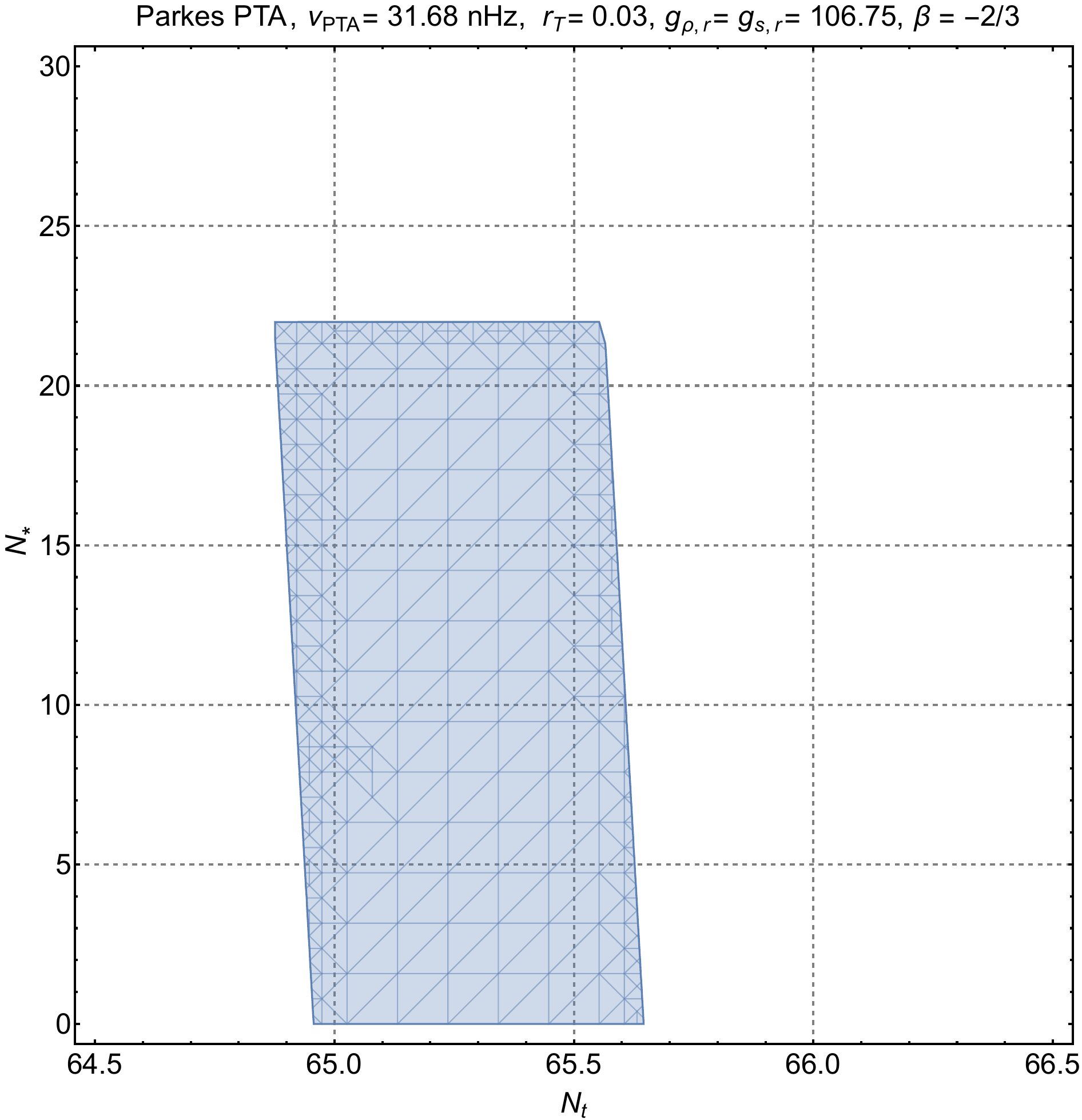}
\caption[a]{The regions pinned down by the Parkes PTA are illustrated in the left plot in terms of the 
parameters appearing in Eqs. (\ref{STHR7}) and (\ref{STHR10}). The 
shaded areas follow by imposing the range of Eq. (\ref{STHR12}) for different values of $N_{t}$ (denoting, as repeatedly mentioned, 
the total number of $e$-folds). We also remind that $N_{\ast}$ controls the length of the phase where 
the refractive index is effectively dynamical. In both plots we traded 
$\alpha$ for $\beta$ at a fixed value of $\epsilon_{k}=0.0018$ (see Eq. (\ref{STHR11}) and discussion therein); the value of $\epsilon_{k}$ is related to $r_{T}=0.03$ by the consistency relations. While in the left plot we illustrated $q_{0}(\beta, \,N_{\ast})$ for different values of $N_{t}$,  in the right plot $\beta$ has been fixed to 
$-2/3$; in this case the allowed range of $q_{0}(N_{t}, N_{\ast})$  is slightly different from the one of Eq. (\ref{STHR12}) and it is given by $1.82< q_{0}(N_{t}, N_{\ast}) < 2.29$. Finally, the shaded regions in both plots are consistent with the higher-frequency bounds coming 
from the audio band.}
\label{FIG6}      
\end{figure}

\subsection{Accounting for the PTA excesses?}
Equations (\ref{STHR5}), (\ref{STHR7}) and (\ref{STHR10}) are now compared with the parametrizations of the PTA signal 
given in Eqs. (\ref{NOTT2}) and (\ref{STWO6}). Since, by definition, the intermediate spectral index is given as 
$2 + 2\beta = n_{T}$  Eq. (\ref{STHR9}) implies a relation that determines 
$\alpha$ as a function of $\epsilon_{k}$ (or $r_{T}$) and $\beta$:
\begin{equation} 
\alpha = \frac{2 [\beta(\epsilon_{k} -1)-1]}{2 \beta -1}.
\label{STHR11}
\end{equation}
Moreover, given that $q_{0}$ depends on all the other parameters
determining the amplitude of $\Omega_{gw}(\nu,\tau_{0})$ (see Eqs. (\ref{STHR7}) and (\ref{STHR10})), we can demand that $\beta$ and $q_{0}$ fall within the phenomenologically 
allowed ranges and check if the results of Eqs. (\ref{STHR7}) and (\ref{STHR10}) 
are compatible with the empirical determinations of the PTA. According to the Parkes PTA  the values of $\beta$ and $q_{0}$ fall, respectively,  in the following intervals:
\begin{equation}
-0.65 \leq \beta \leq -0.25, \qquad \qquad 2.2 < q_{0} < 4.4.
\label{STHR12}
\end{equation}
Equation (\ref{STHR12}) constrains the spectral energy density and 
the corresponding region of the theoretical parameters is illustrated in the left plot of Fig. \ref{FIG6} where we report $q_{0}(\beta,\, N_{\ast})$ for different values of $N_{t}$; the shape of each shaded region  directly follows by requiring $ 2.2 < q_{0}(\beta,\,N_{\ast}) < 4.4$ for the various $N_{t}$ mentioned in the plot. On a technical side we note that Eq. (\ref{STHR11}) has been used with the purpose of trading directly $\alpha$ for $\beta$ at a fixed value of $\epsilon_{k}$. In the right plot of Fig. \ref{FIG6}  $\beta$ is however fixed (i.e. $\beta \to -2/3$) and, for such a choice, also the range of $q_{0}$ must be adapted following the observational determinations (i.e. $q_{0}= 2.04_{-0.22}^{+0.25}$ \cite{PPTA2}). For this reason in the right plot of Fig. \ref{FIG6} the constraints can be directly examined in the plane $(N_{t}, \, N_{\ast})$. 
\begin{figure}[!ht]
\centering
\includegraphics[height=7.5cm]{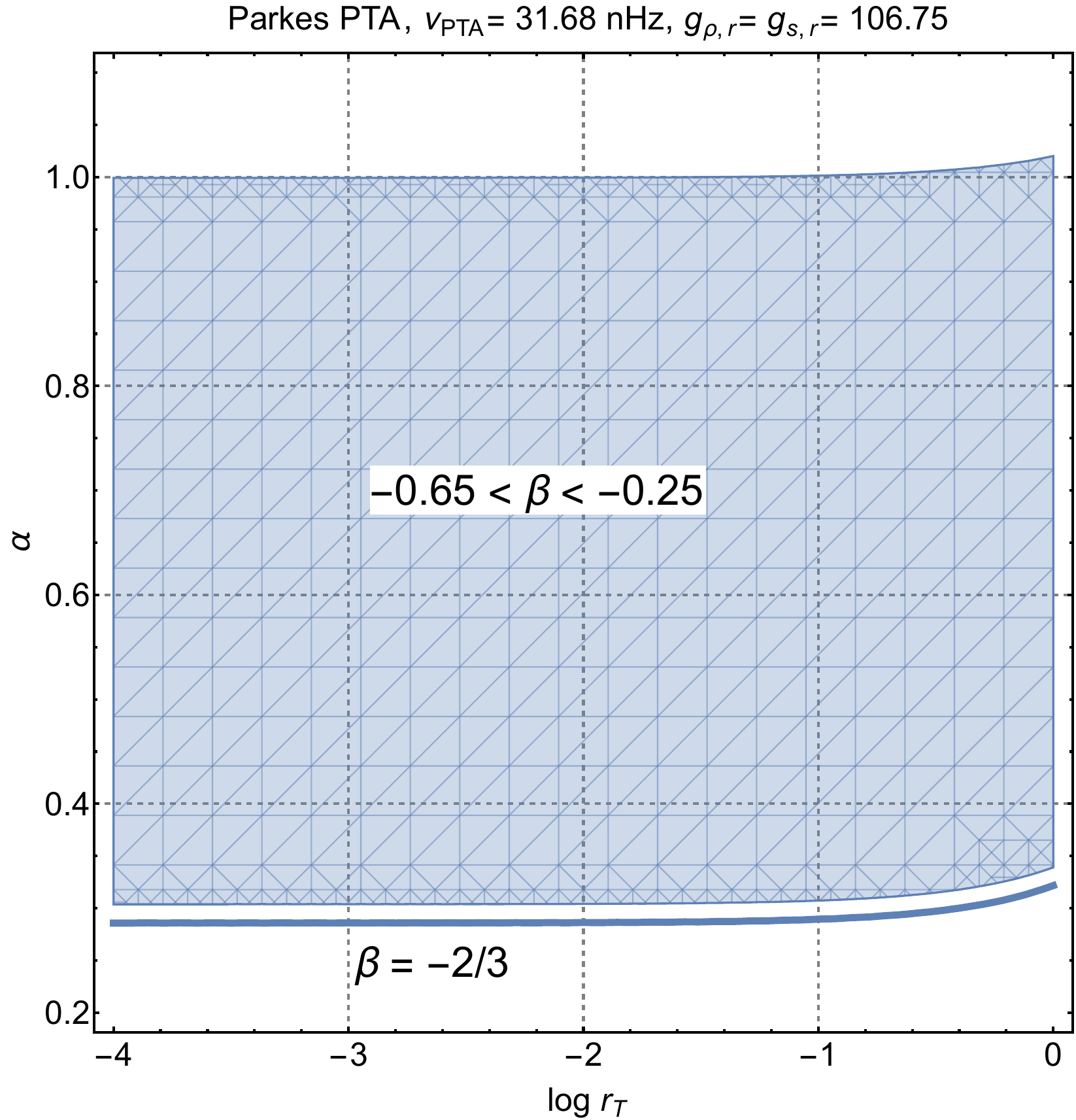}
\includegraphics[height=7.5cm]{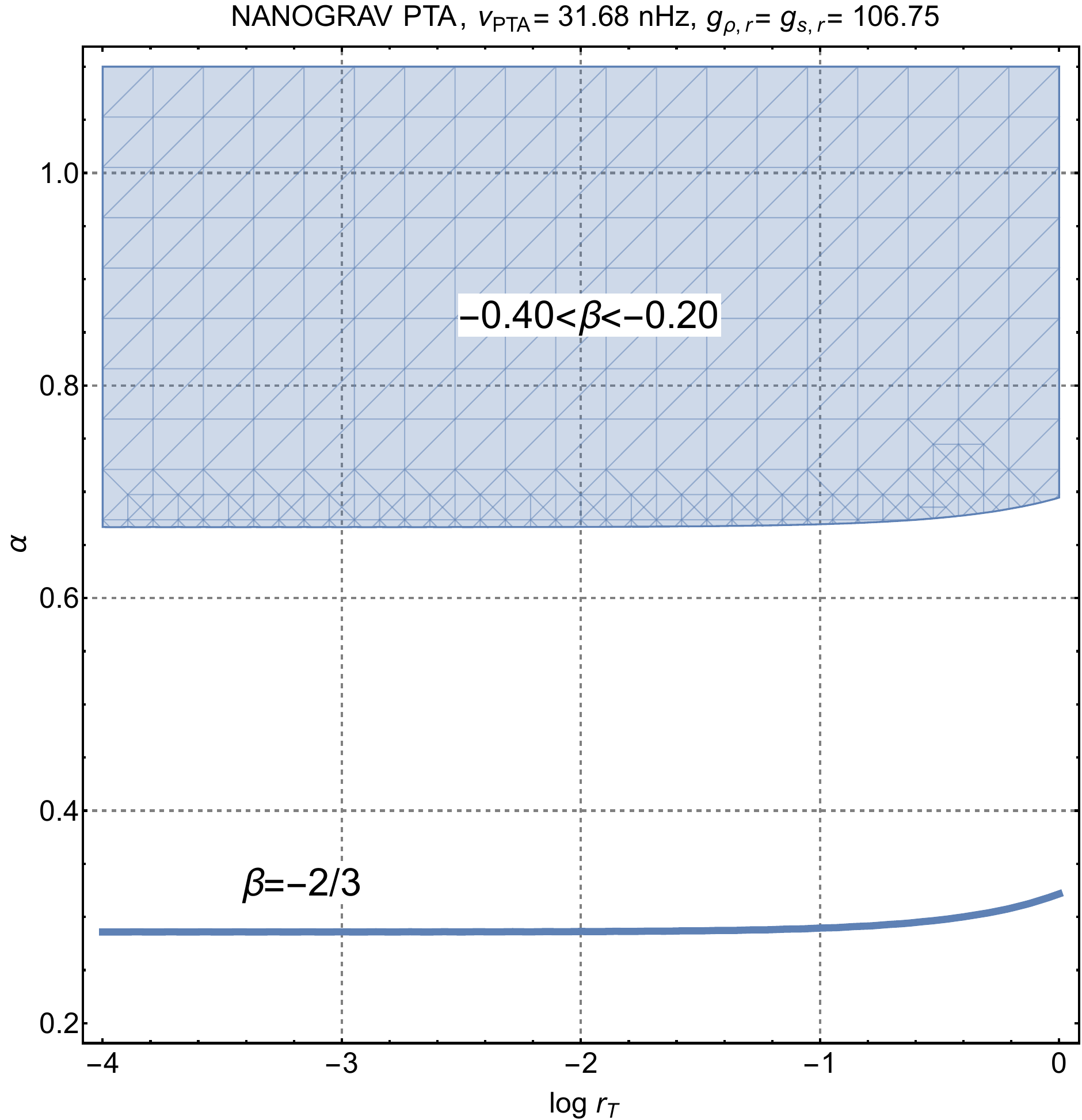}
\caption[a]{The shaded region follows by requiring $-0.65< \beta< -0.25$ (as implied by Eq. (\ref{STHR12})) 
and by imposing the relation of Eq. (\ref{STHR11}). The thick line outside the shaded region 
corresponds to $\beta = -2/3$. If the consistency relations are imposed $\beta$ is determined in the $(\alpha, r_{T})$ plane. However, as the plot shows, the smallness of $r_{T}$ implies that $\epsilon_{k}$ can be neglected for the determination of the spectral index in the region $\nu<\nu_{\ast}$.}
\label{FIG7}      
\end{figure}
For the sake of accuracy, in Fig. \ref{FIG7} we illustrated the region pinned down in the plane ($\alpha$, $\log{r_{T}}$) by the different values of $\beta$. The same analysis illustrated in the case of the Parkes PTA can be repeated for the NANOgrav determinations with slightly different results; the analog of Eq. (\ref{STHR12}) is now given by \cite{NANO2}
\begin{equation}
-0.40 \leq \beta \leq -0.20, \qquad \qquad 3.7 < q_{0} < 10.6.
\label{STHR13}
\end{equation}
\begin{figure}[!ht]
\centering
\includegraphics[height=7.5cm]{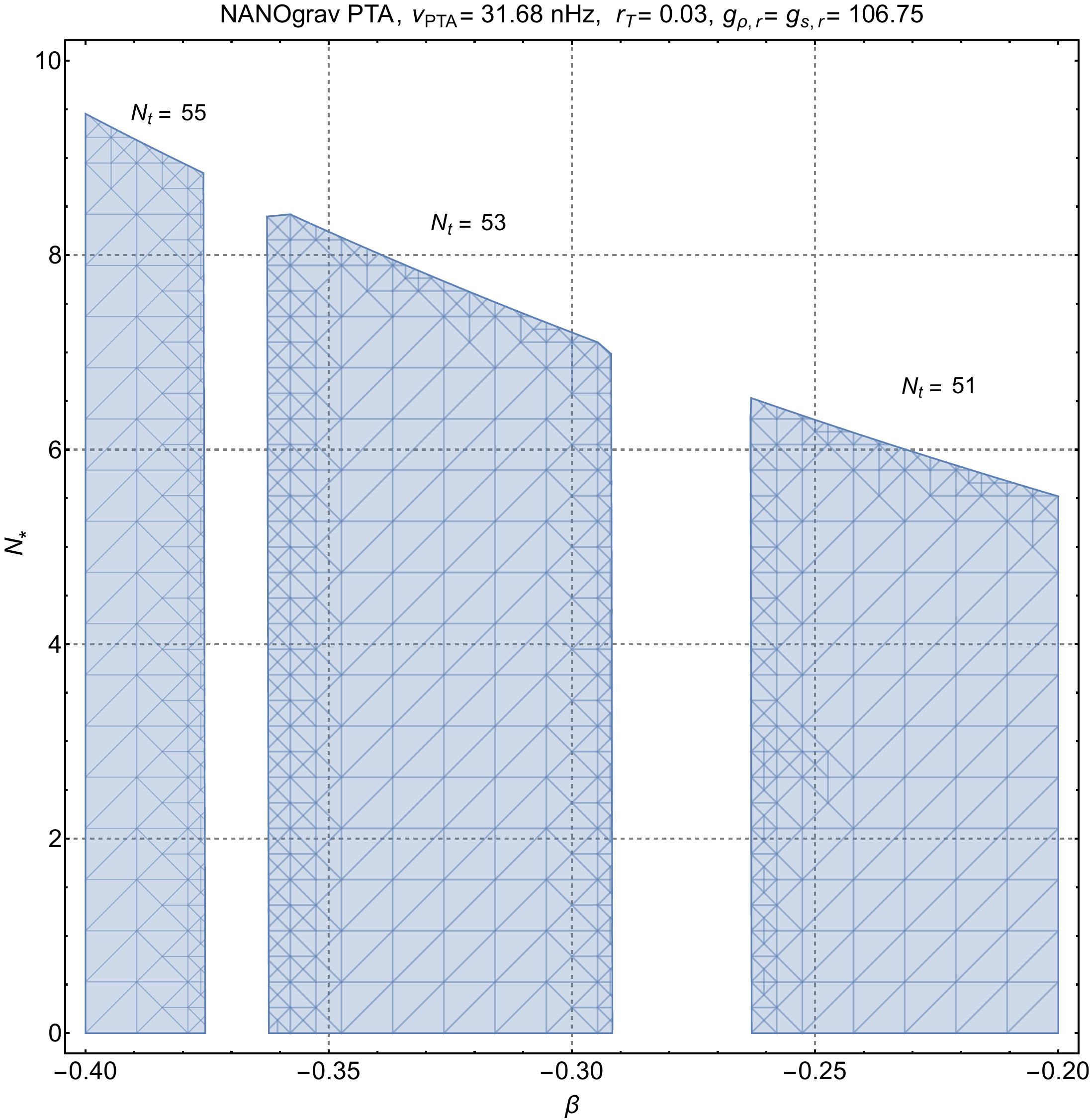}
\includegraphics[height=7.5cm]{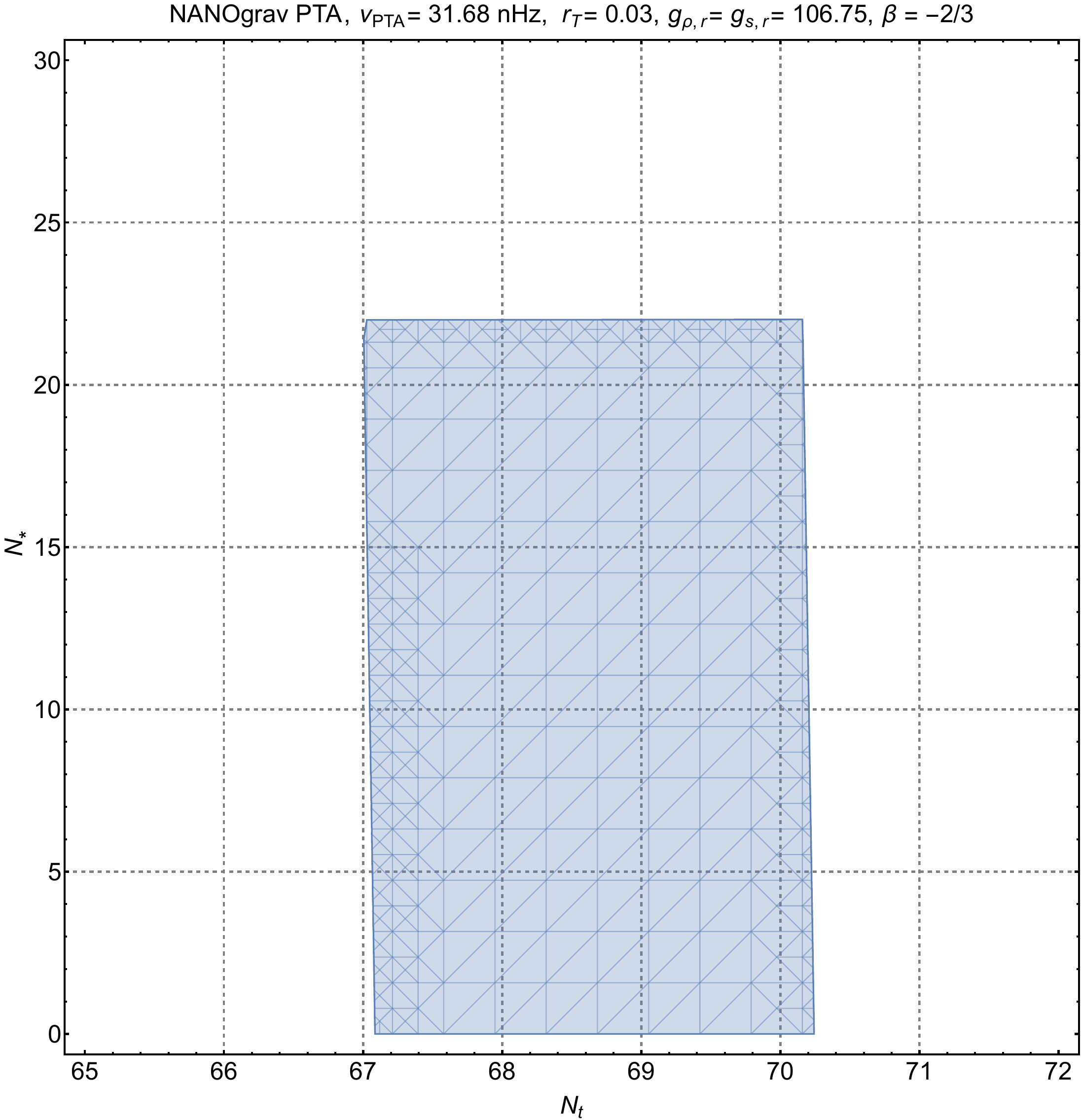}
\caption[a]{In the left plot we illustrate the region pinned down by the NANOgrav PTA in the plane 
$(\beta,\, N_{\ast})$ for different values of $N_{t}$. Both plots can be  compared with the ones of Fig. \ref{FIG6}
where we discussed the case of the Parkes PTA. The 
shaded regions in the left plot follow by imposing Eq. (\ref{STHR13}) for different values of $N_{t}$. The right plot is obtained 
by fixing $\beta\to -2/3$ and the shaded region corresponds to the range $1.8< q_{0}(N_{t},N_{\ast}) < 3.1$; this is because 
for $\beta\to -2/3$ the NANOgrav collaboration suggests $q_{0}= 2.4_{-0.6}^{0.7}$. As in Fig. \ref{FIG6} the high-frequency 
bounds from the audio band have been also imposed.}
\label{FIG8}      
\end{figure}
While the range of $\beta$ given in Eq. (\ref{STHR13}) is narrower than in Eq. (\ref{STHR12}), in the case of $q_{0}$  we observe the opposite: the allowed values of $q_{0}$ of Eq. (\ref{STHR13}) are comparatively larger than the ones of Eq. (\ref{STHR12}). These differences slightly modify the shaded regions of Figs. \ref{FIG6} and \ref{FIG8} and the related parameters:
both the ranges of $N_{t}$ and  $N_{\ast}$ get narrower and $N_{t}$ is at most 
${\mathcal O}(55)$ while $N_{\ast} \leq {\mathcal O}(10)$.

A second class of constraints determining the shaded regions of Figs. \ref{FIG6} and \ref{FIG8} is related to the direct bounds from the operating wide-band detectors. In particular we remind that the LIGO, Virgo and Kagra collaborations (LVK) reported a constraint \cite{LIGO1} implying:
\begin{equation}
\Omega_{gw}(\nu, \tau_{0}) < 5.8 \times 10^{-9}, \qquad\qquad 20 \,\, \mathrm{Hz} < \nu_{L} < 76.6 \,\, \mathrm{Hz},
\label{CONS2}
\end{equation}
in the case of a flat spectral energy density; in the present notations $\nu_{L}$ indicates the LIGO-Virgo-Kagra frequency.  The limit of Eq. (\ref{CONS2}) improves on a series of bounds previously deduced by the wide-band interferometers (see Ref. \cite{MG1} for a review of the older results); in particular 
in Ref. \cite{LIGO2} the analog of Eq. (\ref{CONS2}) implied $ \Omega_{gw}(\nu, \tau_{0}) < 6 \times 10^{-8}$ for a comparable frequency interval and always in the case of a flat spectral energy density. The bound of Eq. (\ref{CONS2}) can be
used also in Eq. (\ref{STHR7}) since at high-frequency the spectral energy density is 
nearly scale-invariant. The results of Ref. \cite{LIGO1} report however a threefold bound for a handful 
of spectral slopes; in particular, if the spectral energy density is parametrized as
\begin{equation}
\Omega_{gw}(\nu, \tau_{0}) = \overline{\Omega}(\sigma) \biggl(\frac{\nu}{\nu_{L}}\biggr)^{\sigma}, \qquad \qquad \nu_{L} = 25 \,\, \mathrm{Hz},
\label{NOT1}
\end{equation}
the limits of Ref. \cite{LIGO1} read 
$\overline{\Omega}(0) < 5.8 \times 10^{-9}$ (valid in the case $\sigma =0$),
$\overline{\Omega}(2/3) < 3.4 \times 10^{-9}$ (when $\sigma= 2/3$) 
and $\overline{\Omega}(3) < 3.9 \times 10^{-10}$ (when $\sigma= 3$).
As the value of $\sigma$ increases the bound becomes more restrictive 
for a fixed reference frequency and the three previous results are summarized by the following interpolating formula:
\begin{equation}
\log{\overline{\Omega}}(\sigma) < -\,8.236 -\, 0.335\, \sigma- 0.018\, \sigma^2.
\label{NOT2}
\end{equation}
 Since in the present case the bound (\ref{NOT2}) should be applied at high-frequencies we will have $ \sigma =- 2 \epsilon_{k}/ (1- \epsilon_{k})$ with $\epsilon \ll 0.1$; to leading order in $\epsilon_{k}$, 
Eq. (\ref{NOT2}) implies that $\log{\overline{\Omega}}(\epsilon_{k}) < -\,8.236 -\, 0.335\, \epsilon_{k} -0.393 \epsilon_{k}^2$.

\subsection{Shapes of the spectra and further possibilities}
Instead of using the approximations employed above it is instructive to compute numerically the spectral energy density  from the exact form of the mode functions\footnote{As we saw the pivotal parameters that determine the spectrum are $\alpha$, $N_{\ast}$ and $N_{t}$. Recalling Eq. (\ref{STHR11}) we trade $\alpha$ for $\beta$ at a fixed value of $\epsilon_{k}$ (or $r_{T}$) and express the spectral index directly with the notations preferred by the experimental collaborations. }.

\begin{figure}[!ht]
\centering
\includegraphics[height=5.7cm]{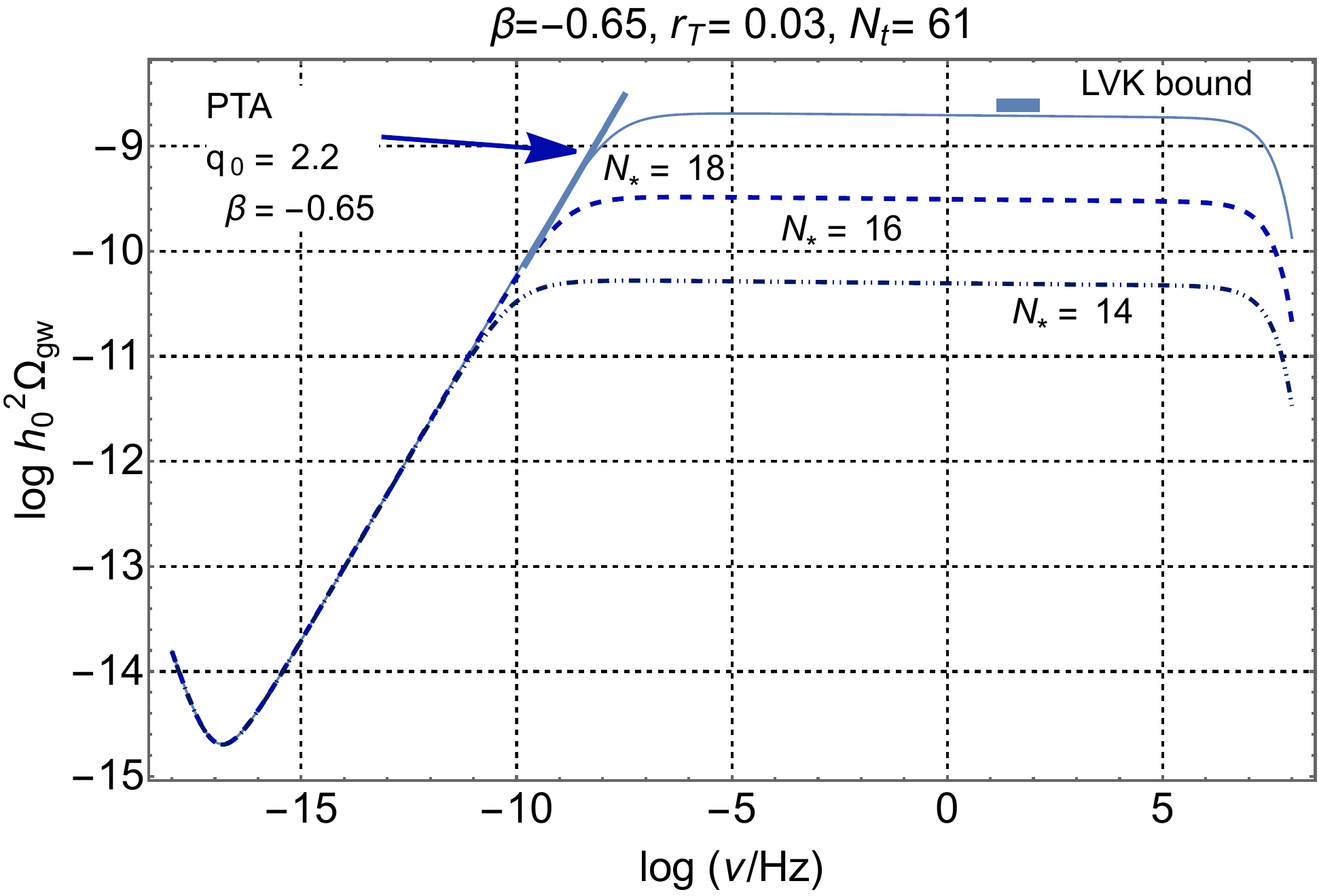}
\includegraphics[height=5.7cm]{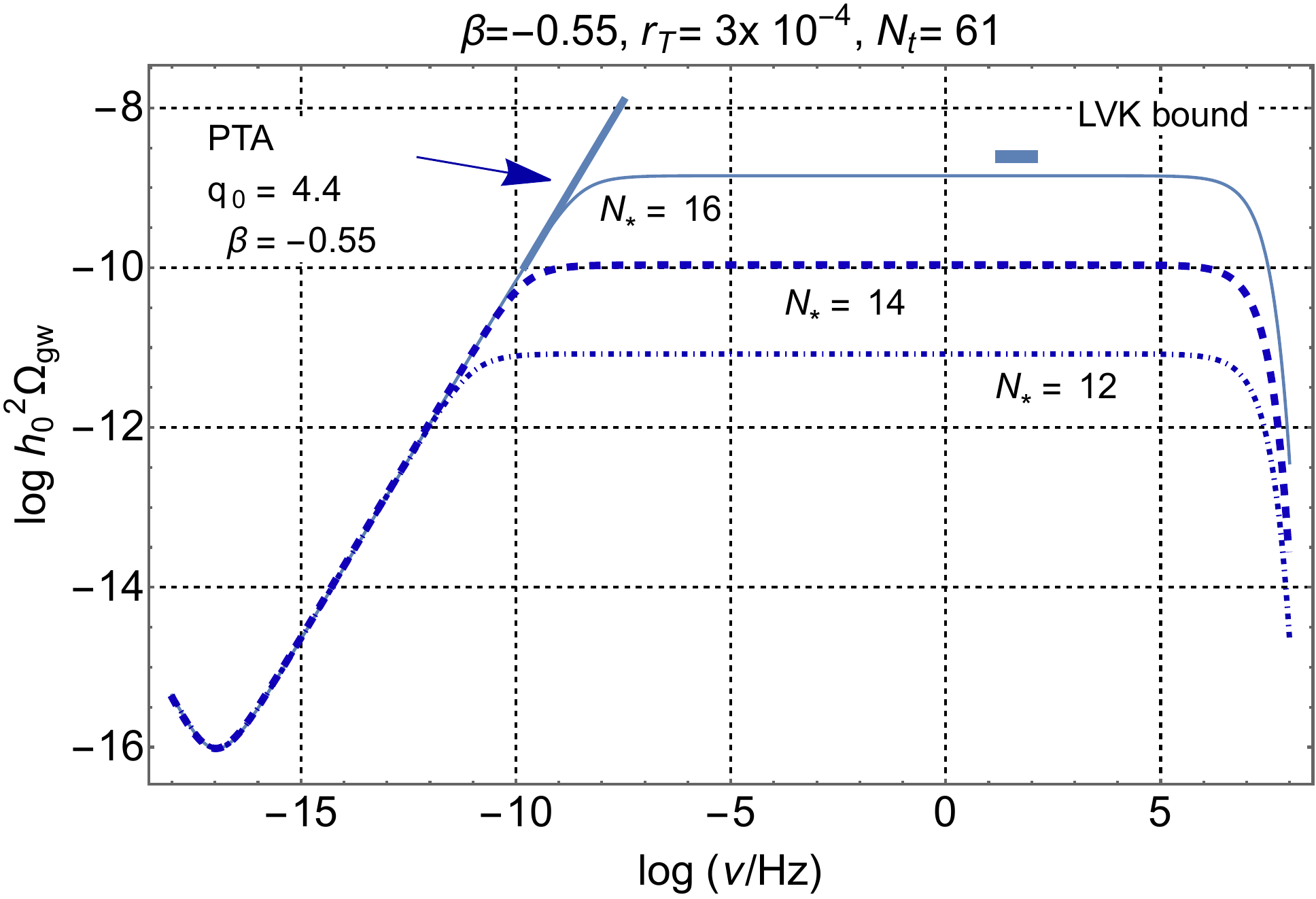}
\caption[a]{We illustrate the common logarithm of the spectral energy density in critical units as a function of the common logarithm of the comoving frequency. In both plots $N_{t} =61$ but the values of $\beta$ and $r_{T}$ do not coincide and they are indicated above each of the two cartoons. The arrows indicate the PTA signal for the spectral indices corresponding 
to the ones selected in each of the plots. The high-frequency region labeled by LVK refers to the Ligo-Virgo-Kagra bound that applies in the audio band. The increasing branch and the flat plateau corresponds, respectively, to the analytic estimate of Eqs. (\ref{STHR5}) and (\ref{STHR7}).}
\label{FIG8a}      
\end{figure}

As an example in the two plots of Fig. \ref{FIG8a} we considered two different values of $\beta$ (i.e. $\beta = -0.65$ and $\beta= -0.55$). Given a specific value of $\beta$ within the observational 
range, the previous results (see, in particular, Figs. \ref{FIG6} and \ref{FIG8}) lead directly to the 
allowed values of $N_{\ast}$ and $N_{t}$. If $N_{\ast}$ and $N_{t}$ are of the 
same order the refractive index stops evolving when inflation approximately ends and,
in this case, it is impossible to get a 
large signal in the nHz range without jeopardizing the big-bang nuclosynthesis constraint  \cite{CC2,CC3}.  
Conversely,  when $N_{*} <N_{t}$ the refractive index stops evolving well before the onset of the post-inflationary stage, i.e. when the background is still inflating deep inside the quasi-de Sitter stage of expansion. In both plots of Fig. \ref{FIG8a}, to ease the comparison, we selected $N_{t} = 61$ while different values of $N_{\ast}$ are illustrated. In both plots, for the same choice of the parameters, we also illustrated (with an arrow) the PTA excess and the Ligo-Virgo-Kagra bound \cite{LIGO1,LIGO2}. The PTA signal occurs for typical frequencies 
${\mathcal O}(\nu_{ref})$ while the LVK bound applies approximately between
$25$ and $100$ Hz. 

\begin{figure}[!ht]
\centering
\includegraphics[height=5.4cm]{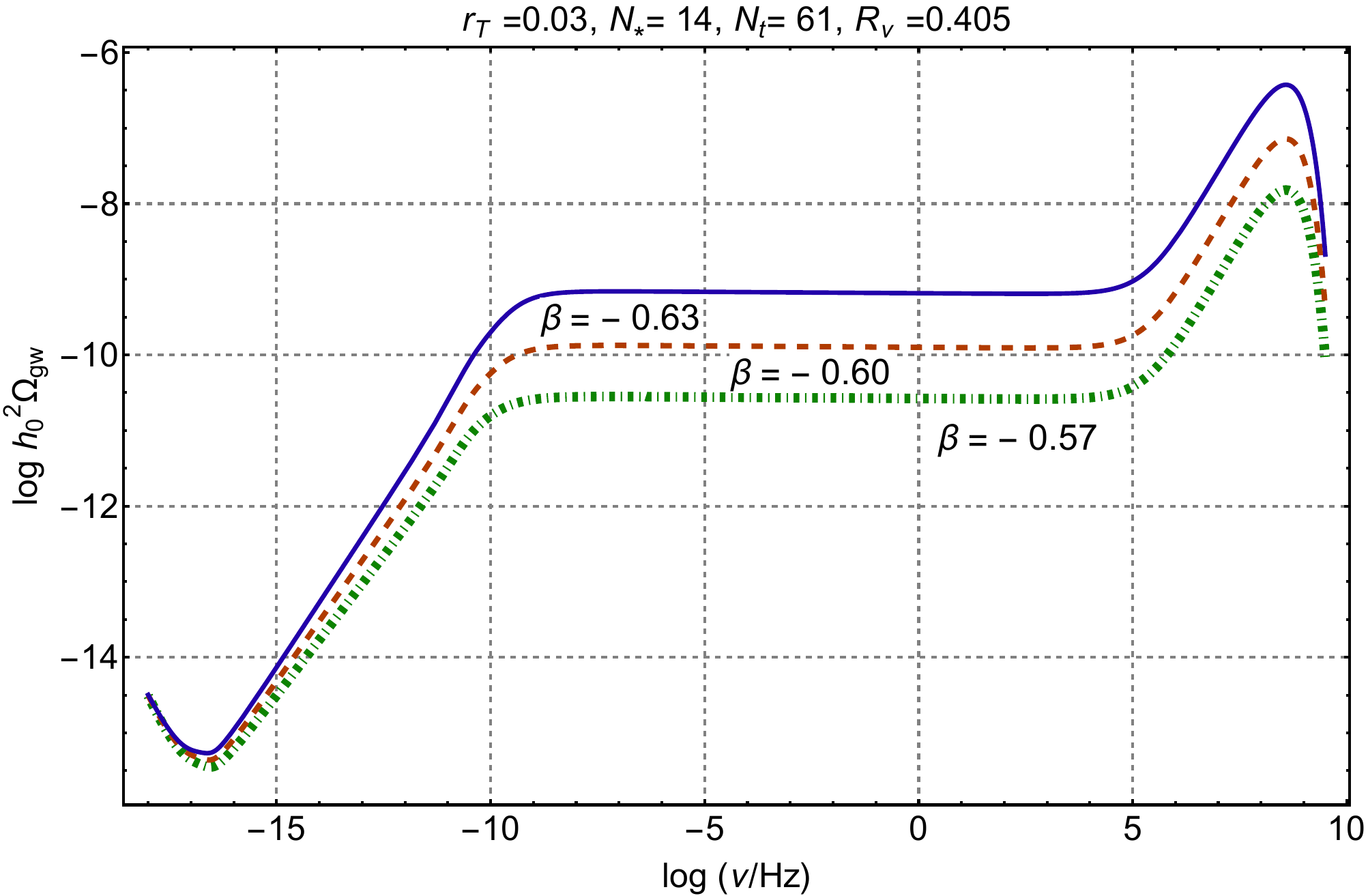}
\includegraphics[height=5.4cm]{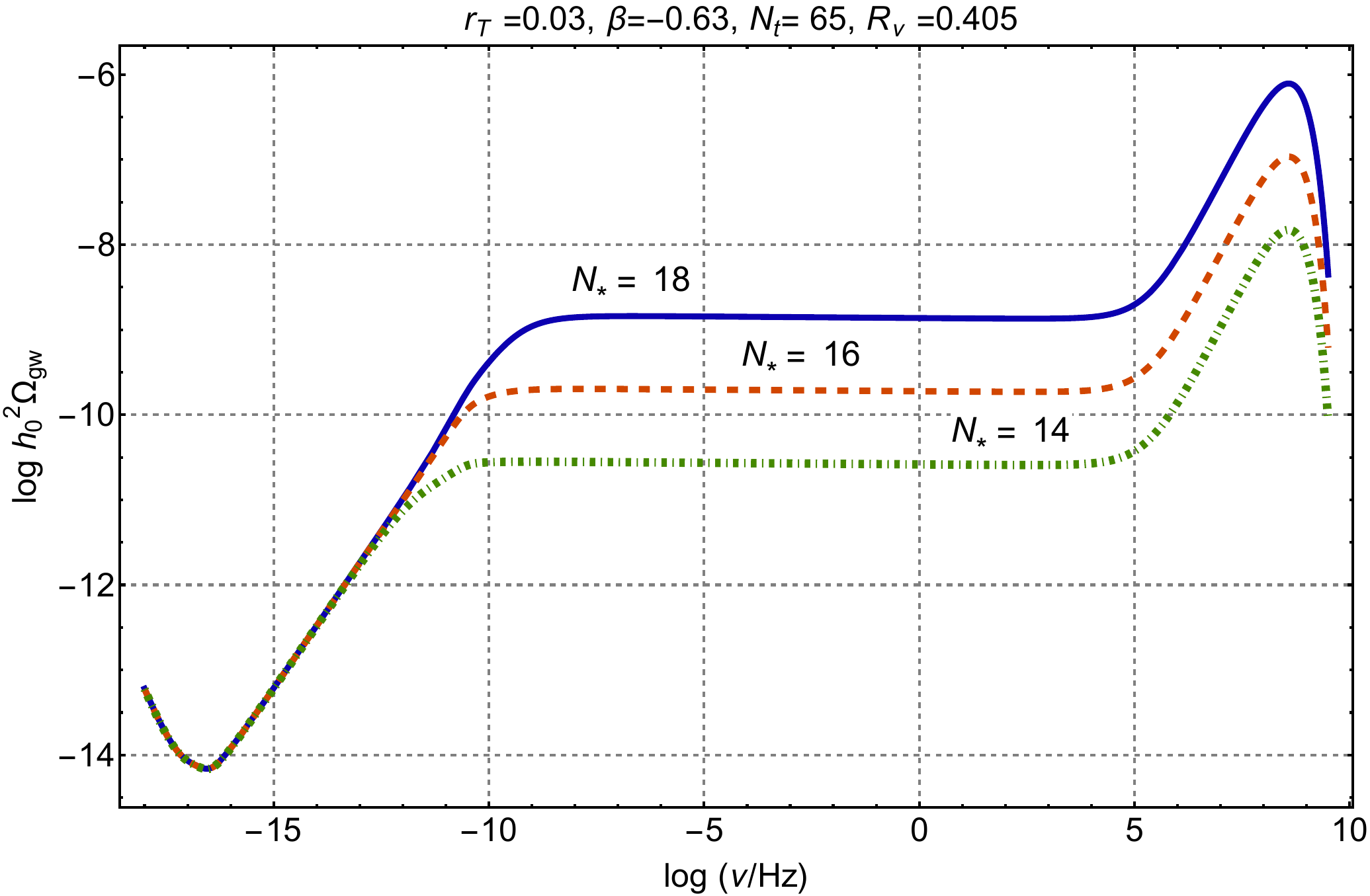}
\caption[a]{As in Fig. \ref{FIG8a} we illustrate the common logarithm of the spectral energy density as a function of the common logarithm of the 
comoving frequency. In the two plots the value of $r_{T}$ is the same but the values of $N_{t}$
are slightly dissimilar. In the plot at the left $N_{\ast}= 14$ while the three spectra correspond 
slightly different values of $\beta$. In the plot at the right $\beta = -0.63$ and the three curves 
illustrate the variation of $N_{\ast}$. Since the effect of neutrino free-streaming has been included, in both plots $R_{\nu}$ denotes the neutrino 
fraction.} 
\label{FIG8b}      
\end{figure}

The discussion of section \ref{sec2} does not exclude the possibility of two 
concurrent modifications of the comoving horizon operating before and after 
the end of inflation \cite{CC4,CC5}. This viewpoint is explored in Fig. 
\ref{FIG8b} where we consider the possibility that the refractive index 
stops its evolution well before the end of inflation (i.e. $N_{\ast} \ll N_{t}$);
however, unlike the case of Fig. \ref{FIG8a}, the post-inflationary evolution 
includes a  long phase expanding at a rate slower than radiation.
The spectral energy density in critical units will therefore have three 
different slopes for $\nu > \nu_{eq}$. In both plots of Fig. \ref{FIG8b}, at intermediate 
frequencies $h_{0}^2 \Omega_{gw}(\nu,\tau_{0})$ 
has the same intermediate slopes appearing in Fig. \ref{FIG8a} (see also Eqs. (\ref{STHR9})--
and (\ref{STHR11})). However, after the quasi-flat plateau, the spectral energy density 
exhibits a further increasing branch before the maximal frequency. The corresponding 
wavelengths left the comoving Hubble radius during inflation and reentered
in the post-inflationary stage before radiation dominance. In Fig. \ref{FIG8b} the high-frequency 
spectral slope is ${\mathcal O}(1)$ since during the post-inflationary stage 
the evolution is described by a stiff fluid with $\delta \simeq 1/2$ implying 
that $(a\, H)^{-1} \propto a^2$.
The main difference between the plots of Figs. \ref{FIG8a} and \ref{FIG8b}
ultimately comes from the high-frequency shape. While 
in the case of Fig. \ref{FIG8a} the most relevant constraint 
comes from the operating interferometers \cite{LIGO1,LIGO2}, 
in the case of Fig. \ref{FIG8b} the bounds coming from big-bang nucleosynthesis \cite{bbn1,bbn2,bbn3} must be taken into account since they imply:
\begin{equation}
h_{0}^2  \int_{\nu_{bbn}}^{\nu_{max}}
  \Omega_{\mathrm{gw}}(\nu,\tau_{0}) d\ln{\nu} = 5.61 \times 10^{-6} \Delta N_{\nu} 
  \biggl(\frac{h_{0}^2 \Omega_{\gamma0}}{2.47 \times 10^{-5}}\biggr),
\label{BBB1}
\end{equation}
where $\Omega_{\gamma0}$ is the (present) critical fraction of CMB photons.
As it is well known, the limit  of Eq. (\ref{BBB1}) also sets an indirect constraint  on all the extra-relativistic species (and, among others, on the relic gravitons). If applied to massless fermionic species, the limit is often expressed for practical reasons  in terms of $\Delta N_{\nu}$ representing the contribution of supplementary neutrino species. The actual bounds on $\Delta N_{\nu}$ range from $\Delta N_{\nu} \leq 0.2$ to $\Delta N_{\nu} \leq 1$;  the integrated spectral density in Eq. (\ref{BBB1}) is thus between $10^{-6}$ and $10^{-5}$. 
It is interesting to point out that the spectra of Fig. \ref{FIG8b} are sensitive 
both to the interferometric bounds \cite{LIGO1,LIGO2} and to the 
the limits of Eq. (\ref{BBB1}): the  wide-band detectors 
constrain the height of the intermediate plateau while Eq. (\ref{BBB1}) sets a 
bound on the integrated $h_{0}^2 \, \Omega_{gw}(\nu,\tau_{0})$ and, ultimately, on the 
(global) maximum of the spectral energy density.

\renewcommand{\theequation}{4.\arabic{equation}}
\setcounter{equation}{0}
\section{Bounces of the scale factor and curvature bounces}
\label{sec4}
\subsection{Basic considerations}
In bouncing scenarios the spectral index can 
be positive between the fHz and the Hz  \cite{MG1} as it happens in the presence of a dynamical refractive index. 
However, while in the previous section both the amplitudes and the slopes could be predicted 
with reasonable accuracy thanks to the underlying inflationary dynamics, 
for the bouncing case it is comparatively easier to estimate $h_{0}^2 \, \Omega_{gw}(\nu,\tau_{0})$ in the intermediate frequency region rather than in the high-frequency domain which is often associated 
with a regime of strong curvatures. Even in the absence of a detailed theoretical derivation of the corresponding slopes, 
the high-frequency normalization can be disambiguated by employing the constraints of the audio band \cite{LIGO1,LIGO2}
and the big-bang nucleosynthesis limits \cite{bbn1,bbn2,bbn3}.

The bouncing dynamics is sometimes associated with a contracting stage (i.e. $\dot{a} <\,0$  and $H<0$), as historically suggested, along slightly different perspectives, by Tolman and Lema\^itre \cite{TL1,TL2} (see also \cite{PP1,PP2}). For the present purposes it is useful to distinguish (at a purely kinematical level) the bounces of the scale factor from the ones involving the extrinsic (Hubble) curvature.  While $H$ and $\dot{H}$ change sign (at least) once in the bounces of the scale factor, for the curvature bounces $H$ is always positive but $\dot{H}$ changes sign (at least) once. Broadly speaking the spectral slope at intermediate frequencies (which is the relevant one 
for the discussion of the PTA excesses) is related to the wavelengths 
that crossed the comoving Hubble radius before either $H$ or $\dot{H}$ 
changed their sign the first time\footnote{The distinction between the bounces of the scale factor and the curvature bounces rests on the Einstein frame where the gravity part of the action appears in its canonical form. It is however possible rephrase the same distinction, mutatis mutandis, in any conformally 
related frame since the effective action of the relic gravitons (and the spectral 
energy density) ultimately coincide after the gauge-invariant fluctuations 
of the metric and the corresponding backgrounds are correctly expressed in the new frame (see, for instance, 
Ref. \cite{MG1} and discussion therein).}. In both situations we can express the spectral energy density in terms of a common template
\begin{equation}
h_{0}^2 \,\Omega_{gw}(\nu, \tau_{0}) = \overline{\Omega}_{\ast} \,\, (\nu/\nu_{\ast})^{m_{T}}, \qquad \qquad 
\nu_{\ast} < \nu <\nu_{max},
\label{SFOU1}
\end{equation}
where $m_{T}$ is the high-frequency spectral index and $\overline{\Omega}_{\ast}$ 
accounts for the corresponding normalization. The notations employed in Eq. (\ref{SFOU1}) are purposely similar to the ones of the previous section 
since, in both cases, $\overline{\Omega}_{\ast}$  controls the high-frequency normalization. The physical meaning of the two quantities 
is however slightly different since $\overline{\Omega}_{\ast}$  now depends on the maximal curvature scale at the bounce (of the order of $H_{1}$) and on a number of other late-times parameters; in the simplest situation (where all the 
wavelengths reenter during a radiation dominated stage) the value of $\overline{\Omega}_{\ast}$ is given by
\begin{equation}
\overline{\Omega}_{\ast} = \frac{4 \, h_{0}^2 \Omega_{R0}}{3 \pi} \, \biggl(\frac{H_{1}}{M_{P}}\biggr)^2\, \biggl(\frac{g_{\rho, \, r}}{g_{\rho, \, eq}}\biggr) \biggl(\frac{g_{s,\, eq}}{g_{s,\, r}}\biggr)^{4/3} \, 
\biggl(\frac{\Omega_{M0}}{\Omega_{\Lambda}}\biggr)^2.
\label{SFOU4}
\end{equation}
In inflationary scenarios the analog of $(H_{1}/M_{P})$ is fixed by the amplitude of the (adiabatic and Guassian) curvature inhomogeneities 
whereas in the case of Eq. (\ref{SFOU4}) it is more productive to determine $\overline{\Omega}_{\ast}$ directly from the available phenomenological constraints and to confront the obtained templates with the PTA observations. In this 
respect the most relevant limits are the ones coming from the audio band (see Eqs. (\ref{NOT1})--(\ref{NOT2})) and the big-bang 
nucleosynthesis constraint (see Eq. (\ref{BBB1}) and discussion thereafter).

\subsection{Growing spectral slopes at intermediate frequencies}
While Eq. (\ref{SFOU4}) applies in the high-frequency regime where the amplitude of $h_{0}^2 \, \Omega_{gw}(\nu,\tau_{0})$ 
is constrained by the current phenomenological bounds, in the intermediate frequency range (i.e. for $\nu< \nu_{\ast}$) the slope of the spectral energy density 
is instead denoted by $n_{T}= n_{T}(\gamma, \delta)$:
\begin{eqnarray}
h_{0}^2\, \Omega_{gw}(\nu,\tau_{0}) = \overline{\Omega}_{\ast} \,\, (\nu/\nu_{\ast})^{n_{T}(\gamma, \delta)},\qquad\qquad \nu_{eq} < \nu <\nu_{\ast}.
\label{SFOU2}
\end{eqnarray}
In Eq. (\ref{SFOU2}) $\gamma$ and $\delta$ control, respectively, 
the expansion (or contraction) rates at early and late times; the general form of the spectral index 
can be expressed as:
\begin{equation}
n_{T}(\gamma, \delta) = 4 - 2 \biggl|\delta - \frac{1}{2}\biggr| - 2\biggl|\frac{\gamma}{1 - \gamma} - \frac{1}{2} \biggr|.
\label{SFOU5}
\end{equation}
where the value of $\gamma$ is related to the expansion (or contraction)
rate in the cosmic time coordinate as $a(t) = a_{1}(- t/t_{1})^{\gamma}$ and this parametrization is valid before  the first zero of $H$ or $\dot{H}$. As in the previous sections also in Eq. (\ref{SFOU5}) $\delta$ denotes the expansion rate in the conformal time coordinate during the decelerated stage after the bounce. For $0 < \gamma < 1$ Eq. (\ref{SFOU5}) gives the spectral slope when, prior to the bouncing regime (taking place for $|t | < t_{1}$), the background experiences a stage 
of accelerated contraction (i.e. $\dot{a} < 0$ and $\ddot{a} <0$). Conversely when $\gamma < 0$ 
the background expands and accelerates (i.e. $\dot{a} >0$ and $\ddot{a} >0$) with growing Hubble rate (i.e. $\dot{H} >0$). 
If $\delta > 1/2$ the contribution to $n_{T}(\gamma,\delta)$ only comes from the 
term $k^4\, |a_{re}/a_{ex}|^2$ in Eq. (\ref{STWO1}) 
while ${\mathcal Q}_{k}(\tau_{ex}, \tau_{re})\to 1$. Conversely, when $\delta< 1/2$, 
${\mathcal Q}_{k}(\tau_{ex}, \tau_{re})\propto 1 + |k \, \tau_{1}|^{-1 + 2 \delta}$ which is much larger 
than $1$ for all the amplified modes of the spectrum (i.e. for $k \tau_{1} \ll 1$). Since the second term in ${\mathcal Q}_{k}(\tau_{ex}, \tau_{re})$ dominates for $\delta <1/2$ and for $k \tau_{1} < 1$ we have 
that the spectral energy density 
$\Omega_{gw}(k, \tau) \propto k^4\, |a_{re}/a_{ex}|^2 \bigl|{\mathcal Q}_{k}(\tau_{ex}, \tau_{re})\bigr|^2$ has overall a spectral slope (valid for a generic value of $\delta$) 
proportional to $| \delta -1/2|$. After matter-radiation equality the same analysis implies: 
\begin{equation}
h_{0}^2 \, \Omega_{gw}(\nu, \tau_{0}) = \overline{\Omega}_{\ast} \,\, (\nu/\nu_{\ast})^{n_{T}(\gamma, \delta)}\,\,(\nu/\nu_{eq})^{-2},\qquad\qquad \nu <\nu_{eq}.
\label{SFOU3}
\end{equation}
\begin{figure}[!ht]
\centering
\includegraphics[height=8cm]{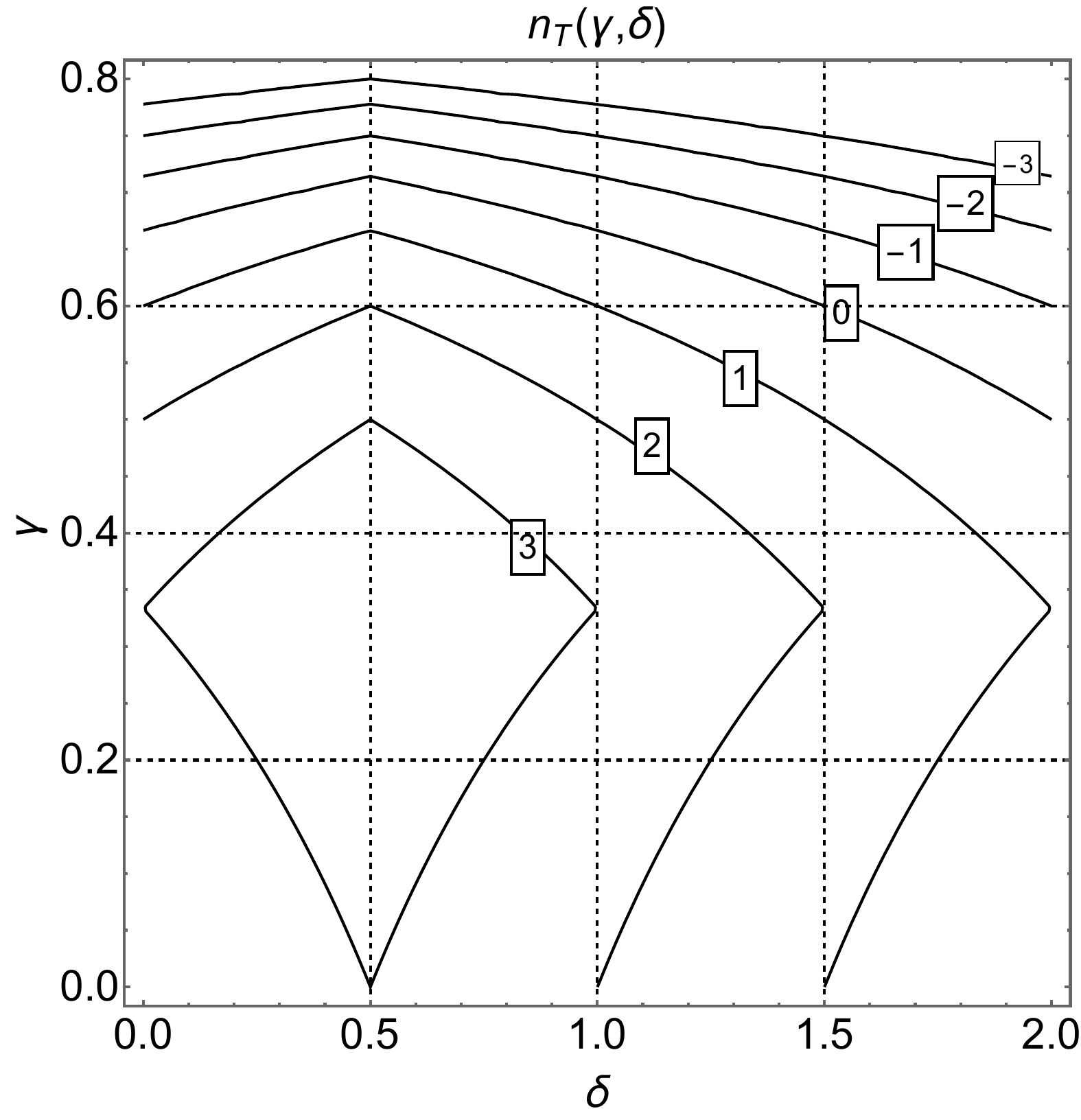}
\includegraphics[height=8cm]{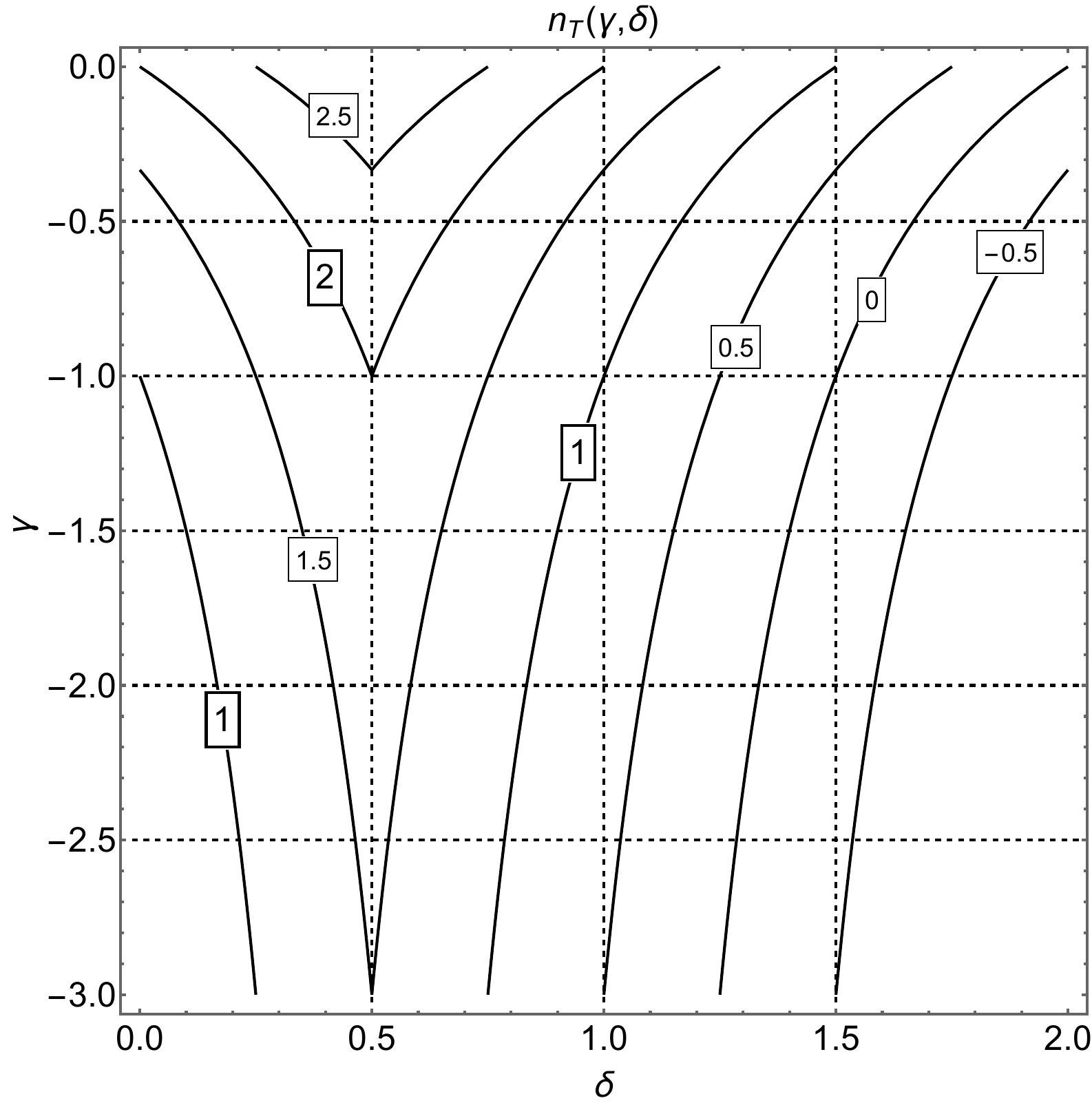}
\caption[a]{We illustrate the intermediate spectral index $n_{T}(\gamma,\delta)$ appearing in Eqs. (\ref{SFOU2})--(\ref{SFOU5}). 
In both plots $\gamma$ indicates the expansion (or 
contraction) rates before the bounce; $\delta$ denotes instead the expansion 
rate at late times (after the bounce) when the intermediate wavelengths of the spectrum reenter the Hubble radius.  In the left plot $0\leq \gamma< 1$ and the evolution before the bouncing regime is characterized by 
an accelerated contraction. In the right plot $\gamma < 0$ implying a phase of accelerated expansion (with growing Hubble rate) before the bouncing regime.}
\label{FIG9}      
\end{figure}
Below ${\mathcal O}(100)$ aHz $h_{0}^2 \, \Omega_{gw}(\nu, \tau_{0})$ 
is much smaller than in the case of the concordance paradigm. In particular 
for typical frequencies $ \nu = {\mathcal O}(\nu_{p})$ (where $\nu_{p} =k_{p}/(2\pi) = 
3.092 \, \mathrm{aHz}$) the spectral energy density of Eq. (\ref{SFOU3}) is further 
suppressed by the term  $(\nu_{p}/\nu_{\ast})^{n_{T}(\gamma, \delta)}$ in comparison with 
the case of the concordance paradigm where this contribution is ${\mathcal O}(1)$.
\begin{figure}[!ht]
\centering
\includegraphics[height=8cm]{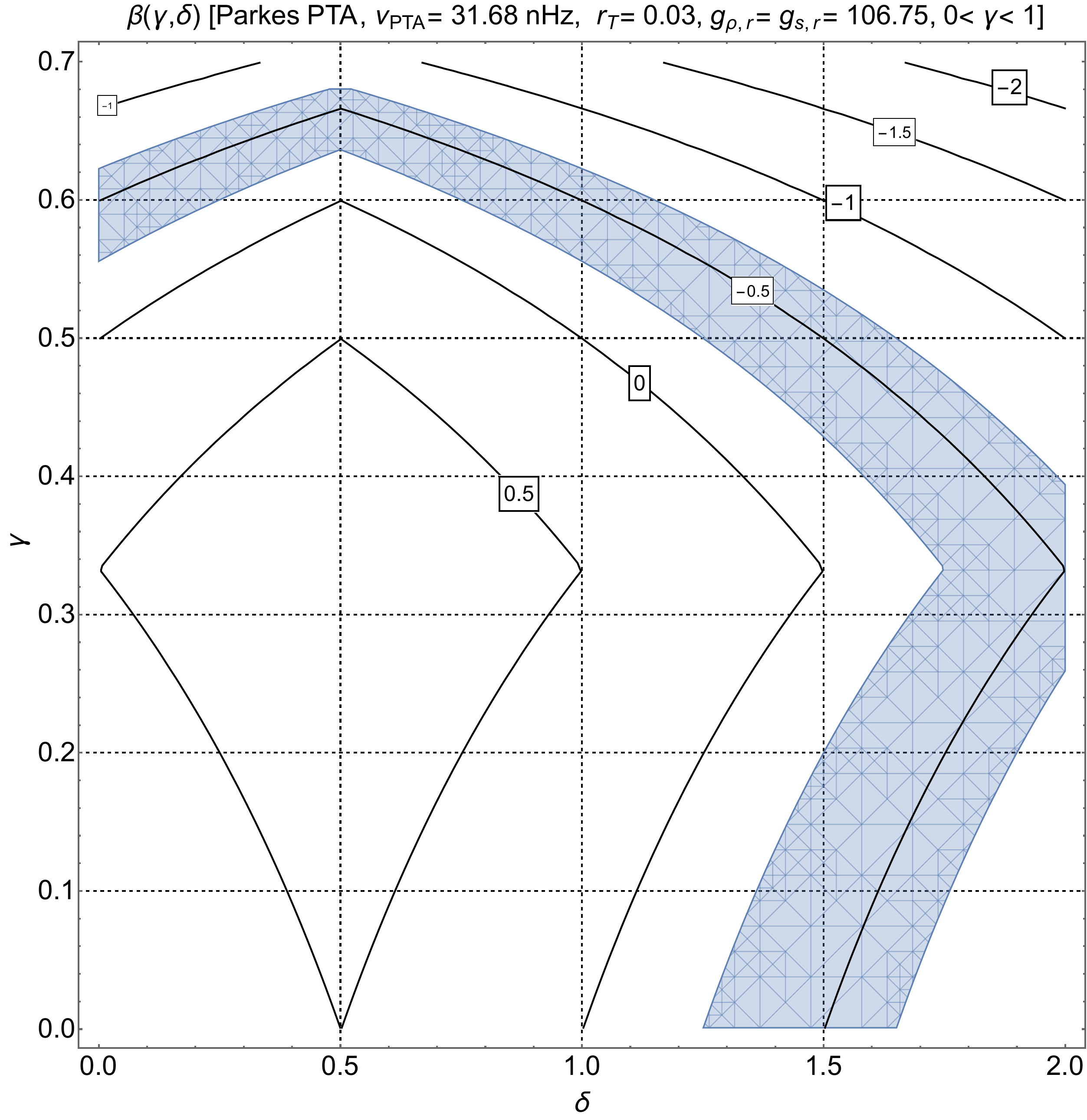}
\includegraphics[height=8cm]{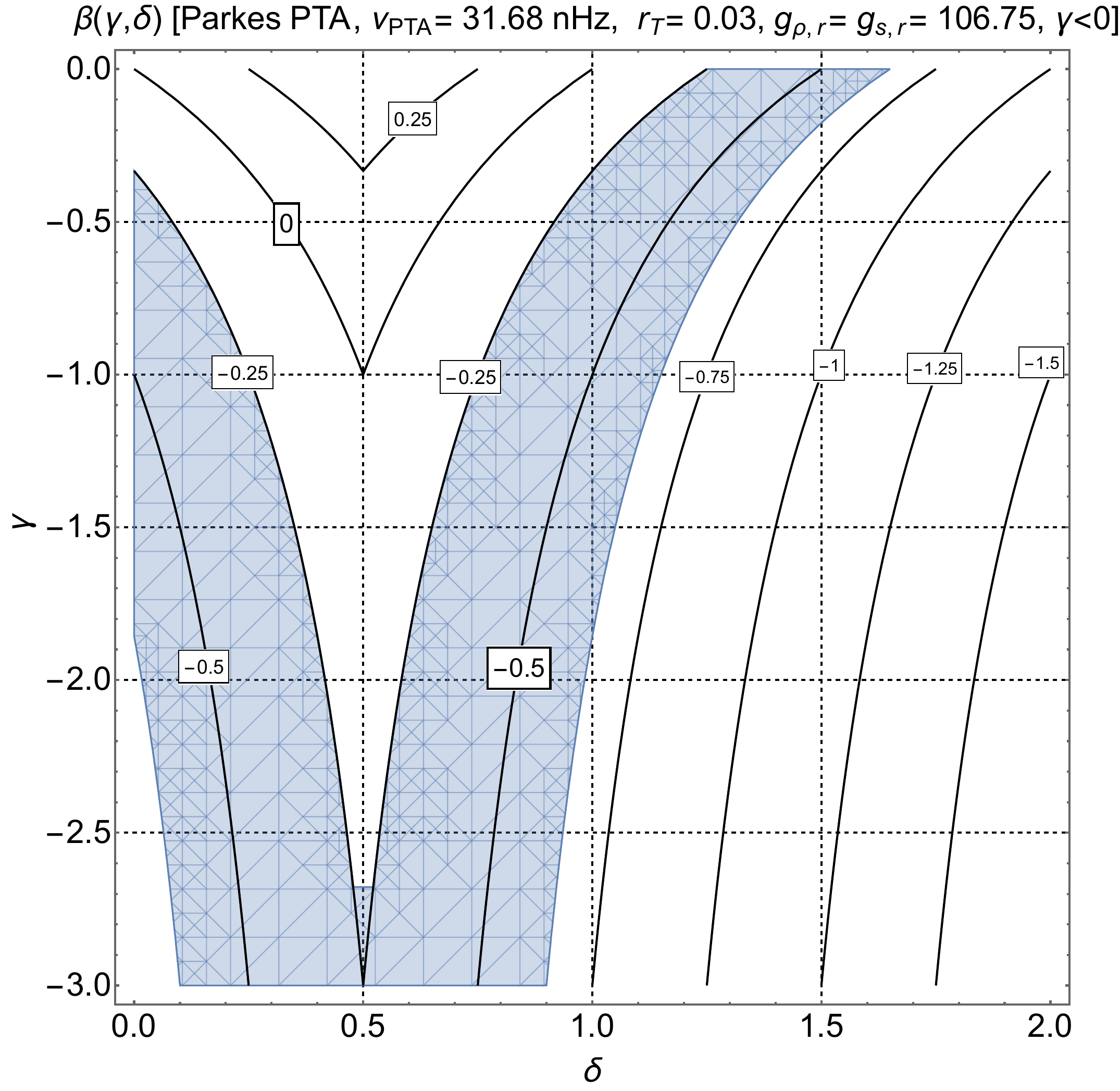}
\caption[a]{The values of $\beta(\gamma,\delta)$ are illustrated together with 
the observational constraints derived from the Parkes PTA data. The shaded region 
denotes, in both plots, the range $-0.65< \beta(\gamma,\delta) < -0.25$ which is 
pinned down by the Parkes PTA \cite{PPTA2}. In the left plot $0\leq \gamma <1$ and this values  
describe a stage of accelerated contraction, as it happens for the bounces of the scale factor. In the right plot we discuss the case $\gamma <0$ (i.e. accelerated expansion with growing expansion rate) 
that occurs for the bounces of the scale factor.}
\label{FIG10}      
\end{figure}
\subsection{The slopes and amplitudes of the PTA excesses}
\begin{figure}[!ht]
\centering
\includegraphics[height=8cm]{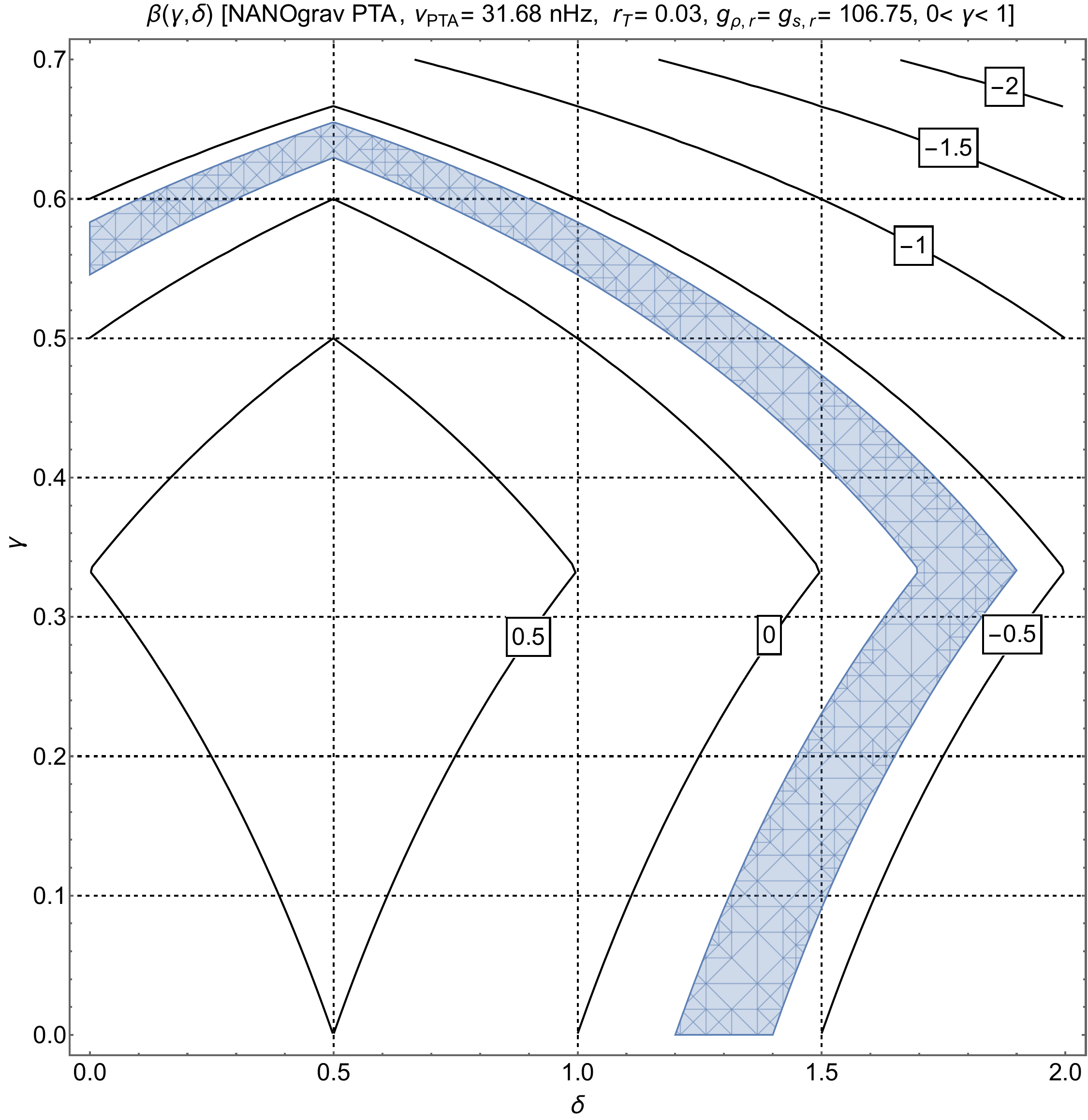}
\includegraphics[height=8cm]{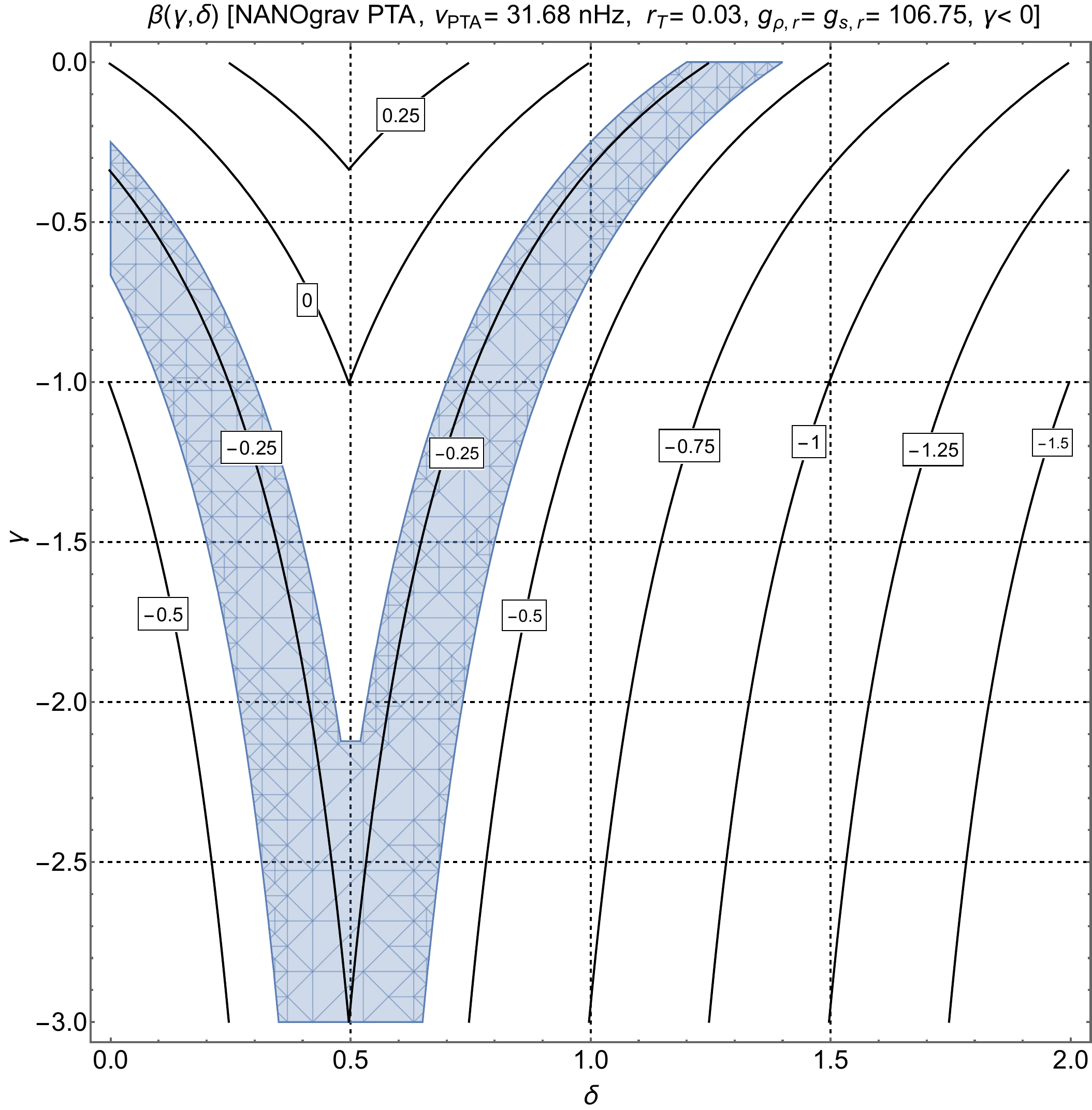}
\caption[a]{The contours are determined exactly as in the case of Fig. \ref{FIG10} but in the present situation  
we consider the NANOgrav intervals \cite{NANO2} for $\beta$ rather than the ones of the Parkes PTA. Since the interval of $\beta$ is narrower 
in the case of Ref. \cite{NANO2}, the shaded regions of both plots are narrower than the ones 
illustrated in Fig. \ref{FIG10}. As usual, in both figures the values of $\beta(\gamma,\delta)$ remains the same on each contour.}
\label{FIG11}      
\end{figure}

If all the wavelengths of the order of $\lambda_{PTA}$ do the second 
crossing in a radiation-dominated stage (i.e. when $\delta \to 1$) the spectral index of Eq. (\ref{SFOU5}) becomes:
\begin{equation}
\lim_{\delta \to 1} n_{T}(\gamma, \delta) = 3 - 2 \biggl|\frac{\gamma}{1- \gamma} - \frac{1}{2} \biggr|.
\label{SFOU6}
\end{equation}
When $\gamma \to 1/3$ in Eq. (\ref{SFOU6}) the spectral index $n_{T}(1/3, 1) \to 3$; in this case the 
stage of accelerated contraction may be driven, in the simplest situation, by the kinetic energy of a scalar field as it 
happens, for instance, in the dilaton-driven phase of pre-big bang scenarios \cite{PP3,PP4,PP5,PP6}. Similarly when $\gamma \to 0$ we would have, from Eq. (\ref{SFOU6}),
that $n_{T}(\gamma, 1) \to 2$; this is the situation preferred in the context of ekpyrotic models \cite{PP7,PP8}. 
 Whenever the expansion rate at the second 
crossing is eventually different from radiation (i.e. $\delta \neq 1$) also the form of Eq. (\ref{SFOU6}) 
is different. For this reason it is relevant to illustrate the general situation and this has been done in Fig. \ref{FIG9} for the two complementary ranges $0< \gamma < 1$ and $\gamma<0$. Grossly speaking the results of Fig. \ref{FIG9} imply that the spectral slopes suggested 
by the PTA collaborations can be reproduced either when $ \delta >1$ (if 
$0 \leq \gamma < 1$) or when $ 0< \delta < 1$ (provided $\gamma< 0$).
To scrutinize this point in further detail it is easier to express $\beta$ directly as a function of $\gamma$ and $\delta$:
\begin{equation}
\beta(\gamma, \delta) = 1 - \biggl|\delta - \frac{1}{2}\biggr| - \biggl|\frac{\gamma}{1 -\gamma} - \frac{1}{2} \biggr|.
\label{SFOU7}
\end{equation}
Using Eq. (\ref{SFOU7}) we can therefore limit the range of variation of $\beta(\gamma,\delta)$ and deduce the allowed domain of the parameters. Along this perspective, the shaded areas of Fig. \ref{FIG10} are determined by requiring $-0.65 < \beta(\gamma,\delta) <- 0.25$ as suggested 
by the Parkes PTA \cite{PPTA2}. In Fig. \ref{FIG10} the line $\delta = 1$ intersects the allowed region only if $\gamma > {\mathcal O}(0.6)$. A similar conclusion follows from Fig. \ref{FIG11} where we examine the NANOgrav observations \cite{NANO2}; in this case, as already stressed in the previous sections, the allowed range of $\beta$ is 
narrower (even if the corresponding range of $q_{0}$ is larger). 
The shaded area in both plots of Fig. \ref{FIG11} correspond to the interval $ -0.40 < \beta < -0.20$ and
the correct values of $\beta$ seem to be reproduced for $0 < \gamma < 1$ but only when $\delta > 1$. The opposite is true in the case $\gamma < 0$ where the region $\delta < 1$ seems comparatively 
wider.

\begin{figure}[!ht]
\centering
\includegraphics[height=8cm]{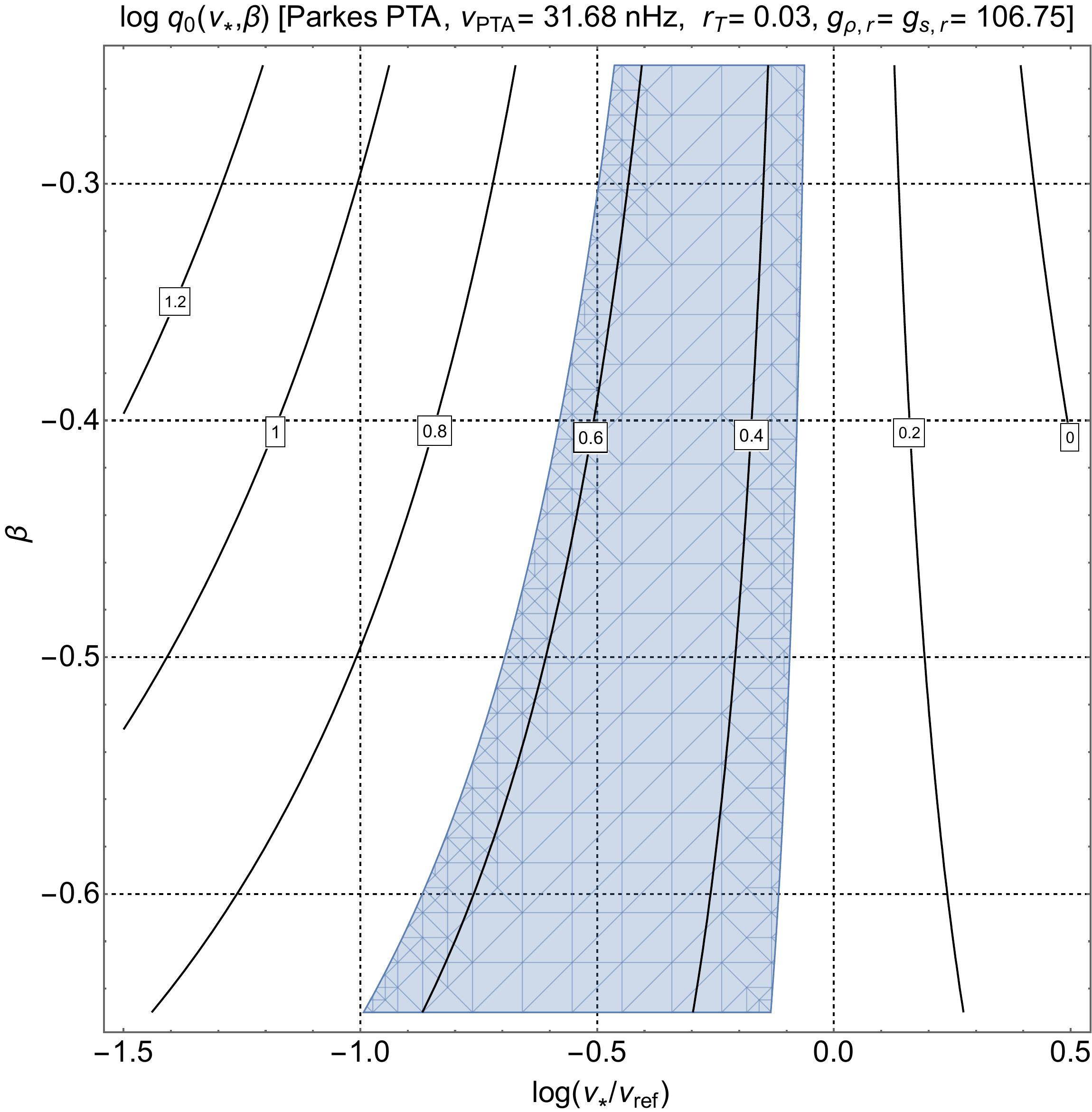}
\includegraphics[height=8cm]{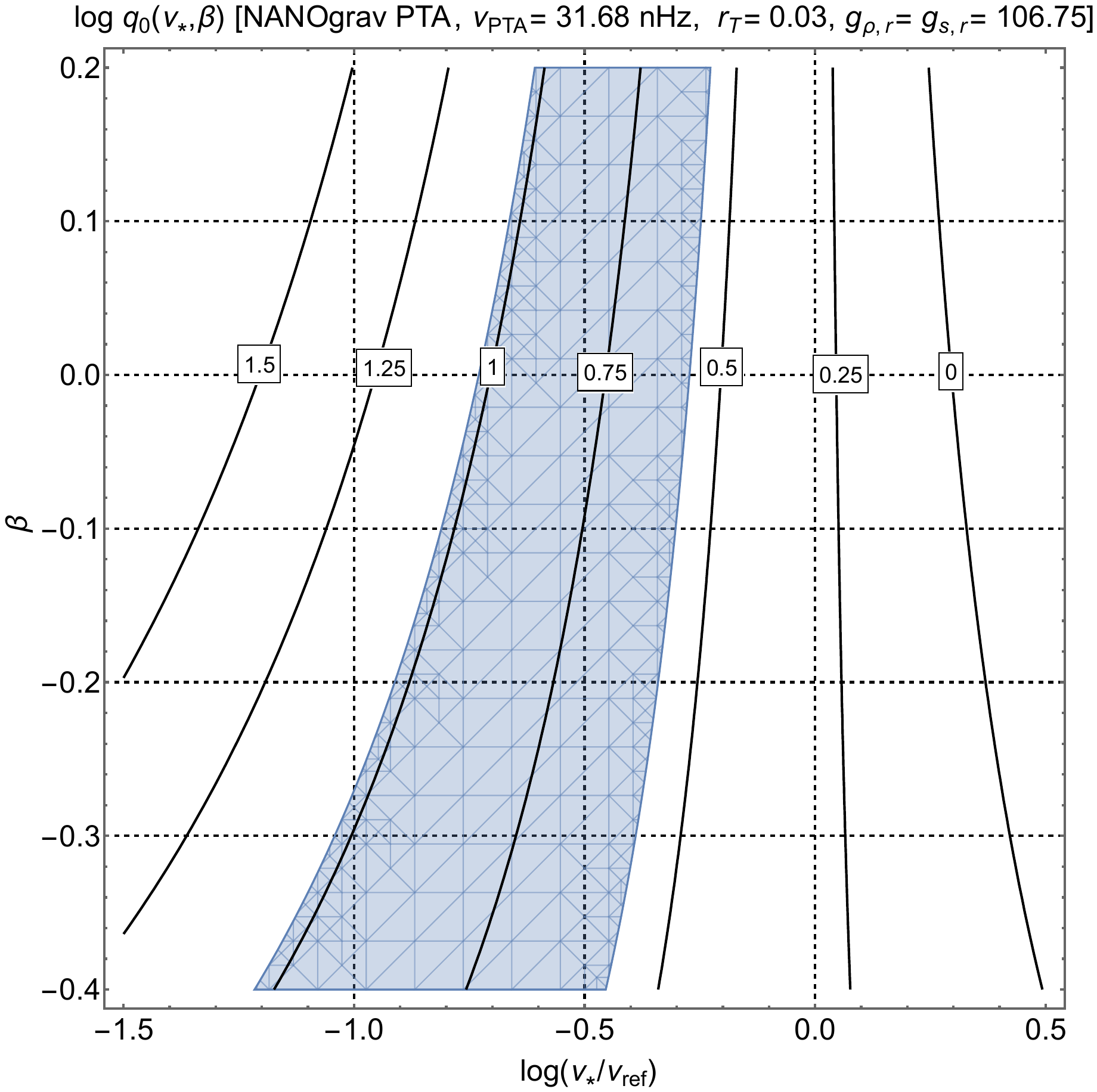}
\caption[a]{We illustrate the contours of constant $q_{0}(\nu_{\ast},\beta)$ for a 
fixed value of $\overline{\Omega}_{\ast}$. The labels appearing on the various contours correspond to the 
common logarithms of $q_{0}(\nu_{\ast},\beta)$. The intervals of $\beta$ match the ones of the corresponding observations. The shaded regions correspond, as indicated in each of the plots, to the ranges of $q_{0}$ determined by the Parkes \cite{PPTA2} and by the NANOgrav \cite{NANO2} PTA. }
\label{FIG12}      
\end{figure}

So far we discussed the spectral slopes and we must now consider 
the corresponding amplitudes. For this purpose we then examine Eq. (\ref{SFOU1}) and 
express it in terms of the parametrization of Eq. (\ref{NOTT1}):
\begin{equation}
6.287\times 10^{-10} q_{0}^2 (\nu/\nu_{ref})^{2 + 2 \beta} = \overline{\Omega}_{\ast} \, (\nu/\nu_{\ast})^{n_{T}}.
\label{SFOU8}
\end{equation}
Since the slope $n_{T}(\gamma,\delta)$ coincides  
with $2 + 2 \beta(\gamma,\delta)$ the amplitude $q_{0}$ (which 
is determined experimentally) not only depends on $\beta$ but also on $(\nu_{\ast}/\nu_{ref})$ and 
$\overline{\Omega}_{\ast}$. The high-frequency amplitude of the spectral energy density appearing in Eqs. (\ref{SFOU1})--(\ref{SFOU2})
can then be fixed to the largest value compatible with the limits of wide-band 
interferometers in the audio band  \cite{LIGO1,LIGO2}. From the theoretical viewpoint $q_{0}$ becomes  
a function of $\beta$ and of $(\nu_{\ast}/\nu_{ref})$:
\begin{equation}
q_{0}(\beta, \nu_{\ast}) = 1.97 (\nu_{ref}/\nu_{\ast})^{1 + \beta}.
\label{SFOU9}
\end{equation}
Since we know from Figs. \ref{FIG9} and \ref{FIG10} that there exist
regions in the $(\gamma,\,\delta)$ plane where $\beta$ falls within 
the observed range, from Eq. (\ref{SFOU9}) we can determine the range of variation 
of $\nu_{\ast}$. A careful analysis reported in Fig.  \ref{FIG12} shows 
that $\nu_{\ast}$ must be slightly smaller than $\nu_{ref}$. More specifically in Fig. \ref{FIG12} we illustrate the different contours of $q_{0}(\beta, \nu_{\ast})$ and the shaded areas correspond, respectively, to the regions pinned down, respectively, by the Parkes PTA (i.e. $3.7\leq q_{0} \leq 10.6$ in the left plot of Fig. \ref{FIG12}) and by the NANOgrav PTA (i.e. $2.2\leq q_{0} \leq 4.4$ in the right plot of Fig. \ref{FIG12}).

\subsection{Excluding a single blue slope}
\begin{figure}[!ht]
\centering
\includegraphics[height=8cm]{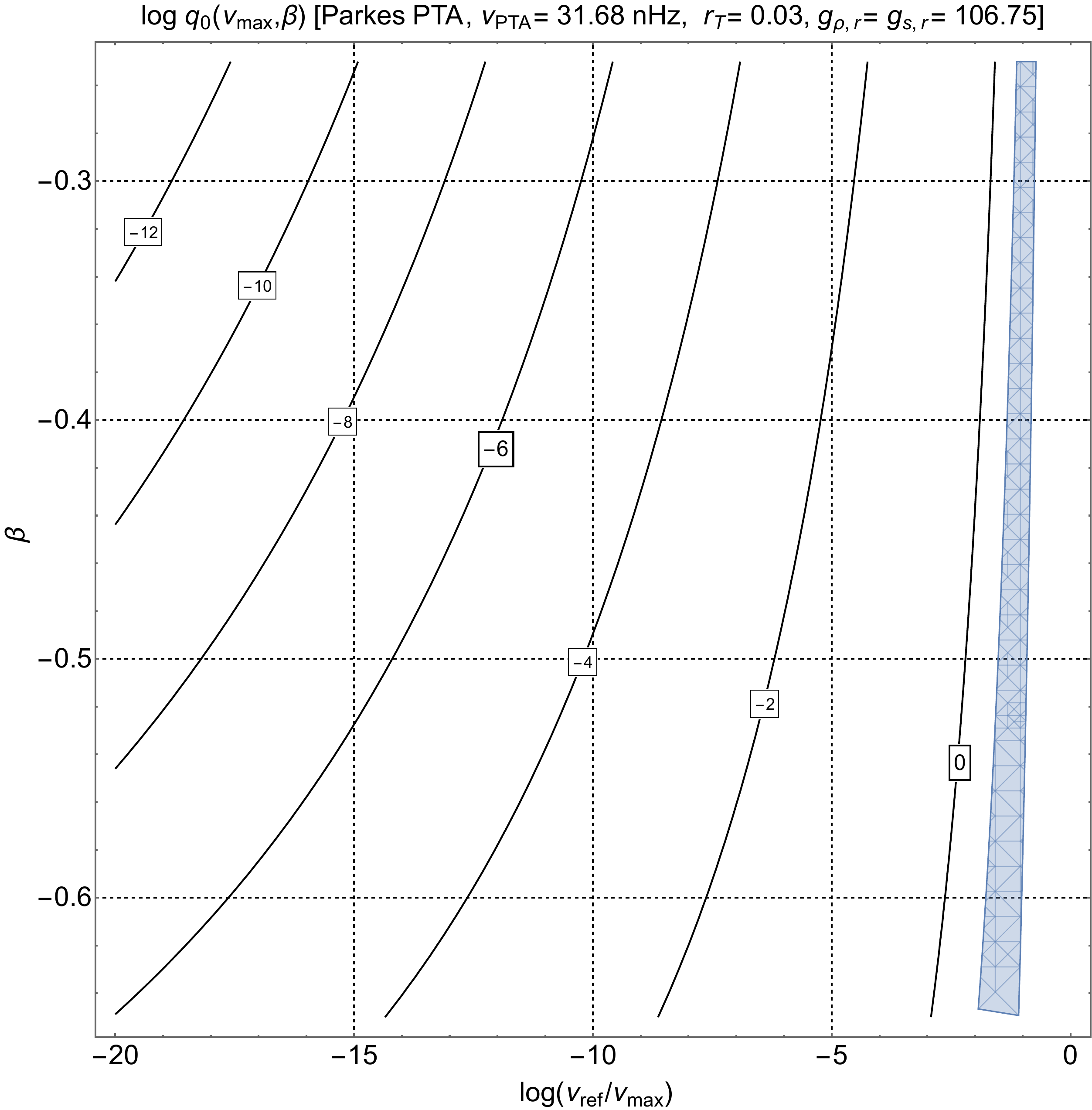}
\includegraphics[height=8cm]{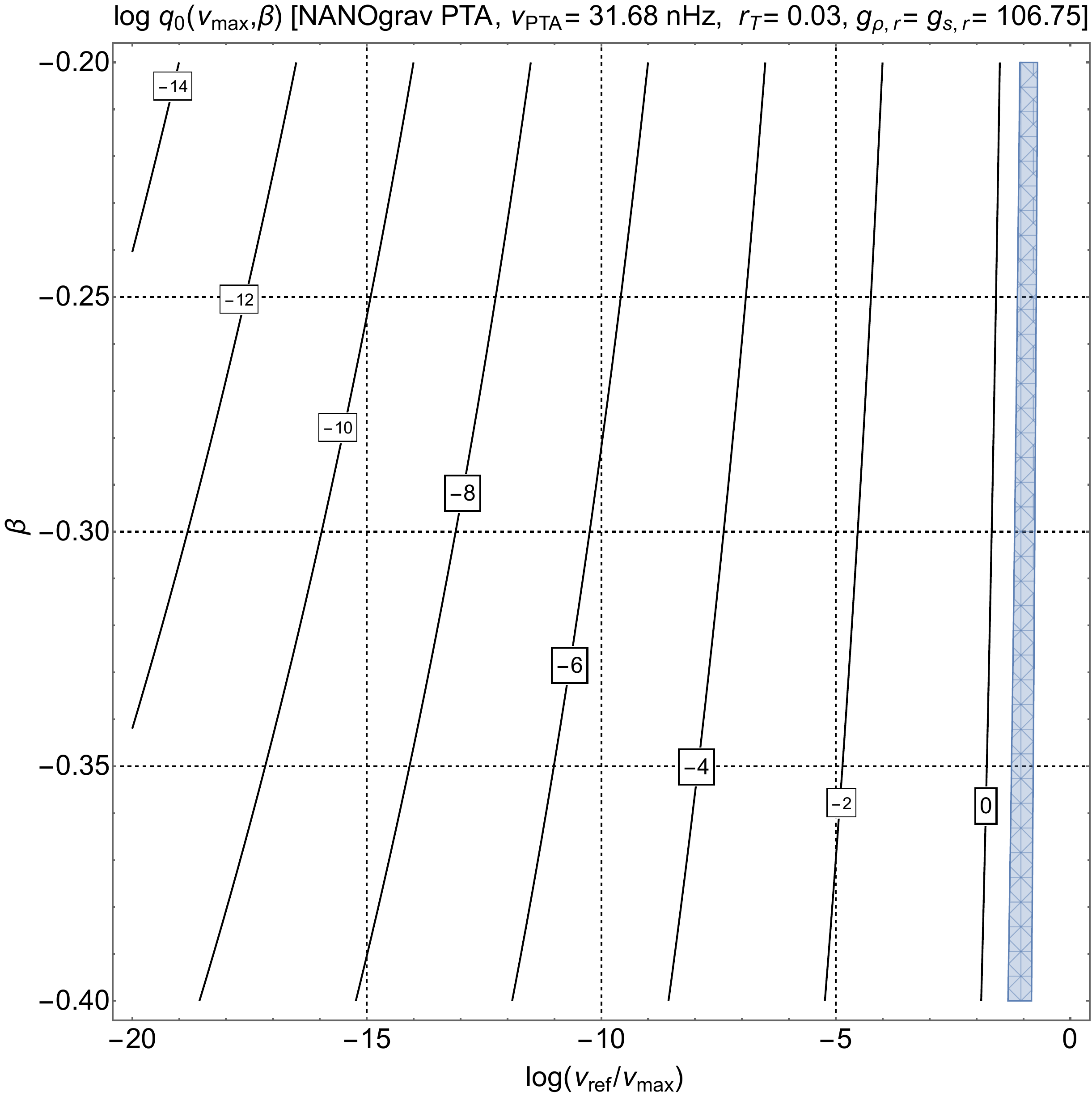}
\caption[a]{After imposing the big-bang nucleosynthesis constraints (see Eq. (\ref{SFOU11})) we  illustrate the contours of constant $q_{0}(\beta,\nu_{ref}/\nu_{max})$. The intervals of $\beta$ match the ones of the corresponding observations and the shaded regions correspond, as indicated in each of the plots, to the ranges of $q_{0}$ determined by the Parkes \cite{PPTA2} and by the NANOgrav \cite{NANO2} PTA. }
\label{FIG13}      
\end{figure}

The allowed region of the parameter space pins down a range $\nu_{\ast} = {\mathcal O}(\nu_{ref})$ 
but there are theoretical models where the transition to the decelerated regime is very short 
(i.e. $\nu_{\ast} \gg \nu_{ref}$) so that $\nu_{\ast}$ can eventually be of the order of $\nu_{max}$. To investigate this situation
we can therefore write the spectral energy density in the following form:
\begin{equation}
h_{0}^2 \Omega_{gw}(\nu, \tau_{0}) = \overline{\Omega}_{max} \, \biggl(\frac{\nu}{\nu_{max}}\biggr)^{n_{T}(\gamma, \delta)}, \qquad\qquad \nu_{eq} < \nu\leq \nu_{max},
\label{SFOU10}
\end{equation}
and  demand that $n_{T}(\gamma, \delta)$ falls in the interval of slopes 
associated with the PTA; if this is the case, then $n_{T}(\gamma, \delta) = 2 (1 + \beta) >0$.
If we combine Eq. (\ref{SFOU10}) with the requirements of Eq. (\ref{BBB1})
we also have that $\overline{\Omega}_{max}$ is constrained as:
\begin{equation}
\overline{\Omega}_{max} \biggl[ 1 - \biggl(\frac{\nu_{bbn}}{\nu_{max}}\biggr)^{2 (1 + \beta)}\biggr] \leq 
2 ( 1 + \beta)\, \overline{\Omega}_{bbn}, \qquad \qquad \overline{\Omega}_{bbn}= {\mathcal O}(10^{-7}).
\label{SFOU11}
\end{equation}
Thanks to Eq. (\ref{SFOU11}) we can first trade $\overline{\Omega}_{max}$ for $\overline{\Omega}_{bbn}$ 
and then we can also express $q_{0}$ in terms of $\beta$ and $y = (\nu_{ref}/\nu_{max})$:
\begin{equation}
q_{0}(y, \beta)= 17.83 \, \biggl(\frac{\overline{\Omega}_{bbn}}{10^{-7}}\biggr)^{1/2} \frac{ y^{\beta+1}}{ \sqrt{1 - w_{bbn} \,\,y^{ 2 \beta+ 2}}}, \qquad\qquad w_{bbn} = (\nu_{bbn}/\nu_{ref}) = 0.011.
\label{SFOU12}
\end{equation}
Equation (\ref{SFOU12}) is illustrated in Fig. \ref{FIG13} where $q_{0}$ is viewed as a function 
of $\beta$ and $\nu_{max}$; on the vertical axis of both plots we report 
the range of $\beta$ compatible with each of the corresponding experiment (i.e. the Parkes \cite{PPTA2} and the 
NANOgrav \cite{NANO2} PTA). The shaded regions define the allowed range of $q_{0}$ for each 
of the two experiments. The results of Fig. \ref{FIG13} are incompatible with Eq. (\ref{SFOU10}) since, grossly speaking, we can expect $\nu_{max}$ between few MHz and the THz \cite{MG1}. In the parametrization of Eq. (\ref{SFOU10}) 
the common logarithm of $(\nu_{ref}/\nu_{max})$ (reported on the horizontal axes of the plots 
of Fig. \ref{FIG13}) should be between $-20$ and $-15$. On the contrary the allowed region 
in Fig. \ref{FIG13} is located for $y = {\mathcal O}(10^{-2})$.

If the PTA excesses are the result of a bouncing stage it is necessary that 
the wavelengths of the order of $\lambda_{PTA}$ do their second crossing during 
a decelerated stage not dominated by radiation. Moreover a bouncing model leading to a spectral energy density with a single blue slope (possibly 
ranging between the equality frequency and $\nu_{max}$) is unable to account for the
 PTA excesses if all the phenomenological bounds are concurrently satisfied.  Various bouncing scenarios  have been constructed with the aim of either complementing or even challenging the conventional inflationary ideas. Bouncing scenarios 
appear by physical premises that are quite different and various reviews are currently available (see, for instance,  \cite{PP9,PP10,PP11,PP12,PP13}). It is not uncommon that the solutions obtained in a given  framework are recycled in an entirely different situation. Recently it has been argued that conventional inflationary models are generically in tension with the swampland criteria \cite{PP14} and this is often seen as a further motivation for bouncing dynamics. The indications 
of the PTA can be very precious in this context and they might even contribute to the long standing problem of bouncing scenarios, i.e. the origin of a Gaussian and adiabatic mode of large-scale curvature inhomogeneities \cite{MG1}.
\newpage
\renewcommand{\theequation}{5.\arabic{equation}}
\setcounter{equation}{0}
\section{Concluding remarks}
\label{sec5}
During the last two decades a series of significant limits from the millisecond pulsars at intermediate frequencies (roughly corresponding to the inverse of the time-scale along which the various pulsars have been monitored) severely constrained the isotropic and random backgrounds of gravitational radiation. More recently different ensembles of millisecond pulsars have been scrutinized  with regular cadence by the pulsar timing arrays (PTA in the bulk of the paper).
Despite the different conclusions on the expected correlations in the pulse arrival times between pairs of pulsars (and despite the slightly dissimilar determinations of the spectral parameters),  the competing experiments seem to suggest concurrent evidences of gravitational radiation with stochastically distributed Fourier amplitudes at a benchmark frequency ${\mathcal O}(30)$ nHz and with $h_{0}^2 \Omega_{gw}(\nu,\tau_{0})$ ranging between $10^{-8}$ and $10^{-9}$. While the origins of the PTA excesses are still perplexing, in this paper we speculate that the relic gravitons 
are responsible of the observed signal. Even if a collection of late-time
sources may eventually lead to a diffuse background of gravitational radiation, the relic gravitons 
are instead produced, by definition, at much earlier times and solely because the rapid variation 
of the space-time curvature. 

The theoretical perspective explored in this investigation strongly suggests that the problem is not yet to fit (more or less reliably) the existing data in terms of a series of preferred scenarios but to understand preliminarily whether or not the observed excesses in the nHz range are compatible with a modified evolution of the comoving horizon since this is the only way the spectrum of relic gravitons at intermediate frequencies can be affected. The goal of this study is therefore not to endorse a specific model (or to pin down the likely values of a hypothetical spectral index) but to see, more modestly, if and how the relic gravitons could be associated with the nHz excesses. All in all the systematic approach developed in this paper propounds {\em three complementary physical possibilities} that have been carefully perused and that should be further analyzed in the near future.
\begin{itemize}
\item{} The most conventional option stipulates that the timeline of the comoving horizon is not modified during inflation so that the nHz excess is caused by the drastic change of the post-inflationary 
expansion rate prior to big-bang nucleosynthesis.  
\item{} A second alternative implies a modified evolution of the tensor modes during a conventional inflationary stage as it happens, for instance, when the gravitons inherit an effective refractive index from the interactions with the geometry. 
\item{} We finally consider the possibility of an epoch of increasing curvature prior to the conventional decelerated stage of expansion and argue that this option is only reconcilable with the observed excesses provided the wavelengths crossing the comoving horizon at early times do not reenter in an epoch dominated by radiation. 
\end{itemize}
In connection with these three complementary options the results obtained in this analysis are, in short, the following.
\begin{itemize}
\item{} A late-time modification of the comoving horizon may indeed alter the spectral energy density of the relic gravitons in the nHz range but the observed amplitudes and slopes are unfortunately neither compatible nor comparable with this minimal explanation. 
\item{} Conversely, if a dynamical refractive index evolves during a conventional inflationary phase of nearly minimal duration the corresponding $h_{0}^2 \, \Omega_{gw}(\nu,\tau_{0})$ compares well with the nHz hump even if the observational data coming from competing experiments pin down slightly different regions of the parameter space. 
\item{} In the context of the bouncing scenarios, it is finally possible to exclude a nHz excess associated with a single slope of the spectral energy density between $100$ aHz and the GHz range. However,  if the spectral energy density has a break in the nHz region the direct limits of the wide-band detectors constrain the high-frequency amplitude and, in this case, we could account for the hump provided the bunch of PTA wavelengths reenter the comoving horizon during a decelerated stage not yet dominated by radiation. 
\end{itemize}
The three possibilities discussed here are therefore not mutually exclusive: in the bouncing case, the simplest way to match the amplitudes and the slopes of the spectra is to modify the comoving horizon also at late times. The potential nHz excesses make even more relevant the 
high-frequency determinations of the spectral energy density. In particular the direct bounds in the audio band are essential even if the largest signal of the relic gravitons is expected in the MHz and GHz domains where smaller detectors may play a crucial r\^ole, as repeatedly suggested in the past. While the observational aspects of the problem cannot be addressed with the theoretical approach reported here, it is nonetheless true that the physical interpretation of the results probably demands a 
conceptual framework (such as the one pursued in this analysis) that could clarify (and even exclude)  the primeval origin of the nHz excesses. 

\section*{Acknowledgements}
The author is indebted to A. Gentil-Beccot, L. Pieper, S. Reyes, S. Rohr and J. Vigen of the CERN Scientific Information Service for their kind assistance during the various 
stages of the preparation of this manuscript.

\newpage 

\begin{appendix}

\renewcommand{\theequation}{A.\arabic{equation}}
\setcounter{equation}{0}
\section{Basic conventions and notations}
\label{APPA}
With the purpose of making the presentation self-contained, in this and in the following 
appendix we illustrate the main notations employed in the bulk of the paper.
We start by reminding that the tensor modes of the geometry are related to the first-order traceless and solenoidal fluctuations of the four-dimensional metric
\begin{equation}
g_{\mu\nu}(\vec{x},\tau) = a^2(\tau) \eta_{\mu\nu} + \delta_{t}^{(1)} g_{\mu\nu}(\vec{x}, \tau), \qquad 
\delta_{t}^{(1)} g_{i\,j}(\vec{x}, \tau) = - a^2(\tau) h_{ij}(\vec{x},\tau),
\label{APP1}
\end{equation}
where $a(\tau)$ denotes the scale factor of a conformally flat background 
geometry, $\tau$ is the conformal time coordinate and $\eta_{\mu\nu}$ is the 
Minkowski metric with signature $(+,\, -,\,-,\, -)$; Greek indices are four-dimensional while the Latin (lowercase) indices are three-dimensional; in Eq. (\ref{APP1}) the first-order 
(tensor) fluctuations $h_{i\,j}(\vec{x}, \tau)$ are, by definition, solenoidal and traceless. As a general rule throughout the text 
the prime denotes a derivation with respect to $\tau$ while the overdot denotes a derivation 
with respect to the cosmic time coordinate $t$. Since the relation between the two is given 
by $a(\tau) d \tau = d\,t$ it also follows, as usual, that 
\begin{equation}
a \, H = {\mathcal H} =  a^{\prime}/a, \qquad \qquad H= \dot{a}/a.
\label{APP1a}
\end{equation}
 Different parametrizations of the time coordinate are also discussed in appendix \ref{APPB} 
and are relevant in the discussion of section \ref{sec3}. Within these notations the effective action of the tensor modes of the geometry, originally due to Ford and Parker \cite{par1} can be written as:
\begin{equation}
S = \frac{\overline{M}_{P}^2}{8} \int d^{4} x \, \, \sqrt{- \overline{g}} \,\,\overline{g}^{\alpha\beta} \partial_{\alpha} h_{i\,j} \, \partial_{\beta} h^{i\,j}, \qquad \overline{M}_{P} = M_{P}/\sqrt{8\, \pi}.
\label{APP2}
\end{equation}
In the framework of a specific scenario Eq. (\ref{APP2}) may be complemented by other terms. For instance, in the case single-field inflationary models the effective action
contains all the terms that contain four space-time derivatives 
and that are suppressed by the inverse power of a large mass scale $M$ \cite{AAAW}; 
among all these terms there are combinations that break parity and that are 
neglected here even if their inclusion does not substantially alter the conclusions of the 
discussion given the smallness of the corresponding effects \cite{FOUR}.
From Eq. (\ref{APP2}) the effective energy-momentum pseudo-tensor is given by: 
\begin{equation}
{\mathcal T}_{\mu}^{\nu} = \frac{\overline{M}_{P}^2}{4} \biggl( \partial_{\mu} h_{i\, j} \partial^{\nu} h^{i\, j} 
- \delta_{\mu}^{\,\,\, \nu} \overline{g}^{\alpha\beta} \partial_{\alpha} h_{i\,j} \partial_{\beta} h^{i\, j} \biggr),
\label{APP3}
\end{equation}
where the indices are raised and lowered by means of the background metric $\overline{g}_{\mu\nu}(\tau)$.
While there exist different strategies to assign the energy-momentum pseudo-tensor, the one obtained from the functional variation (with respect to the background metric) of the effective action (\ref{APP2}) leads 
to a positive semi-definite energy density both inside and outside the Hubble 
radius \cite{qq}; from the temporal components of Eq. (\ref{APP3}) we obtain the explicit expression 
of the energy density:
\begin{equation}
\rho_{gw} = \frac{\overline{M}_{P}^2}{8\, a^2} \biggl(\partial_{\tau} h_{i\,j} \partial_{\tau} \, h^{i\,j} + 
\partial_{k} h_{i\,j} \partial_{k} \, h^{i\,j}\biggr).
\label{APP4}
\end{equation}
In the present investigation the tensor amplitude in Fourier space is defined as:
\begin{equation}
h_{i\,j}(\vec{x},\tau)= \frac{1}{(2\pi)^{3/2}} \, \int d^{3} k \, e^{-i \vec{k}\cdot\vec{x}} \, h_{i\, j}(\vec{k}, \tau), \qquad\qquad 
h_{i\, j}^{\ast}(\vec{k}, \tau) = h_{i\, j}(- \vec{k}, \tau),
\label{APP5}
\end{equation}
where $k^{i} \, h_{i\,j} = k^{j} \,h_{i\,j} =0$ and $h_{i}^{i} =0$. For a stochastic 
random field the Fourier amplitudes of Eq. (\ref{APP5}) and their time derivatives 
are associated, respectively, with the two tensor power spectra $P_{T}(k,\tau)$ and $Q_{T}(k,\tau)$
\begin{eqnarray}
\langle h_{i\,j}(\vec{k},\tau) \, h_{m\, n}(\vec{p}, \, \tau) \rangle = \frac{2 \pi^2}{k^3} \,\,P_{T}(k,\tau) \,\,
\delta^{(3)}(\vec{k} + \vec{p}) \,\,S_{i\,j\,m\,n}(\hat{k}),
\label{APP6}\\
\langle \partial_{\tau} h_{i\,j}(\vec{k},\tau) \, \partial_{\tau} h_{m\, n}(\vec{p}, \, \tau) \rangle = \frac{2 \pi^2}{k^3} \,\,Q_{T}(k,\tau) \,\,
\delta^{(3)}(\vec{k} + \vec{p}) \,\,S_{i\,j\,m\,n}(\hat{k}),
\label{APP7}
\end{eqnarray}
where $S_{i\,j\,m\,n}(\hat{k}) = [p_{i\,m}(\hat{k})\, p_{j\,n}(\hat{k}) + p_{j\,m}(\hat{k})\, p_{i\,n}(\hat{k}) -
p_{i\, j}(\hat{k}) p_{m\, n}(\hat{k})]/4$ is given in terms of the transverse projectors $p_{i\,j}(\hat{k}) = (\delta_{i\, j} - \hat{k}_{i} \, \hat{k}_{j})$. After inserting Eq. (\ref{APP5}) into Eq. (\ref{APP4}) we can 
average the obtained result term by term and deduce: 
\begin{equation}
\langle \rho_{gw} \rangle = \frac{\overline{M}_{P}^2}{8\,\, a^2 } \int \frac{d k}{k} \biggl[ k^2 P_{T}(k,\tau) + Q_{T}(k,\tau)\biggr].
\label{APP8}
\end{equation}
If the result of Eq. (\ref{APP8}) is divided by the critical energy density $\rho_{crit} = 3 \, H^2\, \overline{M}_{P}^2$ we obtain the spectral energy density in critical units conventionally denoted by $\Omega_{gw}(k,\tau)$:
\begin{equation}
\Omega_{gw}(k,\tau) = \frac{1}{\rho_{crit}} \, \frac{ d \, \langle \rho_{gw} \rangle}{d \ln{k}} = 
\frac{1}{24 \, H^2 \, a^2} \biggl[ k^2 \, P_{T}(k,\tau) + Q_{T}(k,\tau)\biggr].
\label{APP9}
\end{equation}
When the wavelengths are shorter than the comoving horizon the corresponding wavenumbers 
exceed the expansion rate and, in this regime, $Q_{T}(k,\tau)$ and $k^2 \, P_{T}(k,\tau)$ 
are roughly comparable so that Eq. (\ref{APP9}) can also be written as: 
\begin{equation}
\Omega_{gw}(k,\tau) = \frac{k^2}{12 \, H^2 \, a^2} \, P_{T}(k, \tau), \qquad \frac{k}{{\mathcal H}} = \frac{k}{a\, H} >1.
\label{APP10}
\end{equation}
The tensor power spectra $P_{T}(k,\tau)$ and $Q_{T}(k,\tau)$ depend on the evolution of the mode functions $F_{k}(\tau)$ 
and $G_{k}(\tau) = F_{k}^{\prime}(\tau)$ where, as already mentioned above, the prime denotes throughout a derivation with respect to the 
cosmic time coordinate $\tau$: 
\begin{eqnarray}
P_{T}(k, \tau) &=& \frac{4 k^3 }{\overline{M}_{P}^2\,\pi^2} \bigl| F_{k}(\tau) \bigr|^2 =  \frac{4 k^3}{\pi^2\,\overline{M}_{P}^2\, a^2} \bigl| f_{k}(\tau) \bigr|^2, 
\label{PPSS1}\\
Q_{T}(k, \tau) &=& \frac{4 k^3}{\overline{M}_{P}^2 \, \pi^2} \bigl| G_{k}(\tau) \bigr|^2 =  \frac{4 k^3}{\pi^2\, \overline{M}_{P}^2 a^2} \bigl| g_{k}(\tau) \bigr|^2, 
\label{PPSS2}
\end{eqnarray}
In Eqs. (\ref{PPSS1})--(\ref{PPSS2}) the rescaled mode functions $f_{k}(\tau)$ and $g_{k}(\tau)$ obey: 
\begin{equation}
f_{k}^{\prime} = g_{k} + {\mathcal H} f_{k}, \qquad\qquad g_{k}^{\prime} = - k^2 \, f_{k} - {\mathcal H} f_{k}.
\label{PPSS3}
\end{equation}
The evolution of the mode functions can be solved in various regimes and a detailed discussion 
on these issues can be found elsewhere. If we eliminate $g_{k}(\tau)$ from Eq. (\ref{PPSS3}) 
the evolution of $f_{k}(\tau)$ can be expressed as 
\begin{equation}
f_{k}^{\prime\prime} + \bigl\{ k^2 - a^2 \, H^2 [ 2 - \epsilon(a)]\bigr\} f_{k}=0, \qquad \qquad \epsilon = - \dot{H}/H^2.
\label{PPSS3a}
\end{equation}
As stressed in the bulk of the paper, the condition $k^2 = a^2 \, H^2 [ 2 - \epsilon(a)]$ defines two complementary 
turning points where the analytical evolution of the mode functions gets modified; moreover, since 
$(a\, H)^{-1}$ gives approximately the comoving horizon, they 
are associated, respectively, to the exit and to the reentry of a given wavelength. The use of the term
horizon to identify the inverse of the Hubble rate (and its comoving counterpart) is rather 
common even if potentially inaccurate; we shall not try to correct here this terminology. 
When the wavelengths are shorter than the comoving horizon the expressions of 
$F_{k}(\tau)$ and $G_{k}(\tau)$ are: 
\begin{eqnarray}
F_{k}(\tau) &=& \frac{e^{- i k \tau_{ex}}}{\sqrt{2 k}\, a} \biggl(\frac{a_{re}}{a_{ex}}\biggr) {\mathcal Q}_{k}(\tau_{ex}, \tau_{re}) 
\biggl[ \frac{{\mathcal H}_{re}}{k} \sin{ k \Delta \tau} + \cos{k \Delta\tau} \biggr],
\label{PPSS4}\\
G_{k}(\tau) &=& \frac{e^{- i k \tau_{ex}}}{a} \,\sqrt{\frac{k}{2}}\, \biggl(\frac{a_{re}}{a_{ex}}\biggr) {\mathcal Q}_{k}(\tau_{ex}, \tau_{re}) \biggl\{
\biggl[ \frac{{\mathcal H}_{re}}{k} \cos{ k \Delta \tau} - \sin{k \Delta\tau} \biggr]
\nonumber\\
&-& \frac{{\mathcal H}}{k} \biggr[\frac{{\mathcal H}_{re}}{k} \sin{ k \Delta \tau} + \cos{k \Delta\tau} \biggr]\biggr\},
\label{PPSS5}
\end{eqnarray}
where ${\mathcal Q}_{k}(\tau_{ex}, \tau_{re})$ has been already introduced in Eq. (\ref{STWO2}) and $\Delta\tau = (\tau - \tau_{re})$. If Eqs. (\ref{PPSS4})--(\ref{PPSS5}) are inserted first into Eqs. (\ref{PPSS1})--(\ref{PPSS2}) and then into Eq. (\ref{APP9}) we can obtain the expression of Eq. (\ref{STWO1}). To match the notations of the present paper with the ones of the PTA collaborations we recall that the relation between the chirp amplitude 
$h_{c}(\nu,\tau)$ and the tensor power spectrum $P_{T}(k,\tau)$ is  $ 2 \, h_{c}^2(\nu,\tau) = P_{T}(\nu, \tau)$.
Similarly the connection between the spectral amplitude $S_{h}(\nu,\tau)$ and the chirp 
amplitude is given by $\nu \, S_{h}(\nu, \tau) = h_{c}^2(\nu, \tau)$. Putting everything together 
we then have 
\begin{equation}
\Omega_{gw}(k,\tau_{0}) = \frac{2 \pi^2 \nu^2}{3 \, H_{0}^2 } \,h_{c}^2(\nu, \tau_{0}), 
\label{APP11}
\end{equation}
where we used $a_{0} =1$; this means, within our conventions, that 
the comoving and the physical frequencies coincide at the present time.
If we now parametrize the signal as in Eq. (\ref{NOTT1}) a more explicit relations 
can be obtained from Eq. (\ref{APP1}) and it is given by:
\begin{equation}
\Omega_{gw}(\nu,\tau_{0}) = \frac{2 \pi^2}{3} \, Q^2 \, \biggl(\frac{\nu_{ref}}{H_{0}}\biggr)^2 \,\, \biggl(\frac{\nu}{\nu_{ref}}\biggr)^{2 + 2 \beta}.
\label{APP12}
\end{equation}
If we now multiply Eq. (\ref{APP12}) by $h_{0}^2$ (where $h_{0}$ denotes the indetermination 
in the present value of the Hubble rate), take into account the explicit value of $\nu_{ref}$ and parametrize 
$Q$ as $Q= q_{0} \times 10^{-15}$ we obtain the relations already discussed in Eq. (\ref{NOTT2}) (for $\nu\to \nu_{ref}$) and, more generally, in Eq. (\ref{STWO6}). With the same logic we can also deduce 
the explicit relation between the spectral amplitude and the chirp amplitude:
\begin{equation}
S_{h}(\nu, \tau_{0}) = 3.15 \, \times 10^{-23} \, \,  q_{0}^2\,\,\bigl(\nu/\nu_{ref}\bigr)^{2 \beta-1} \, \, \mathrm{Hz}^{-1}.
\label{APP13}
\end{equation}
It is often customary to employ the square root of Eq. (\ref{APP13}) so that $\sqrt{S_{h}(\nu, \tau_{0})}$ is measured in units 
$1/\sqrt{\mathrm{Hz}}$, i.e. $\sqrt{S_{h}(\nu, \tau_{0})} =5.61\times10^{-12} \, \, q_{0} (\nu/\nu_{ref})^{\beta -1/2} \,\, 1/\sqrt{\mathrm{Hz}}$.

\renewcommand{\theequation}{B.\arabic{equation}}
\setcounter{equation}{0}
\section{The general form of the effective action}
\label{APPB}
A more general form of the effective action of the tensor modes of the geometry suggests to weight the various terms 
of Eq. (\ref{APP2}) with time-dependent coefficients:
\begin{equation}
S_{g} = \frac{\overline{M}_{P}^2}{8} \int \, d^{3} x \int \, d \tau \biggl[ c_{1}(\tau) \,\partial_{\tau} h_{i\, j}\, \partial_{\tau} h^{i\, j} - c_{2}(\tau) \, \partial_{k} h_{i \, j} \,\partial^{k}h^{i\, j} 
- c_{3}(\tau)\, m_{c}^2\, h_{i \, j} \, \partial h^{i \, j} \biggr].
\label{APPB1}
\end{equation}
As before, in Eq. (\ref{APPB1})  the parity-breaking terms associated 
with quadratic combinations involving either the dual Riemann or the dual Weyl tensors 
have been neglected; both terms would appear in the effective action 
and can polarize the backgrounds of relic gravitons \cite{FOUR} (see also \cite{FOURa}). Among the three function $c_{i}(\tau)$ 
(with $i =1,\, 2,\, 3$) we have that $c_{1}(\tau)$ and $c_{2}(\tau)$ are related to the expanding 
dimensions while $c_{3}(\tau)$ may appear in the case of compact extra-dimensions \cite{MGextra}.
If we factor $c_{1}(\tau)$ in Eq. (\ref{APPB1}) the resulting expression will be given by:
\begin{equation}
S_{g} = \frac{\overline{M}_{P}^2}{8} \int \, d^{3} x \int \, d \tau \,c_{1}(\tau) \biggl[\partial_{\tau} h_{i\, j} \partial_{\tau} h^{i\, j} - \frac{1}{n^2(\tau)} \,\partial_{k} h_{i \, j} \partial^{k} h^{i\, j} - \frac{1}{\overline{n}^2(\tau)} m_{c}^2  h_{i \, j} h^{i \, j} \biggr],
\label{APPB2}
\end{equation}
where $n(\tau)$ and $\overline{n}(\tau)$ denote, respectively, the refractive indices associated with the expanding and with the compact dimensions \cite{MGextra}, i.e. $n(\tau) = \sqrt{c_{1}(\tau)/c_{2}(\tau)}$ and $\overline{n}(\tau) = \sqrt{c_{1}(\tau)/c_{3}(\tau)}$. Equation (\ref{APPB2}) simplifies after a rescaling of the background dependence and its final form becomes\footnote{Once the new parametrization has been introduced 
we can also rescale the background dependence so that 
$b(\eta) = \sqrt{c_{1}(\eta)/n(\eta)}$ and $r_{c}(\eta) = n(\eta)/\overline{n}(\eta)$.
In the absence of a contribution from the internal dimensions (i.e. $m_{c} \to 0$)
Eq. (\ref{APPB3}) reproduces exactly Eq. (\ref{APP2}) for when $n \to 1$ and 
$c_{1}(\tau) = a^2(\tau)$.} :
\begin{equation}
S_{g} = \frac{\overline{M}_{P}^2}{8} \int \, d^{3} x \int \, d \eta \, \, b^2(\eta) \biggl[\partial_{\eta} h_{i\, j} \partial_{\eta} h^{i\, j} -  \, \, \partial_{k} h_{i \, j} \partial^{k} h^{i\, j} 
- r_{c}^2(\eta)  m_{c}^2 \, h_{i \, j} h^{i \, j} \biggr].
\label{APPB3}
\end{equation}
Equation (\ref{APPB3}) follows from Eq. (\ref{APPB2}) by first changing the time 
parametrization from $\tau$ (the conformal time coordinate) to $\eta$;
the relation between the two time parametrizations is simply given by$n(\eta) d\eta = d\tau$.  Let us therefore consider the simplest situation 
where the refractive index increases during inflation as suggested in Eq. (\ref{STHR3}); in this case for $a< a_{\ast}$ we would have $n(a) = n_{\ast} (a/a_{\ast})^{\alpha}$ (with $\alpha >0$) so that the relation between the conformal time coordinate $\tau$ and the $\eta$-time can be swiftly worked since 
$d \eta = d\tau/n(a)$. From the definition of $\eta$ we therefore have:
\begin{equation}
\eta = \int \frac{da}{a^2 \, H\, n} = - \frac{1}{a\, H\, n} + (\epsilon - \alpha)\int \frac{d a}{a^2 \, H\, n},
\label{APPB4}
\end{equation}
where, as in Eq. (\ref{APP1a}), $H = \dot{a}/a$ and the overdot denotes a derivation 
with respect to the cosmic time coordinate. The second equality in Eq. (\ref{APPB4}) follows after integration by parts since $\dot{\epsilon} \ll 1$ and $\dot{\alpha} =0$. Equation 
(\ref{APPB4}) also implies that $a \, H\, n= - 1/[(1 -\epsilon + \alpha) \eta]$; note 
once more that when $n \to 1$ we also have $\alpha \to 0$ and the standard 
relation $a \, H= -1/[(1-\epsilon)\tau]$ is immediately recovered.
In the $\eta$-time parametrization the evolution of the mode functions simplifies 
and it is given by
\begin{equation}
\partial_{\eta}^2 f_{k} + \biggl[ k^2 - \frac{\partial_{\eta}^2 b}{b}\biggr] f_{k} =0, \qquad 
g_{k} = \partial_{\eta} f_{k} - \frac{\partial_{\eta}b}{b} f_{k}.
\label{APPB5}
\end{equation}
From Eq. (\ref{APPB5}) it follows that the crossing of a given 
wavelength occurs when $k^2 = (\partial_{\eta}^2 b)/b$. This expression 
generalizes therefore the notion of the comoving horizon 
during the refractive phase. More specifically we may recall the connection between the derivations 
in the various time parametrizations introduced so far, namely
\begin{equation}
\partial_{\eta} = n \, \partial_{\tau} = n\, a\,\partial_{t}, \qquad \partial_{X} = \frac{\partial}{\partial X},
\label{APPB6}
\end{equation}
where $\eta$, $\tau$ and $t$ denote, once more, the $\eta$-time, the conformal time and the cosmic time coordinates. 
By using Eq. (\ref{APPB6}) in the condition $k^2 =  (\partial_{\eta}^2 b)/b$ we obtain, after simple algebra, a condition similar to Eq. (\ref{STHR0}): 
\begin{equation}
k^2 = n^3 \, b^2 \, F^2 \biggl[ 2 - \overline{\epsilon}+ \frac{3 \dot{n}}{2 n F}\biggr], \qquad F = \frac{\dot{b}}{b}, \qquad \overline{\epsilon} = - \frac{\dot{F}}{F^2}.
\label{APPB6a}
\end{equation}
In the limit $n\to 1$ we get $b \to a$, $F\to H$ and $\overline{\epsilon}\to \epsilon$. This is why it is appropriate to consider $(b\, F)^{-1}$ as the generalization of the comoving horizon during inflation. Different choices in Eq. (\ref{APPB1}) are in fact artificial since they are ultimately equivalent to the 
one of Eq. (\ref{APPB2}). For instance instead of factoring $c_{1}(\tau)$ we may factor $c_{2}(\tau)$. If we rescale $c_{2}(\tau)$ we simply get the analog of Eq. (\ref{APPB2}):
\begin{equation}
S_{g} = \frac{\overline{M}_{P}^2}{8} \int \, d^{3} x \int \, d \tau \,\,c_{2}(\tau)\, \biggl[ n^2(\tau)\, \partial_{\tau} h_{i\, j} \partial_{\tau} h^{i\, j} -  \, \partial_{k} h_{i \, j} \partial^{k} h^{i\, j} 
- m_{c}^2 \frac{c_{3}(\tau)}{c_{2}(\tau)}\,h_{i \, j} h^{i \, j} \biggr].
\label{APPB7}
\end{equation}
We may now introduce the $\eta$-time defined as 
$n(\eta) \, d\eta = d\tau$;  Eq. (\ref{APPB7}) then takes the same form 
of Eq. (\ref{APPB2}) with the difference that $b^2(\eta)$ is now replaced 
by $\overline{b}^2(\eta)$ where $\overline{b}(\eta) = \sqrt{c_{2}(\eta)\, n(\eta)}$. It turns out, however, that  $\overline{b}(\eta)$ and  $b(\eta)$ coincide since, ultimately, $\overline{b}(\eta) = b(\eta) = [ c_{1}(\eta)\, c_{2}(\eta)]^{1/4}$. In the $\eta$-time parametrization the evolution of the mode functions of Eq. (\ref{APPB5})
is exactly solvable in various situations of practical interest. For instance if $b(\eta)$ 
is given by Eq. (\ref{STHR6}) the solution of Eq. (\ref{APPB5}) is simply given by
\begin{equation}
f_{k}(\eta) = \frac{{\mathcal M}}{\sqrt{ 2 k}} \, \sqrt{- k \eta} \, H_{\mu}^{(1)}(- k\, \eta), \qquad \qquad g_{k}(\eta) = - {\mathcal M}\,\sqrt{\frac{k}{2}}  \,\sqrt{- k\,\eta} \, H_{\mu-1}^{(1)}(- k \eta), 
\label{APPB8}
\end{equation}
where $\mu = \zeta +1/2$ and $H_{\mu}^{(1)}(-k\eta)$ is the Hankel function of first kind \cite{abr1} 
and  $|{\mathcal M}| = \sqrt{\pi/2}$.  From Eq. (\ref{APPB8}) the tensor power spectrum becomes:
\begin{equation}
P_{T}(k, \eta) = \frac{4\, k^{3}}{\pi^2 \, \overline{M}_{P}^2\,b^2(\eta)} \bigl| f_{k}(\eta) \bigr|^2 =
\biggl(\frac{H_{1}}{M_{P}}\biggr)^2\,\, {\mathcal B}(n_{T}, N_{*}, N_{t}, \epsilon) \,\, \biggl(\frac{k}{a_{1} H_{1}}\biggr)^{n_{T}},
 \label{APPB9}
 \end{equation}
 where we remind that $\overline{M}_{P} = M_{P}/\sqrt{8 \,\pi}$; in Eq. (\ref{APPB9}) we also introduced, for practical reasons,  the following rescaled amplitude 
 \begin {equation}
{\mathcal B}(n_{T}, N_{*}, N_{t}, \epsilon) = \frac{ 2^{6 - n_{T}}}{\pi^2} \biggl| 1 + \frac{\alpha}{1 -\epsilon}\biggr|^{2 - n_{T}}\, \Gamma^2\biggl(\frac{3 -n_{T}}{2}\biggr)e^{\alpha\, N_{\ast}( 3 - n_{T}) - n_{T} (N_{\ast} - N_{t})}.
 \label{APPB10}
 \end{equation}
The result of  Eq. (\ref{APPB9}) corresponds 
to the limit $| k \eta| \ll 1$ where $P_{T}(k, \eta)$ becomes constant in the $\eta$-time with $r_{T}(\nu)$ given by
\begin{equation}
r_{T}(\nu) = \frac{\epsilon}{\pi} \, {\mathcal B}(n_{T}, N_{*}, N_{t}, \epsilon) \biggl(\frac{\nu}{\nu_{max}}\biggr)^{n_{T}}, \qquad\qquad \lim_{\alpha\to 0} {\mathcal C}(n_{T}, N_{*}, N_{t}, \epsilon) = 16\pi.
\label{APPB11}
\end{equation}
For $\alpha\to 0$ we have that $r_{T}\to 16 \epsilon$ and the standard consistency 
relation is recovered; note that $n_T$ is given exactly by the expression already mentioned in Eq. (\ref{STHR9}); therefore in the limit $\alpha\to 0$ we have that $ n_{T}\to- 2\epsilon$.  The notations employed here imply that the definition of the power spectrum $P_{T}(k,\eta)$ 
(and equally for $P_{T}(k,\tau)$) directly follow from the quantum mechanical normalization of the corresponding 
field operators. In particular we have that 
\begin{equation}
\widehat{h}_{i\,j}(\vec{x}, \, \eta) = \frac{\sqrt{2}}{\overline{M}_{P}(2\,\pi)^{3/2}\,b^2(\eta)} \sum_{\alpha= \oplus,\otimes} e^{(\alpha)}_{i\, j}(\hat{k}) \int\, d^{3} k\, \biggl[ \widehat{a}_{\vec{k},\,\alpha}\, f_{k,\alpha}(\eta) e^{- i \vec{k}\cdot\vec{x}} +
 \widehat{a}_{\vec{k},\,\alpha}^{\dagger}\, f_{k,\alpha}^{\ast}(\eta) e^{ i \vec{k}\cdot\vec{x}} \biggr],
 \label{APPB12}
 \end{equation}
 where $[ \widehat{a}_{\vec{k},\, \alpha}, \widehat{a}_{\vec{p},\, \beta}^{\dagger} ] = \delta^{(3)}(\vec{k}- \vec{p})$; 
 the two tensor polarizations are $e^{\oplus}(\hat{k}) = (\hat{m}_{i}\, \hat{m}_{j} - \hat{n}_{i} \hat{n}_{j})$ and 
 $e^{\otimes}(\hat{k}) = (\hat{m}_{i}\, \hat{n}_{j} + \hat{n}_{i} \hat{m}_{j})$ where $\hat{m}$, $\hat{n}$ and $\hat{k}$ 
 form a triplet of mutually orthogonal unit vectors. From Eq. (\ref{APPB12}) we can compute 
 \begin{equation}
 \langle \widehat{h}_{i\,j}(\vec{x}, \, \eta)\, \widehat{h}_{i\,j}(\vec{x} + \vec{r}, \, \eta)\rangle =
 \int d \ln{k}\,\, P_{T}(k,\eta)\, j_{0}(k\, r),\qquad j_{0}(k\, r) =  \sin{k\, r}/(k\, r).
 \label{APPB13}
 \end{equation}
As anticipated, the tensor power spectrum of Eq. (\ref{APPB13}) is defined as in Eq. (\ref{APPB9}). For a full quantum mechanical discussion of this class of problems problem see, for instance, Ref. \cite{qq}.

\end{appendix}

\end{document}